\documentclass[twocolumn]{aastex701}

\newcommand{\mystar}{GJ\,9827~}
\newcommand{\myplanet}{GJ\,9827\,d~}

\usepackage[version=4]{mhchem}
\usepackage{subfigure}
\usepackage[T1]{fontenc}

\begin{document}

\title{Missing Carbon on a sub-Neptune: The Carbon-poor, High-Metallicity Atmosphere of GJ\,9827\,d}

\newcommand{\umontreal}{Department of Physics and Trottier Institute for Research on Exoplanets, Universit\'{e} de Montr\'{e}al, Montreal, QC, Canada}
\defcitealias{piaulet-ghorayeb_jwstniriss_2024}{PG24}
\defcitealias{roy_water_2023}{R23}
\newcommand{\citeA}[1]{\hyperlink{cite.#1}{\citealias{#1}}}
\newcommand{\citeAp}[1]{\hyperlink{cite.#1}{(\citealias{#1})}}

\correspondingauthor{Caroline Piaulet-Ghorayeb}
\email{carolinepiaulet@uchicago.edu}

\author[0000-0002-2875-917X]{Caroline Piaulet-Ghorayeb}
\altaffiliation{51 Pegasi b Fellow}
\affiliation{Department of Astronomy \& Astrophysics, University of Chicago, 5640 South Ellis Avenue, Chicago, IL 60637, USA}
\email{carolinepiaulet@uchicago.edu}

\author[0009-0000-1424-7694]{Nicole Gromek}
\affil{Department of Physics \& Astronomy, McMaster University, 1280 Main St W, Hamilton, ON, L8S 4L8, Canada}
\email{gromekn@mcmaster.ca}

\author[0009-0009-1816-1586]{Jacob Weglarz}
\affiliation{Department of Astronomy \& Astrophysics, University of Chicago, 5640 South Ellis Avenue, Chicago, IL 60637, USA}
\email{jweglarz@umd.edu}

\author[0009-0009-6098-296X]{Bo Peng}
\affil{Department of Earth \& Planetary Sciences, Stanford University, Stanford, CA, USA}
\email{bpengeps@stanford.edu}

\author[0000-0002-0692-7822]{Tyler R. Fairnington}
\affiliation{Department of Astronomy \& Astrophysics, University of Chicago, 5640 South Ellis Avenue, Chicago, IL 60637, USA}
\email{tfairnington@uchicago.edu}

\author[0009-0002-2380-6683]{Eshan Raul}
\affil{Department of Astronomy, University of Wisconsin-Madison, 475 N Charter St, Madison, WI 53706, USA}
\email{eraul@wisc.edu}

\author[0000-0002-0747-8862]{Yamila Miguel}
\affil{Leiden Observatory, Leiden University, P.O. Box 9513, 2300 RA Leiden, The Netherlands}
\affil{SRON Netherlands Institute for Space Research, Niels Bohrweg 4, 2333 CA Leiden, The Netherlands}
\email{ymiguel@strw.leidenuniv.nl}

\author[0000-0002-8466-5469]{Collin Cherubim}
\altaffiliation{NHFP Sagan Fellow}
\affiliation{Department of Astronomy \& Astrophysics, University of Chicago, 5640 South Ellis Avenue, Chicago, IL 60637, USA}
\email{collinc@uchicago.edu}

\author[0009-0007-1339-962X]{Emily Mirkin}
\affiliation{The Masters School, 49 Clinton Ave, Dobbs Ferry, NY 10522, USA}
\email{Emily.Mirkin@mastersny.org}

\author[0000-0003-0973-8426]{Eva-Maria Ahrer}
\affil{Max Planck Institute for Astronomy, K\"{o}nigstuhl 17, D-69117 Heidelberg, Germany}
\email{ahrer@mpia.de}

\author[0000-0002-3191-2200]{Amy Louca}
\affil{Leiden Observatory, Leiden University, P.O. Box 9513, 2300 RA Leiden, The Netherlands}
\affil{SRON Netherlands Institute for Space Research, Niels Bohrweg 4, 2333 CA Leiden, The Netherlands}
\email{louca@strw.leidenuniv.nl}

\author[0000-0002-1551-2610]{Jeehyun Yang} 
\affiliation{Department of Astronomy \& Astrophysics, University of Chicago, 5640 South Ellis Avenue, Chicago, IL 60637, USA}
\email{jeehyuny@uchicago.edu}

\author[0000-0001-8236-5553]{Matthew C.\ Nixon}
\altaffiliation{51 Pegasi b Fellow}
\affiliation{School of Earth \& Space Exploration, Arizona State University, Tempe, AZ 85287, USA}
\affiliation{Department of Astronomy, University of Maryland, College Park, MD, USA}
\email{mcnixon@umd.edu}

\author[0000-0001-5383-9393]{Ryan Cloutier}
\affiliation{Department of Physics \& Astronomy, McMaster University, 1280 Main St W, Hamilton, ON, L8S 4L8, Canada}
\email{ryan.cloutier@mcmaster.ca}

\author[0000-0002-1337-9051]{Eliza Kempton}
\affiliation{Department of Astronomy \& Astrophysics, University of Chicago, 5640 South Ellis Avenue, Chicago, IL 60637, USA}
\email{ekempton@uchicago.edu}

\author[0000-0003-4816-3469]{Ryan J. MacDonald}
\affiliation{School of Physics and Astronomy, University of St Andrews, North Haugh, St Andrews, KY16 9SS, UK}
\email{ryan.macdonald@st-andrews.ac.uk}

\author[0000-0003-0156-4564]{Luis Welbanks} 
\affil{School of Earth \& Space Exploration, Arizona State University, Tempe, AZ 85287, USA}
\email{luis.welbanks@asu.edu}

\author[0000-0002-0298-8089]{Hilke E. Schlichting}
\affil{Department of Earth, Planetary, and Space Sciences, University of California, Los Angeles, Los Angeles, CA 90095, USA}
\email{hilke@g.ucla.edu}

\author[0000-0003-0605-0263]{Marie-Luise Steinmeyer}
\affil{Institute for Particle Physics and Astrophysics, ETH Zurich, CH-8093 Zurich, Switzerland}
\email{msteinmeyer@phys.ethz.ch}

\author[0009-0005-1133-7586]{Aaron Werlen} 
\affil{Department of Earth, Planetary, and Space Sciences, University of California, Los Angeles, Los Angeles, CA 90095, USA}
\email{werlen@ucla.edu}

\author[0000-0001-6110-4610]{Caroline Dorn}
\affil{Institute for Particle Physics and Astrophysics, ETH Zurich, CH-8093 Zurich, Switzerland}
\email{dornc@ethz.ch}

\author[0000-0002-1299-0801]{Edward D. Young} 
\affil{Department of Earth, Planetary, and Space Sciences, University of California, Los Angeles, Los Angeles, CA 90095, USA}
\email{eyoung@epss.ucla.edu}

\author[0000-0002-2195-735X]{Louis-Philippe Coulombe}
\affiliation{Plan\'{e}tarium de Montr\'{e}al, Espace pour la Vie, 4801 av. Pierre-de Coubertin, Montr\'{e}al, Canada}
\affiliation{Trottier Institute for Research on Exoplanets, Department of Physics, Universit\'{e} de Montr\'{e}al, Montr\'{e}al, Canada}
\email{louis-philippe.coulombe@umontreal.ca}

\author[0000-0001-6809-3520]{Pierre-Alexis Roy} 
\affil{Department of Earth, Planetary, and Space Sciences, University of California, Los Angeles, Los Angeles, CA 90095, USA}
\email{paroy@g.ucla.edu}

\author[0000-0001-5578-1498]{Bj\"{o}rn Benneke} 
\affil{Department of Earth, Planetary, and Space Sciences, University of California, Los Angeles, Los Angeles, CA 90095, USA}
\email{bbenneke@ucla.edu}

\author[0009-0003-1219-2193]{Phoebe Burgis} 
\affiliation{Department of Astronomy \& Astrophysics, University of Chicago, 5640 South Ellis Avenue, Chicago, IL 60637, USA}
\email{pburgis@uchicago.edu}

\author[0000-0001-9291-5555]{Charles Cadieux} 
\affiliation{Observatoire de Gen\`eve, D\'epartement d’Astronomie, Universit\'e de Gen\`eve, Chemin Pegasi 51, 1290 Versoix, Switzerland}
\email{Charles.Cadieux@unige.ch}

\author[0000-0001-9137-9818]{Daria Kubyshkina}
\affil{Space Research Institute, Austrian Academy of Sciences, Schmiedlstrasse 6, 8042 Graz, Austria}
\affil{Space Research and Planetary Sciences, Physics Institute, University of Bern, Gesellschaftsstrasse 6, 3012 Bern, Switzerland}
\email{daria.kubyshkina@unibe.ch}

\author[0000-0001-5442-1300]{Thomas M. Evans-Soma}
\affiliation{School of Science, University of Newcastle, University Drive, Callaghan NSW 2308, Australia}
\affiliation{Max Planck Institute for Astronomy, K\"{o}nigstuhl 17, D-69117 Heidelberg, Germany}
\email{Tom.Evans-Soma@newcastle.edu.au}

\author[0000-0002-5887-1197]{Raymond Pierrehumbert}
\affil{University of Oxford, Department of Physics Oxford, OX1 3PW, UK}
\affil{Department of Earth and Planetary Science, Massachusetts Institute of Technology, Cambridge, MA 02139}
\email{raymond.pierrehumbert@physics.ox.ac.uk}

\author[0000-0002-3328-1203]{Michael Radica} 
\affiliation{Department of Astronomy \& Astrophysics, University of Chicago, 5640 South Ellis Avenue, Chicago, IL 60637, USA}
\email{radicamc@uchicago.edu}

\author[orcid=0009-0007-9356-8576,gname=Cyril,sname=Gapp]{Cyril Gapp}
\affiliation{University Observatory Munich, Faculty of Physics, Ludwig-Maximilians-Universität München, Scheinerstr. 1, 81679 Munich, Germany}
\affiliation{Munich Center for Geoastronomy, Ludwig-Maximilians-Universität München, Theresienstrasse 41, 80333, Munich, Germany}
\email{cyril.gapp@lmu.de}

\author[0000-0002-1199-9759]{Romain Allart}
\affiliation{Department of Physics and Trottier Institute for Research on Exoplanets, Universit\'{e} de Montr\'{e}al, Montréal, QC, Canada}
\email{romain.allart@umontreal.ca}  

\author[0000-0002-5113-8558]{Daniel P. Thorngren}
\affiliation{Department of Physics and Astronomy, Johns Hopkins University}
\email{dpthorngren@jhu.edu}



\begin{abstract}
Sub-Neptunes are keystone planets, with atmospheres carrying the imprints of evolution through interior interactions and atmospheric escape. The warm ($T_{\rm eq}=675$\,K) planet GJ 9827 d is the smallest ($R_p\simeq2\,R_\oplus$) sub-Neptune characterized to date, with abundant water in a high-metallicity atmosphere. Here we present constraints on carbon-bearing species in the atmosphere of GJ\,9827\,d using two transit observations with JWST/NIRSpec G395H, and elemental abundance measurements for its host star with CFHT/SPIRou. We confirm a high metallicity ([M/H]$>2.4$ to $2.9$; 95\% confidence depending on model assumptions), and constrain a low, substellar C/O ratio (C/O$<0.1$--$0.5$ at 95\% confidence), with no statistically-detected carbon species. This chemical profile stands in stark contrast to the omnipresent carbon-bearing molecules detected on other sub-Neptunes, and is challenging to explain as an outcome of canonical sub-Neptune formation and magma-ocean equilibration. We propose that GJ\,9827\,d's atmosphere could be secondary in nature, with the primary hydrogen-rich atmosphere lost early, leading to the loss of carbon-bearing species entrained in the outflow while water was shielded in the molten mantle. Alternatively, GJ\,9827\,d could have formed carbon-poor, with a large volatile inventory ($\sim 20\%$ atmosphere mass fraction) and an initial atmosphere metallicity high enough for long-term retention ([M/H]$_\mathrm{ini}\sim2.7$, $Z_\mathrm{ini}>0.6$). We also calculate new condensation curves for carbon-bearing species in high-metallicity atmospheres, and explore their potential role in lowering the gas-phase C/O ratio. Further atmospheric characterization of GJ\,9827\,d will be crucial to distinguish between these scenarios, particularly through the abundances of sulfur-bearing species such as SO$_2$, only hinted at by existing observations.
\end{abstract}




\section{Introduction} \label{sec:intro}

Understanding which small rocky planets can retain or revive substantial atmospheres is arguably one of the most important questions on the path to discovering habitable worlds \citep{AstroDecadal2020}. The scientific community has invested tremendous theoretical effort and observational time in answering this question (e.g. \citealp{Dong2018,Schlichting2018,Wyatt2020,kite_exoplanet_2020,krissansen-totton_erosion_2024,ji_cosmic_2025,nguyen_shoreline_2026}), most notably through JWST's 500-hr ``Rocky Worlds'' Director's Discretionary Time program \citep{redfield_report_2024}, aimed at observationally constraining the physical conditions under which atmospheres can be found on temperate exoplanets.

Fundamentally, the fate of small planet atmospheres is dictated by a few key conditions: (1) the initial inventory of gas and volatile ices accumulated by the planet at formation and through later delivery (see \citealp{Wordsworth2022}), (2) the planet's history of stellar irradiation and (potentially fractionated) atmospheric mass-loss \citep{owen_atmospheres_2016,owen_evaporation_2017,rogers_photoevaporation_2021,gupta_sculpting_2019,Gupta_2020,cherubim_oxidation_2025,valatsou_oxygen_2026}, and (3) the balance between the dissolution of volatiles in the planet's interior and their release through outgassing from the mantle into the atmosphere \citep{dorn_hidden_2021,postolec_atmospheric_2026}, interwoven with the planetary thermal evolution history \citep{nicholls_volatile-rich_2025}. These processes are coupled, and largely observationally unconstrained since only the present-day conditions can be probed for each individual system. Rocky-planet studies in the JWST era can so far only provide information regarding the presence or absence of thick atmospheres on rocky planets, while a larger number of systems are required to probe the regulating axes (e.g., irradiation temperature, stellar type, and so on) that likely govern the final atmospheric outcome \citep{ih_jwst_2026}. 

Small sub-Neptunes are shaped by the same set of processes \citep{lichtenberg_constraining_2025} -- but with low densities \textit{ensuring} atmosphere presence \citep{lopez_understanding_2014}. Therefore, they stand out as incredible windows into the underlying physics and chemistry, holding the potential for the identification of the chemical fingerprints of the mechanisms at play through atmosphere characterization (e.g. \citealp{benneke_jwst_2024,nixon_magma_2025,Nicholls2026}). Among the few benchmark members of the small sub-Neptune population where molecular absorption signatures have been measured (Piaulet-Ghorayeb, subm.), the low-density 2R$_\oplus$, $\sim 675$\,K sub-Neptune GJ\,9827\,d \citep{niraula_three_2017,prieto-arranz_mass_2018,rodriguez_system_2018} stands out as the only one where water vapor has been identified from near-infrared HST (\citealp{roy_water_2023}, hereafter \citetalias{roy_water_2023}) and JWST transmission spectroscopy (\citealp{piaulet-ghorayeb_jwstniriss_2024}, hereafter \citetalias{piaulet-ghorayeb_jwstniriss_2024}), in stark contrast with the prevalent methane absorption found on other sub-Neptunes \citep{benneke_jwst_2024,madhusudhan_carbon-bearing_2023,roy_diversity_2025,rigby_jwst_2025}. 

Interior-atmosphere connections are crucial for tracing the formation and evolutionary histories of sub-Neptunes by lifting the composition-bulk density degeneracy between H/He- or volatile-dominated envelopes \citep{rogers_framework_2010,benneke_how_2013,luque_density_2022,rogers_conclusive_2023}. In that context, constraining a planet's water budget from its atmospheric composition is key \citep{yang_chemical_2024,piaulet-ghorayeb_jwstniriss_2024,piaulet-ghorayeb_window_2025}. On GJ\,9827\,d, such inferences on the deep-envelope water budgets can be made because it is hot enough that water should be well mixed with the surrounding hydrogen throughout the entire gaseous-supercritical envelope \citep{piaulet-ghorayeb_window_2025}, following the ``fully-miscible'' sub-Neptune envelope hypothesis \citep{pierrehumbert_runaway_2023,innes_runaway_2023,benneke_jwst_2024}. 

The greenhouse effect from the thick ($\sim 100$\,kbar for GJ\,9827\,d; \citetalias{piaulet-ghorayeb_jwstniriss_2024}) envelopes of sub-Neptunes is expected to result in widespread molten surfaces (\citealp{calder_most_2025,tang_reassessing_2024,nicholls_constraining_2026}, but see also \citealp{breza_not_2025,piaulet-ghorayeb_strict_2025}). The molten state of the mantle impacts interpretations of the planetary structure \citep{dorn2021hidden}, but also provides fertile ground for sustained chemical interactions between the mantle and atmosphere, altering gas abundances \citep{kite_water_2021,schlichting_chemical_2022}. In particular, while water is extremely soluble in molten silicates \citep{lichtenberg_bifurcation_2021}, carbon-bearing species are orders-of-magnitude less susceptible to being stored in the planetary interior -- making them arguably more reliable atmospheric tracers of the planet's volatile inventory and ultimately, its formation building blocks.

The low density and radius-valley-straddling status of GJ\,9827\,d makes it one of the best targets to test the sub-Neptune composition hypothesis, with large mass-loss rates expected from bespoke models assuming a hydrogen-dominated atmosphere \citep{gupta_caught_2021}. Instead, the strict upper limits on H and He escape from ground- and space-based observations (\citealp{kasper_non-detection_2020,carleo_multiwavelength_2021,krishnamurthy_absence_2023}, \citetalias{piaulet-ghorayeb_jwstniriss_2024}) are in line with predictions for a warm metal-rich atmosphere \citep{kubyshkina_escape_2026}. Atmospheric metal enrichment is independently supported by the JWST/NIRISS SOSS transmission spectrum of GJ\,9827\,d, which shows multiple water absorption features whose reduced amplitudes are consistent with a high mean molecular weight atmosphere \citepalias{piaulet-ghorayeb_jwstniriss_2024}. However, the limited wavelength coverage of the NIRISS instrument, combined with strong water opacity overlapping with methane absorption $<3\mu$m, has so far precluded meaningful constraints on the relative amounts of carbon- vs. oxygen-bearing molecules. 

In this study, we analyze two additional transits of GJ\,9827\,d observed with JWST's NIRSpec/G395H instrument which allow us to constrain its carbon chemistry for the first time, and interpret the complete near-infrared transmission spectrum in terms of the atmosphere and interior composition. We describe the atmosphere observations and composition inference in Section \ref{sec:transspec}, the characterization of the stellar abundances in Section \ref{sec:stellar_abundances}, and compare the results to magma ocean-atmosphere chemistry models in Section \ref{sec:structure_modeling}. We discuss the interpretation of our results in Section \ref{sec:discussion}, and summarize our findings in Section \ref{sec:conclusion}.




\section{Transmission spectroscopy} \label{sec:transspec}

\subsection{Observations}\label{sec:obs}

Our dataset combines transit observations of \myplanet with HST/WFC3 (GO 15333: PI Crossfield; \citetalias{roy_water_2023}), and with JWST NIRISS SOSS and NIRSpec/G395H (GO 4098, PIs: Benneke \& Evans-Soma). We analyze two newly-obtained NIRSpec transits observed on June 8 and Nov 28, 2024, with the already-published HST and JWST/NIRISS data \citepalias{piaulet-ghorayeb_jwstniriss_2024}. Taken together, these observations cover the near-infrared regime from 0.6 to 5.2$\mu$m.

For each of the NIRSpec time series observations, we recorded 2753 integrations, reading 4 groups up the ramp and totalling 3.46hr, i.e. approximately 1.22hr in transit and 2.24hr out of transit. Due to the brightness of the target (\mystar has $J_\mathrm{mag}=7.984$), we performed target acquisition on a nearby faint $J_\mathrm{mag}=15.587$ M dwarf with the NRSRAPID readout pattern. The science observations used the G395H/F290LP grating/filter combination, the NRSRAPID readout mode, and the SUB2048 subarray. 

\subsection{Data Reduction}\label{sec:datared}

We reduce the JWST/NIRSpec data with two independent pipelines: \texttt{exoTEDRF} \citep{radica_exotedrf_2024} and \texttt{Eureka!} \citep{bell_eureka_2022}. The details of the data reduction choices made when running each pipeline are described below.

\subsubsection{\texttt{exoTEDRF} reduction}\label{sec:exotedrf}

We reduce the uncalibrated detector products using \texttt{exoTEDRF} \citep{radica_awesome_2023, feinstein_early_2023, radica_exotedrf_2024}. Our reduction procedure follows that of \citet{radica_super-solar_2026} and \citet{Fairnington2026} to which we refer the reader for full details. Briefly, rather than tuning reduction parameters by hand, we employ an automated coordinate-descent optimizer (Fairnington et al., in prep.) that steps through each parameter sequentially over a user-defined grid, selecting values that minimize the point-to-point scatter in provisional spectroscopic light curves. The optimized parameters include the jump-detection threshold, trace and background mask widths used during 1/$f$ correction, spatial and temporal outlier-rejection thresholds in the bad-pixel interpolation, and the extraction aperture width. We apply the \texttt{median} 1/$f$-correction method, flag cosmic rays using the time-domain routine of \citet{radica_muted_2024}, and perform a box aperture extraction with optimized widths of 7 and 5 pixels for NRS1 and NRS2 for the first visit, and 8 and 7 pixels for the second.

\subsubsection{\texttt{Eureka!} reduction}\label{sec:eureka}
For an independent data reduction we use \texttt{Eureka!} \citep[v1.2.1;][]{bell_eureka_2022}. Our reduction procedure follows closely that of \citet{ahrer_escaping_2025}. In short, we use Stages\,1 \& 2 from the \texttt{jwst} pipeline \citep[v1.18.0; pmap 1364;][]{bushouse_2023} with the default steps enabled. We perform column-by-column background subtraction at group-level to reduce 1/$f$ noise. 

We extract the time-series spectra in \texttt{Eureka!}'s Stage\,3. We conduct an additional column-by-column background subtraction by fitting a constant to the area of $>8$\,pixel from the trace center pixel (doing reiterative $>5\times$ Median Absolute Element Difference of consecutive elements masking until there is no outlier in the relevant background column anymore). We then also mask outliers at $>7\sigma$ along the time axis. To get the time-series spectra we use optimal spectral extraction \citep{Horne1986optimalextraction} and a 3\,pixel half-width (7\,pixels full width) for the aperture size.

\subsection{ExoTEP light curve fitting}\label{sec:lc_analysis}

We use ExoTEP \citep{benneke_spitzer_2017,wong_tess_2020} to perform light curve fitting on the time series of extracted spectra obtained as Stage 3 outputs from the \texttt{exoTEDRF} and \textit{Eureka!} reductions. Following previous work \citepalias{piaulet-ghorayeb_jwstniriss_2024}, we apply the same light curve fitting methodology and model to both data reductions. In this approach, any differences in the resulting spectra can be traced down to differences in the processing of the detector images; we find that the spectra agree consistently throughout the entire wavelength range (Fig. \ref{fig:compare_reductions}).

\subsubsection{Broadband light-curve fitting}

\begin{figure*}
    \centering
    \includegraphics[width=0.95\textwidth]{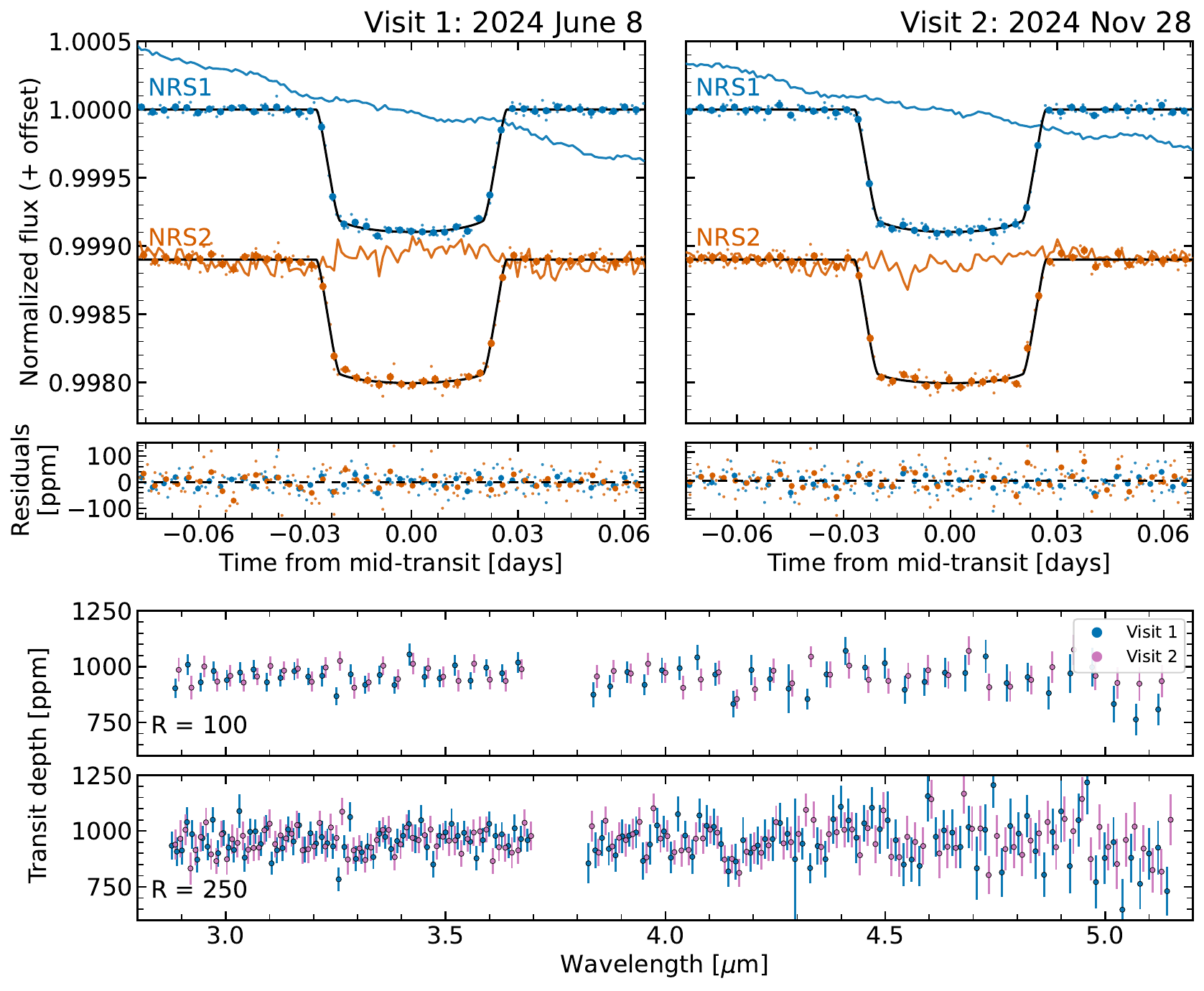}
    \vspace{5mm}
    \vspace{-5mm}\caption{Broadband and spectroscopic light curve fitting results for the NIRSpec/G395H transits of GJ\,9827\,d obtained with the \texttt{exoTEDRF} data reduction. The top panels show systematics-corrected broadband NRS1 (blue) and NRS2 (orange) light curves (binned and offset for clarity), along with the corresponding best-fit astrophysical (black) and systematics (blue and orange lines) models. The residuals are shown under each light-curve panel and show no residual systematics. 
    The bottom two panels compare the spectra measured for each transit (visit 1 in blue, visit 2 in pink), with a small horizontal offset for easier comparison. The spectra are consistent across the two spectral resolutions (top: $R=100$; bottom: $R=250$) and for the two visits.}
    \label{fig:wlc_fit}
\end{figure*}

We first perform broadband light-curve fits to constrain GJ\,9827\,d's orbital parameters, transit times, and time-dependent instrumental and astrophysical systematics (Fig. \ref{fig:wlc_fit}); see Appendix Section \ref{sec:appendix_whitelc}). The broadband light curves span 2.88--3.70 $\mu$m for the NRS1 detector and 3.82--5.16 $\mu$m for NRS2. The measured transit times (Table~\ref{table:wlc_param}) support substantial deviations from a linear ephemeris (Fig. \ref{fig:ttv_data}), which is in line with transit-timing variations (TTVs) already suggested by HST observations (\citetalias{roy_water_2023}). Our dynamical simulations (see Appendix Section \ref{sec:dynamical_analysis}) reveal that such small TTVs on GJ\,9827\,d, which is near second- and fourth-order mean motion resonance (Table~\ref{table:mmr_summary}) with the other two known transiting planets, imply a small eccentricity of $e_d\simeq0.07-0.085$ (Table \ref{tab:eccscan}) -- motivating more detailed future photodynamical analyses of this system. We check for a potential inclination variation between the two NIRSpec visits separated by 173.6 days by leveraging their precise measurements of transit duration, and detect no significant difference. This suggests small mutual inclinations in the system of $\lesssim 2^\circ$, (Appendix Section \ref{sec:dynamical_analysis}), compatible with the population of compact multiplanetary systems revealed by \textit{Kepler} \citep{Fabrycky2014,millholland_split_2021}.

\subsubsection{Spectroscopic light-curve fitting}

We generate two versions of the planetary spectrum, one at a resolving power of 100 which is closest to the version presented for the other sub-Neptunes in this program as well as the NIRISS SOSS observations of GJ 9827 d \citep{benneke_jwst_2024,ahrer_escaping_2025,piaulet-ghorayeb_jwstniriss_2024,coulombe_possible_2025}, and another one at a higher $R=250$ as a sensitivity test in order to confirm that our atmosphere inference results are consistent regardless of the selected resolution. The orbital parameters and transit times are fixed to their best-fitting values in the broadband light curve fit, and we leverage the constraints on instrumental and stellar variability-induced light curve systematics to fit each visit's transmission spectrum (see Appendix \ref{sec:appendix_speclc}).

The two NIRSpec visits produce consistent spectra with no evidence of visit-dependent transit light source effect signatures (see Fig. \ref{fig:wlc_fit}). We therefore show the weighted-average of these spectra throughout the paper when comparing to atmosphere models. Both Eureka! and \texttt{exoTEDRF} produce spectra that are compatible at the $<1\sigma$ level throughout the wavelength range probed (Fig. \ref{fig:compare_reductions}), and use the \texttt{exoTEDRF} spectrum in the following analyses for consistency with the nominal NIRISS data reduction (\citetalias{piaulet-ghorayeb_jwstniriss_2024}). 




\subsection{Atmosphere retrievals} \label{sec:retrievals}

We obtain atmospheric composition constraints from the combined near-infrared transmission spectrum of GJ 9827 d, including the HST+NIRISS data (\citetalias{roy_water_2023}, \citetalias{piaulet-ghorayeb_jwstniriss_2024}) as well as our NIRSpec/G395H spectrum using the SCARLET, \texttt{POSEIDON}, and \textit{Aurora} retrieval frameworks. The details of the atmosphere modeling and retrieval implementation in each model are presented in Section \ref{sec:appendix_retrievals} and summarized below.

\begin{figure*}
    \centering
    \includegraphics[width=0.99\textwidth]{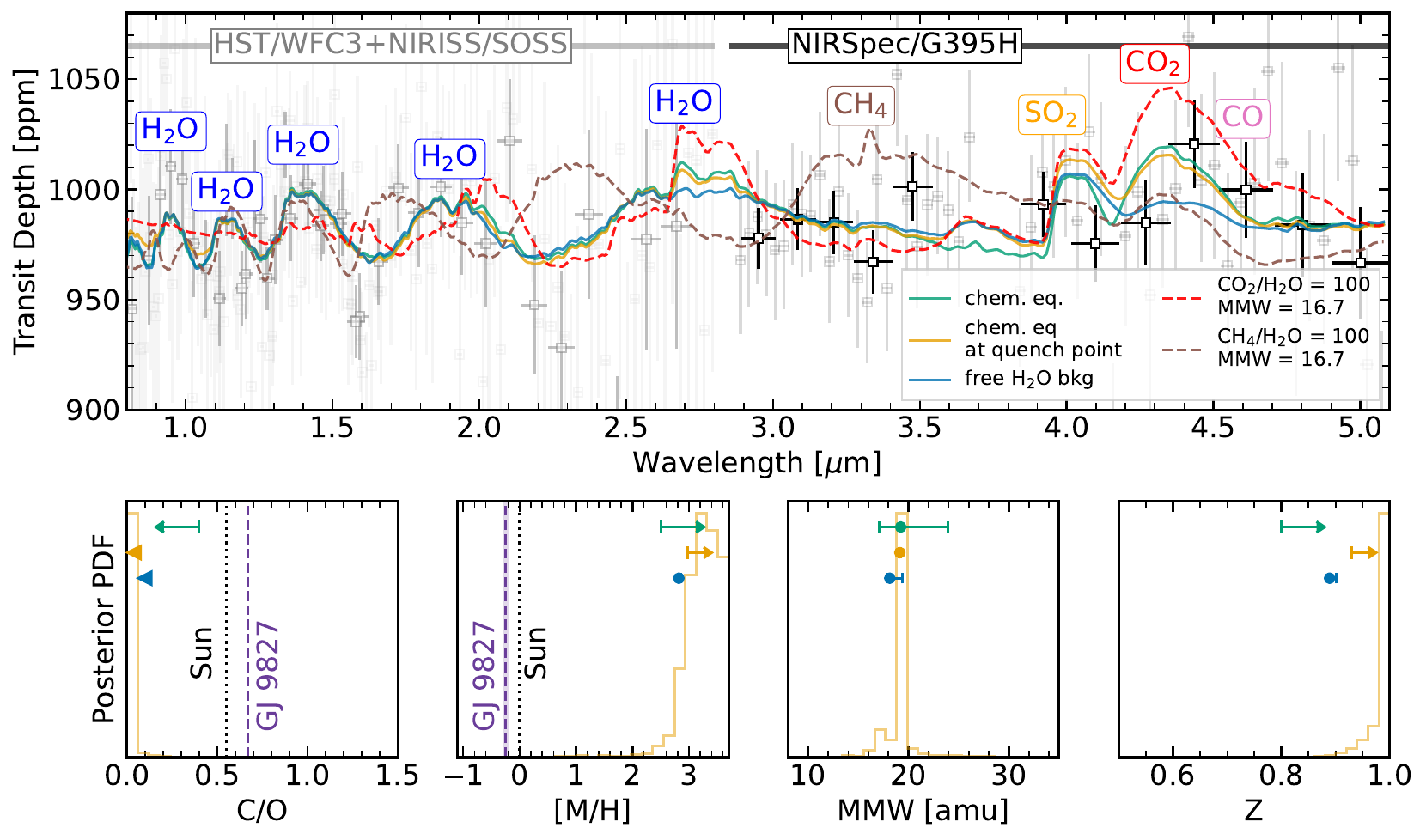}
    \vspace{5mm}
    \vspace{-5mm}\caption{Inference of a low-C/O, high-metallicity composition for GJ\,9827\,d's atmosphere from its 0.6-to-5.2$\mu$m transmission spectrum. \textit{Top panel:} Observed HST/WFC3 and NIRISS/SOSS (gray squares) and \texttt{exoTEDRF} NIRSpec/G395H (black squares), binned to $R\sim25$ for clarity. The $R=100$ transmission spectrum on which the main set of retrievals was run is shown in the background. Prominent molecular features are labeled and representative SCARLET models are overlaid. We show the best-fit models from three retrievals: the chemical equilibrium retrieval (green), the chemically-consistent retrieval including quenching (yellow) and the free retrieval with H$_2$O as the background gas (blue). The dashed lines are models adapted from the best-fit quenched-chemistry retrieval, keeping the mean molecular weight (MMW) constant, but varying the CO$_2$/H$_2$O ratio (red) or the CH$_4$/H$_2$O ratio (brown). \textit{Bottom panels:} Posterior (or derived) probability distributions, and confidence intervals (68\% C.I. limits or 2$\sigma$ upper/lower limits as arrows) on the C/O ratio, metallicity, MMW and metal mass fraction of GJ\,9827\,d's atmosphere. The colors match the models shown in the top panel. The metallicity constraint is shown as [M/H], calculated as [(C+O)/H] for free retrievals, with stellar and solar values annotated in black and purple. Across all tested prescriptions, we robustly recover a low substellar C/O ratio (C/O$<$0.1-0.5 at 95\% confidence across model assumptions; Table \ref{tab:retrieval_results}), and a high $\gtrsim 300\times$solar metallicity. Besides water, no other molecule is statistically detected in the spectrum.}
    \label{fig:retrieval_spectrum_fit}
\end{figure*}

The SCARLET retrievals \citep{benneke_atmospheric_2012,benneke_how_2013,benneke_strict_2015,benneke_sub-neptune_2019, benneke_water_2019, pelletier_where_2021, piaulet_evidence_2023,piaulet-ghorayeb_jwstniriss_2024,piaulet-ghorayeb_strict_2025} are performed assuming either free chemistry (no chemical constraint on the plausible relative amounts of each atmosphere component), chemical equilibrium, or chemical consistency but with quenching occuring deeper than the observations probe (see Section \ref{sec:scarlet} for the detailed description). SCARLET retrieval results across the main set of model variations are presented in Figs. \ref{fig:retrieval_spectrum_fit}, \ref{fig:nominal_retrievals_detailed_constraints}, and \ref{fig:retrieval_variations}, and summarized in Table \ref{tab:retrieval_results}.

As a consistency check and a test of the robustness of our assumptions, we perform additional retrievals with POSEIDON \citep{macdonald_hd_209458b_2017, macdonald_poseidon_2023} and \textit{Aurora} \citep{Welbanks2021,pinhas_retrieval_2018,Welbanks2019a,Nixon2020,Nixon2022,Nixon2024}. POSEIDON retrievals (see Section \ref{sec:poseidon}) use a different prior that allows any of the species included in the atmosphere model to act as the background gas with equal probability, rather than the implied bias toward a user-specified background gas in traditional free-chemistry retrievals. The set of POSEIDON retrievals also explores potential stellar contamination, and tests for the significance of the water detection in the combined spectrum. One free-chemistry retrieval is run with \textit{Aurora}, with H$_2$/He as the background gas, and CS$_2$ additionally included in the set of fitted atmosphere gases (see Section \ref{sec:aurora}).

\subsection{Atmospheric composition inference} \label{sec:retrieval_comparison}

We find remarkable agreement on atmospheric composition inferences across both free and chemically-consistent retrievals, which constrain a low sub-solar and sub-stellar C/O ratio (C/O$<$0.1-0.5 at 95\% confidence across model assumptions), and a high $\gtrsim 300\times$ solar metallicity (Fig. \ref{fig:retrieval_spectrum_fit}).

All prior configurations favor high water abundances, but the low amplitude of potential CH$_4$/CO$_2$ features in such a high metallicity atmosphere leave abundance ratios somewhat dependent on the chemistry prior (Table \ref{tab:retrieval_results}). For free retrievals, water appears to be more abundant than both CH$_4$ and CO$_2$, which have strong absorption features in the near-infrared covered by our observations (Fig. \ref{fig:retrieval_spectrum_fit}). However, chemical equilibrium retrievals predict comparable amounts of CO$_2$, CH$_4$, and water (Fig. \ref{fig:nominal_retrievals_detailed_constraints}), in agreement with the data with indistinguishable best-fit model log-likelihood values. Quenched-chemistry retrievals show intermediate conclusions: when given the added flexibility of following chemical equilibrium at a higher pressure and temperature in the atmosphere, retrieval results converge to a depletion in CH$_4$ and CO$_2$ by $\gtrsim 10\%$ relative to H$_2$O. 

The results from the POSEIDON and Aurora retrievals match the conclusions of high metallicity, but low abundances of carbon-bearing species relative to water. The POSEIDON retrieval setup, using CLR priors on the abundances of individual molecules (i.e. agnostic about which is the background gas), recovers a similar abundance pattern to those inferred from the suite of SCARLET retrievals (Table \ref{tab:retrieval_results}): the inferred metallicity is high ($\log_{10}$ Met [$\times$ solar] $= 2.83^{+0.33}_{-0.02}$, $Z>0.8$, MMW [amu] $18.50^{+7.53}_{-1.58}$), and carbon-bearing species are depleted relative to water ($\log_{10}$ CO$_2$/H$_2$O$=-3.32^{+1.07}_{-6.90}$,  $\log_{10}$ CH$_4$/H$_2$O$=-6.61^{+1.99}_{-4.75}$). Meanwhile, Aurora finds a slightly lower mean molecular weight ($\log_{10}$ Met [$\times$ solar] $= 2.48^{+0.41}_{-0.38}$, $Z=0.75^{+0.13}_{-0.29}$, MMW [amu] $<23$), with the most important difference relative to the other retrieval setups being the lack of N$_2$ in the model, as well as the adopted H$_2$/He-background prior. The retrieved CO$_2$/H$_2$O and CH$_4$/H$_2$O ratios are similarly low ($\log_{10}$ CO$_2$/H$_2$O$=-5.15^{+1.54}_{-4.60}$,  $\log_{10}$ CH$_4$/H$_2$O$=-7.30^{+1.86}_{-2.64}$). Although sulfur-bearing species are not significantly detected, the Aurora posteriors peak at $\log_{10}$ SO$_2=-3.13^{+1.52}_{-2.50}$ (consistent with SCARLET results) and an even higher abundance for CS$_2$ ($\log_{10}$ CS$_2=-1.15^{+0.70}_{-1.50}$), potentially biased by the lack of a spectrally-inactive high mean molecular weight species. 
Since the free retrievals indicate that neither CH$_4$ or CO$_2$ are significantly detected in the spectrum, the CO$_2$/CH$_4$ abundance ratio, proposed to trace deep-envelope H$_2$O/H$_2$ \citep{yang_chemical_2024,roy_diversity_2025}, remains unconstrained by our observations for GJ\,9827\,d. We stress that further chemical modeling of water-enriched atmospheres is necessary to benchmark chemical networks and photochemistry in this regime. 

While SO$_2$ is not detected, our free retrieval constraints favor volume mixing ratios anywhere from a few ppm to a few percent (Table \ref{tab:retrieval_results}). Given the high atmosphere metallicity and low C/O ratio inferred for the planet, even chemically-consistent retrievals predict appreciable amounts of SO$_2$ in the atmosphere (in agreement with \citealp{mukherjee_effects_2024}), which are likely an underestimate if H$_2$S is converted to SO$_2$ via photochemistry \citep{crossfield_mapping_2025,tsai_photochemically_2023}.

\subsection{Plausibility of atmospheric aerosols} \label{sec:aerosols}

Atmospheric retrievals favor a water-rich, carbon-poor, and high-metallicity atmosphere for GJ~9827~d, yet the relatively low signal-to-noise ratio of the NIRSpec spectrum leaves open the possibility of aerosols suppressing molecular absorption features (see Table \ref{tab:retrieval_results}; \citealp{benneke_how_2013}). Similarly to GJ\,1214\,b \citep{kreidberg_clouds_2014,schlawin_possible_2024,ohno_possible_2024}, high metallicity and atmospheric aerosols may coexist in a high mean molecular weight atmosphere. In particular, the range of C/O ratios allowed by our observations of GJ\,9827\,d (C/O$<0.1$--$0.5$) spans vastly distinct regimes in terms of the resulting hydrocarbon haze (or soot) production rates according to the model predictions of \citet{yang_sub-neptunes_2026}. At [M/H]$=2.5$, their C/O$=0.2$ case is unfavorable for haze production, while their C/O$=0.55$ model produces high enough abundances of haze precursor species to mute the transmission spectrum further than what mean molecular weight alone would produce. Future observations with improved spectral precision and broader wavelength coverage will be essential to break this degeneracy and more robustly constrain the atmospheric C/O ratio of GJ~9827~d.

\section{Stellar characterization}\label{sec:stellar_abundances}

Small-planet formation outcomes are sensitively shaped by their stellar birth environments (e.g. \citealp{brinkman_revisiting_2024,brinkman_compositions_2025,Plotnykov2020,Adibekyan2024,plotnykov_evidence_2025,Weisserman2026}). We leverage high-resolution archival observations of GJ 9827 with the SPIRou spectrograph (Spectro-Polarim\`{e}tre InfraRouge; \citealp{Donati2018}) at the Canada France Hawaii Telescope (CFHT) and a custom spectral synthesis method to constrain the elemental abundances in the host star's photosphere and in turn use them to inform the potential interior compositions for GJ 9827 d.

\subsection{SPIRou Observations}\label{sec:spirou}

We analyze 23 spectra of GJ 9827 obtained with CFHT/SPIRou, which provides coverage across YJHK bands from 0.98 to 2.35$\mu$m at $R\sim70,000$. 19 spectra were obtained on Aug 31, 2020 as part of Program ID 20BC27 (PI: Sikora, focused on He absorption searches), and the remaining 4 were obtained on Aug 17 and Dec 9, 2024 as part of Program ID 24BC29 (PI: Gromek). The co-added spectrum achieves a signal-to-noise ratio of $>700$ in the H-band, in agreement with ETC expectations.

\begin{figure}
    \centering
    \includegraphics[width=0.45\textwidth]{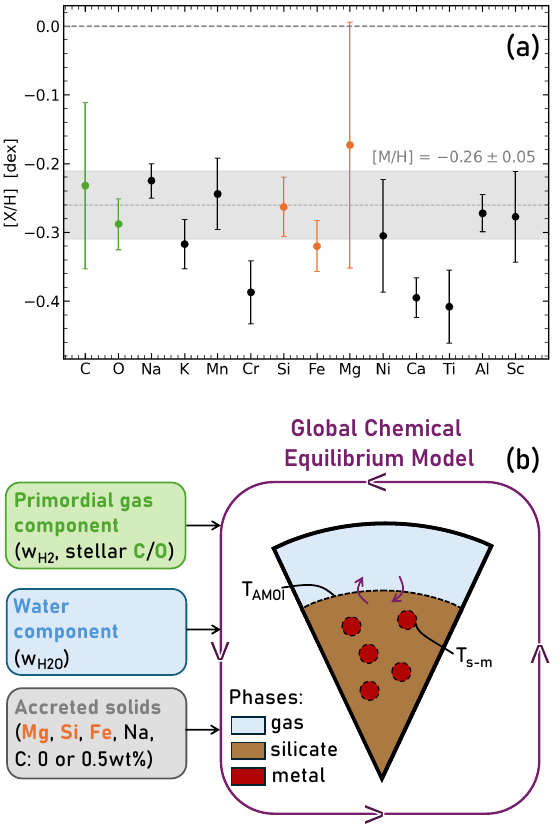}
    \vspace{5mm}
    \vspace{-5mm}\caption{Illustration of the set of stellar abundances derived for GJ 9827 and how they feed into the setup of the global chemical equilibrium model. \textit{Panel (a):} Stellar abundances of GJ\,9827 constrained using the SPIRou high-resolution spectra, ordered by condensation temperature. The semi-transparent band illustrates our reported subsolar metallicity, while the horizontal dashed line corresponds to a solar composition. \textit{Panel (b):} Adopted framework and parameterization for the global chemical equilibrium modeling. The input parameters are indicated in the colored boxes, with the C/O ratio as well as the Mg/Si, Fe/Si elemental ratios informed by the set of measured stellar abundances (matching colors in panel (a). The model solves for the partitioning of elements across three phases, represented in the model on the right. We consider a case where iron is mixed with silicate in the planetary interior, consistent with predictions for sub-Neptunes formed with $\gtrsim 1$wt\%H$_2$.}
    \label{fig:stellar_abundances}
\end{figure}

\subsection{Elemental abundance constraints}

Stellar abundances are notoriously difficult to obtain for late-type stars due to absorption from broad molecular bands which can obscure the continuum, as well as missing or mismatched absorption features in stellar models \citep{jahandar_chemical_2024}. We use a custom pipeline tailored to addressing these outstanding challenges routinely plaguing the elemental abundance measurements of the coolest stars \citep[see][]{Gromek2026}, appropriate for the K7V star GJ\,9827 (\citealp{passegger_compact_2024}; see the full modeling details in Appendix Section \ref{sec:appendix_stellar_abundances}). 

We report the global metallicity and $\alpha$-enhancement of GJ 9827 following the prescription of \citet{hinkel_concise_2022}, and the reference solar abundances are adopted from \citet{asplund_chemical_2021} (see Fig. \ref{fig:stellar_abundances}, Table \ref{tab:stellar_abundances}). We derive a stellar C/O ratio compatible with the solar value (Table \ref{tab:stellar_abundances}), and a subsolar metallicity of [M/H]$=-0.26\pm0.05$. We note that since CO and OH lines are strong and numerous in the near-infrared spectra of late-type stars, CO and OH features are used to estimate the C and O abundances, in place of C I and O I lines (see Appendix Section \ref{sec:appendix_stellar_abundances}). Further, the abundances derived by our pipeline for Ca and Mn are possibly the result of modeling challenges and likely do not reflect the true astrophysical abundances. The wide calcium lines observed for GJ 9827 are poorly matched by models, which leads to unphysically low abundances, while manganese lines are impacted by hyperfine splitting, and line-list fidelity is unfortunately insufficient to provide a good match to the observations. 

Finally, we find that our results are in tension with previous analyses of optical data \citep{sun_jwst_2026}. The previous analysis adopted a much higher $T_\mathrm{eff}$ of 4977\,K, and a lower $\log~g=4.36$ for GJ\,9827, discrepant from literature parameters of this star \citep{kosiarek_physical_2020,passegger_compact_2024,Howard2025}, and the C abundance determination in particular may additionally suffer from the presence of fewer, shallow C$_2$ features in the optical compared to the 176 CO lines used in this work. A more detailed discussion of the difference in approaches and results is presented in \citet{Gromek2026}.

\section{Global interior-atmosphere chemical equilibrium modeling}\label{sec:structure_modeling}

\begin{figure*}
    \centering
    \includegraphics[width=0.95\textwidth]{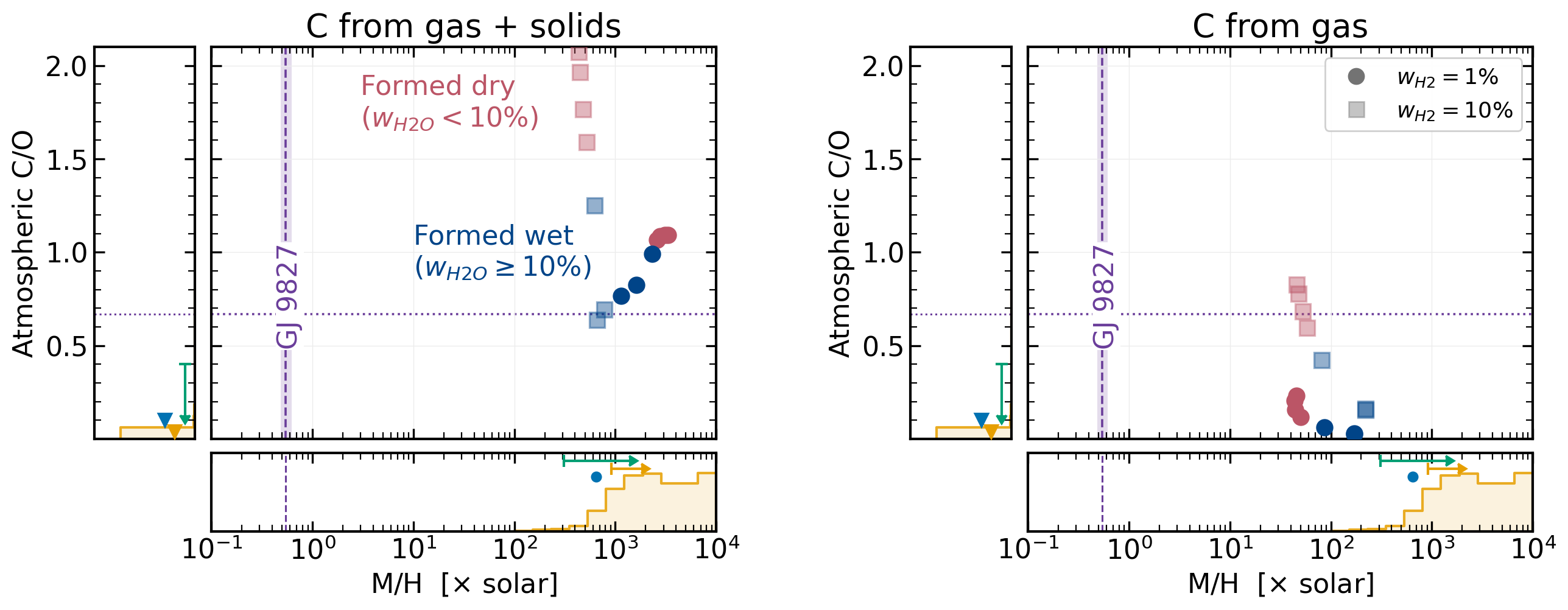}
    \vspace{5mm}
    \vspace{-5mm}\caption{Comparison between global chemical equilibrium (GCE) model results and observational constraints for GJ\,9827\,d, illustrating the challenge of reproducing high-metallicity, low-C/O compositions. The posterior distributions obtained from the SCARLET retrievals performed on the transmission spectrum are reproduced on the left and bottom subpanels surrounding the scatter plot of gas-phase C/O ratios and metallicities predicted by the GCE model, with colors matching Fig. \ref{fig:retrieval_spectrum_fit}. The GCE model results are shown as markers bundled between ``dry'' (water-poor) and ``wet'' (water-rich) building blocks with the transparency and marker type encoding $w_{\rm H2}$. Stellar metallicity and C/O ratio values are annotated as dashed purple lines. The left panel shows the GCE results when carbon is introduced into the elemental makeup of the solids when initializing the model, while the right panel corresponds to a setup where carbon is provided exclusively by the primordial gas component. The detailed breakdown of the $w_{\rm H2O}$-dependence of the GCE results, and their translation to upper-atmosphere molecular abundance ratios, are shown in Appendix Fig. \ref{fig:GCE_2x2panel}.}
    \label{fig:GCE_retrieval_comparison}
\end{figure*}

We use a global chemical equilibrium (GCE) model in order to compare our atmosphere composition inferences from GJ\,9827\,d's transmission spectrum with the expected outcomes of interior-atmosphere chemistry. The model solves for the chemical partitioning of species across the atmosphere and interior in the presence of a magma ocean at the atmosphere-mantle interface. The composition is solved across three phases for the gas, silicate and metallic components (Fig. \ref{fig:stellar_abundances}), largely following previous work \citep{schlichting_chemical_2022,grimm_new_2026}, except for the treatment of carbon partitioning, and the assumed initial carbon budget (see modeling details in Appendix Sections \ref{sec:appendix_modeling_GCE} and \ref{sec:carbon_partitioning}). 

In order to put limits on the plausible volatile inventory of GJ\,9827\,d and guide our GCE modeling setup, we revise previous estimates for the envelope mass fraction of GJ\,9827\,d in the pure-water envelope ``steam world'' hypothesis \citepalias{piaulet-ghorayeb_jwstniriss_2024} using the new implementation of updated sub-Neptune evolutionary models for water worlds \citep{aguichine_evolution_2024} into \texttt{smint}, as described in Section \ref{sec:smint} (\citealp{piaulet_wasp-107bs_2021,piaulet_evidence_2023}, \citetalias{piaulet-ghorayeb_jwstniriss_2024}; see Mirkin \& Piaulet-Ghorayeb in prep. for the latest additions). The smaller radii predicted by the updated models support large water mass fractions of 40 to 70\% (Fig. \ref{fig:water_world_smint}). While formation models likely disfavor the upper end of this range \citep{leger_new_2004, mordasini_planet_2024, burn_radius_2024}, it lends credibility to the exploration of large water inventories of up to 30\% by mass in the GCE modeling. The miscibility of hydrogen in planetary interiors is likely prevalent among sub-Neptunes \citep{young_differentiation_2025, Marcum2026}, and may lead to lower-density interiors than the compositions assumed in the models presented here -- potentially both nudging the true volatile mass fraction closer to the $\lesssim 30$\% adopted in this work, and justifying the exploration of hydrogen mass fractions up to 10\% -- accounting for its interior partitioning.

Informed by this initial exploration, we test compositions ranging across the full spectrum from ``gas-dwarf'' formation with H/He-rich, volatile-poor building blocks, all the way to ``water world'' formation where dozens of percent of the planet's mass in water are available for global partitioning. We perform calculations for $w_\mathrm{H2}$ of 1\% and 10\%, sweeping across the initial water content parameterized by $w_\mathrm{H2O}$ between 0 and 30\% in each case. We test two prescriptions for the initial carbon budget: one with the nominal composition described above (0.5 wt\% C in the overall solids budget), and one with the carbon coming only from the primordial gas (see Fig. \ref{fig:GCE_retrieval_comparison}). Our second prescription mirrors previous work \citep{nixon_magma_2025} which did not consider the solids as a potential source of carbon. This choice could illustrate a scenario where the carbon was either not efficiently accreted, or lost over time (see Section \ref{ssec:formation_lowC}). In both cases, the C/Si is strongly depleted in the initial molar budget that goes into the GCE partitioning code, compared to that of the star: the depletion is about two orders of magnitude when C from the solids is accounted for (consistent with bulk Earth), while it is $<10^{-3}$ of that of the star when C only comes from the gas component. Meanwhile, the O/Si remains within a factor of a few of the solar and stellar values across the range of tested $w_\mathrm{H2O}$.

The GCE model computes the gas-phase composition at the atmosphere-magma ocean interface (AMOI), which has temperature and pressure conditions which strongly deviate from the photosphere probed with JWST in transmission. We therefore use SCARLET to compute chemical-equilibrium compositions corresponding to the elemental abundances obtained after global chemical equilibration in the gas, phase (Fig. \ref{fig:GCE_2x2panel}), adopting $p=1$\,mbar and $T=675$\,K as a representative value for the photosphere, and $p=1$\,bar and $T=1100$\,K for the quenching conditions. The latter choice is an approximative match to the quenched-retrieval conditions, and agrees with more detailed self-consistent chemical-kinetics modeling of sub-Neptune atmospheres \citep{mukherjee_effects_2024}.

We find that across our tested configurations, the predicted atmosphere C/O ratios range from $\sim 0.1$ to 2, while the predicted metallicities span $\sim 40$ to $>1000\times$ solar, qualitatively encompassing the range of observed sub-Neptune compositions \citep{benneke_jwst_2024,wogan_jwst_2024,gao_hazy_2023,barat_jwst_2026,ahrer_escaping_2025}. 

When considering parameters in isolation (Figs. \ref{fig:GCE_retrieval_comparison} and \ref{fig:GCE_2x2panel}), we find that the inclusion of carbon coming from accreted solids in the elemental budget has an order-of-magnitude impact on the atmosphere metallicity (raising it by a factor of 10 to 100). Compositions initialized with a larger $w_{\rm H2}$ produce higher C/O ratios (especially for planets formed dry), and slightly lower metallicities. Formation scenarios that incorporate greater amounts of volatile ice (higher $w_{\rm H2O}$) result in lower C/O ratios. In terms of molecule abundance ratios, the carbon-depleted setup produces distinctly low CH$_4$/H$_2$O ratios that remain below 1 throughout our tested range. On the other hand, the predicted CO$_2$/H$_2$O ratio is more sensitive to the impact of chemical quenching, and can cross over to more water-rich compositions where it becomes indistinguishable from the predictions of the carbon-depleted model (Fig. \ref{fig:GCE_2x2panel}).

Contrary to \citet{grimm_new_2026}, we find that atmospheric compositions are not sensitive to the Fe/Si ratio, while very high Mg/Si ratios can lead to a slight increase in the C/O ratio and a two-to-three-fold decrease in the atmospheric metallicity (Fig. \ref{fig:Mg_Fe_Si_tests}). Finally, while the gas-phase C/O ratio and metallicity are sensitive to $T_{\rm AMOI}$ (especially for dry formation cases), they never impact either quantity to the point where there is crossover between the results for low vs. high carbon or low vs. high water inventories outlined in Fig. \ref{fig:GCE_retrieval_comparison}: achieving low atmosphere C/O ratios require very low carbon abundances, which are associated with atmosphere metallicities ranging from 10 to $200\times$ solar (Fig. \ref{fig:sensitivity_Tamoi}), lower than the observational constraints support. 


\section{Discussion} \label{sec:discussion}

\begin{figure*}
    \centering
    \includegraphics[width=0.95\textwidth]{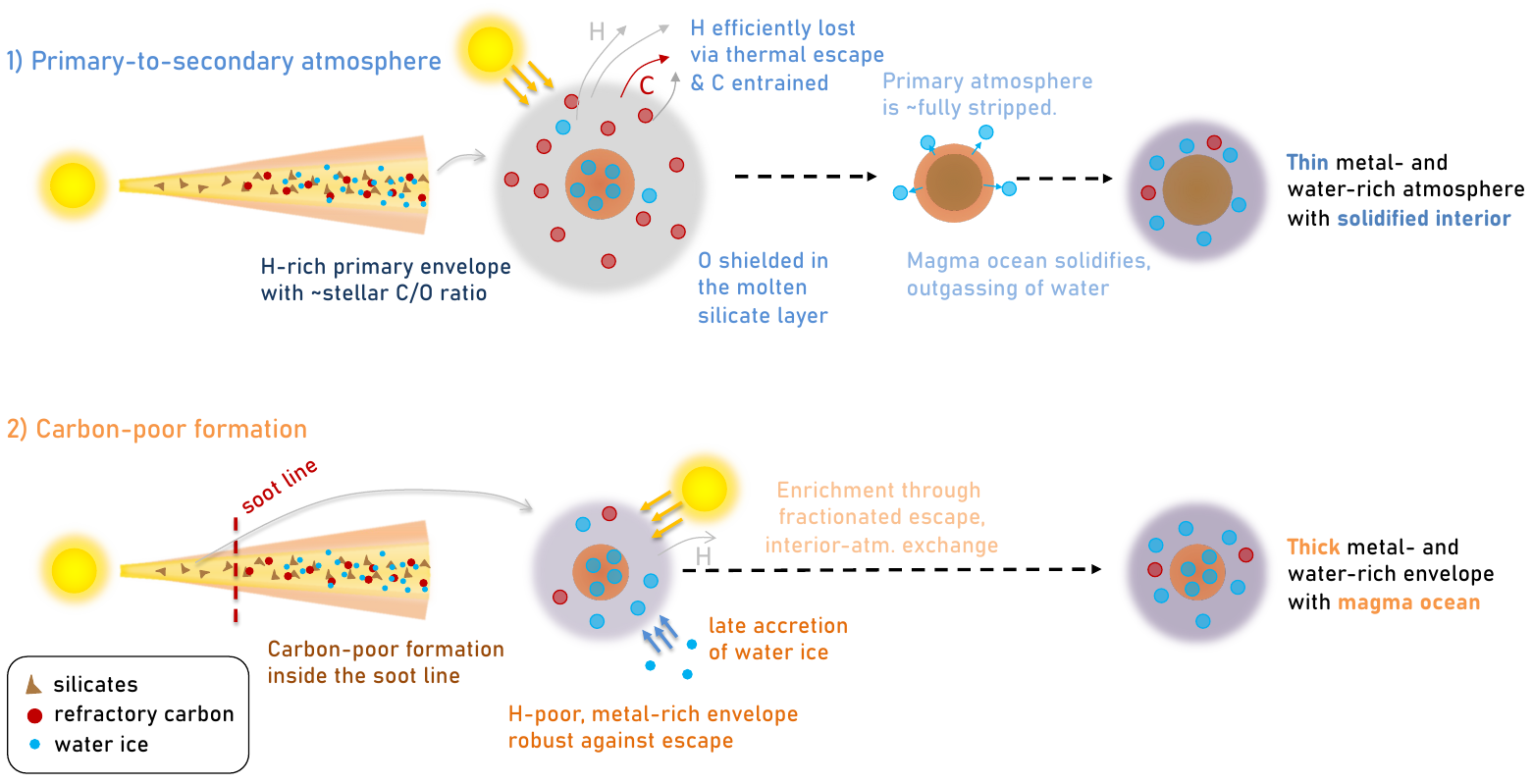}
    \vspace{5mm}
    \vspace{-5mm}\caption{Illustration of the proposed scenarios to explain the low-C/O, high metallicity atmosphere of GJ\,9827\,d. In Scenario 1 (top, shades of blue), the planet forms with a primary, H-rich atmosphere and a stellar C/O ratio. While oxygen/water is efficiently shielded from escape in the magma ocean, carbon is entrained and lost alongside hydrogen until complete atmosphere stripping. The loss of the greenhouse heating from a gaseous envelope leads to gradual solidification of the magma ocean and the revival of a metal-and water-rich atmosphere from outgassing. In Scenario 2 (bottom, shades of orange), the planet forms carbon-poor inside the soot line, and acquires abundant water from late accretion, resulting in a metal-enriched atmosphere robust to stripping from thermal escape. Fractionated escape leads to a moderate amount of further metal enrichment over Gyr timescales. Contrary to Scenario 1 where the outgassing followed magma ocean solidification, the magma ocean persists in Scenario 2 and still holds a substantial fraction of the bulk O, S budget of the planet to this day.}
    \label{fig:scenarios_illustration}
\end{figure*}

We find that GJ\,9827\,d has an unusual low-C/O, high-metallicity atmosphere which sets it apart from the majority of observed sub-Neptunes with detections of abundant carbon-bearing molecules \citep{benneke_jwst_2024,holmberg_possible_2024,madhusudhan_carbon-bearing_2023,schlawin_possible_2024,rigby_jwst_2025,barat_jwst_2026}. The low observed C/O ratio is hard to reconcile with magma ocean-atmosphere chemistry because of the much lower solubility of carbon-bearing species in the melt compared to water \citep{lichtenberg_redox_2021}. The measured value implies an extremely low carbon inventory, and favors water-rich building blocks (Fig. \ref{fig:GCE_retrieval_comparison}). This level of carbon depletion is also reminiscent of Earth's ``Missing Carbon'' problem, whereby the present-day estimate of the bulk carbon content represents less than 1\% of the carbon found in primitive chondritic meteorites \citep{Gail2017,Bergin2018, li_earths_2021}. 

First, we discuss the impact of potential carbon partitioning into the deep interior on our results (Section \ref{ssec:deepinteriorcarbon}). Then, we present two scenarios that can explain our observations (Fig. \ref{fig:scenarios_illustration}). (1) If GJ\,9827\,d initially formed with a low-metallicity hydrogen-dominated atmosphere, its water-rich, carbon-poor present-day composition may be a signature of outgassing, the result of the primary-to-secondary atmosphere transition of this near-radius-valley sub-Neptune (Section \ref{ssec:secondary}). (2) Alternatively, GJ\,9827\,d may have instead formed with an already metal-rich, carbon-poor atmosphere robust against stripping thanks to the combination of high metallicity and sustained magma ocean-atmosphere exchange (Sections \ref{ssec:metalrich_formation} and \ref{ssec:formation_lowC}). Finally, in Section \ref{ssec:condensation}, we address how our conclusions may be impacted by carbon condensing out of the observable atmosphere. 

\subsection{Deep-interior sequestration of carbon} \label{ssec:deepinteriorcarbon}

The siderophile behaviour of carbon, although not described by our model, may play a role in low atmospheric abundances of carbon species. On Earth, 80-90\% of the (already-low) carbon budget is believed to be stored in the lower mantle and metal core \citep{mcdonough_215_2003,dasgupta_carbon_2008,fischer_carbon_2020,davis_carbon_2023,sparrman_when_2026}. The partitioning of carbon in the core is therefore the dominant uncertainty in estimates of the Earth's carbon content, with estimates ranging from 0.2 to 4 wt\% \citep{mcdonough_215_2003,dasgupta_carbon_2008,wood_carbon_2013,dasgupta_carbon_2013,chi_partitioning_2014}. 

Still, we do not expect deep-interior partitioning of carbon to fully explain the observed conditions on GJ\,9827\,d. While GCE predictions appear consistent across a wide range of tested planetary masses \citep{grimm_new_2026}, GJ\,9827\,d's atmospheric composition offers a stark contrast when compared to other sub-Neptunes with higher atmospheric C/O ratios and prominent features from carbon-bearing species \citep{benneke_jwst_2024,holmberg_possible_2024,madhusudhan_carbon-bearing_2023,schlawin_possible_2024,rigby_jwst_2025,barat_jwst_2026}. Further experimental and computational work will be necessary to self-consistently inform both the relevant Gibbs free energies and the nonideal behaviour relevant to the reactions that regulate the partitioning of carbon across the silicate and metal phases in the sub-Neptune pressure-temperature regime (Appendix Section \ref{sec:carbon_partitioning}; \citealp{blanchard_metal-silicate_2022,fischer_carbon_2020,werlen_atmospheric_2025}). 

We also note that as they stand, the range of GCE calculations are limited by the available Gibbs free energy data, which is not only sparse in temperature space, but also treated as pressure-independent \citep{grimm_new_2026}-- a caveat of the results presented here. In what follows, we interpret the chemistry of GJ 9827 d within the context of the modeling performed in this study, and defer further exploration of the impact of these model uncertainties to future work. 

\subsection{Loss of carbon as a result of primary-to-secondary atmosphere transition}\label{ssec:secondary}

Here we explore the first proposed scenario, where GJ 9827\,d's water-rich, carbon-poor atmosphere is a product of interior outgassing following the loss of a primary, hydrogen-rich atmosphere. In this scenario, water remained shielded in the primordial magma ocean while carbon remained largely mixed with and lost alongside the escaping hydrogen-rich gas, until the stripping of GJ\,9827\,d's primary atmosphere. Following the resulting mantle solidification, a water-rich atmosphere was outgassed, matching our observations. We discuss the plausibility of this scenario using models of planetary thermal and compositional evolution assuming hydrogen-dominated initial conditions.

If GJ 9827\,d's envelope initially formed hydrogen-rich, the low planetary mass and close-in orbit are associated with a high propensity for full envelope stripping \citep{gupta_caught_2021}. We first explore this using grid-based composition inference with \texttt{smint} (\citealp{piaulet_wasp-107bs_2021,piaulet_evidence_2023}, \citetalias{piaulet-ghorayeb_jwstniriss_2024}; Mirkin \& Piaulet-Ghorayeb in prep.), adopting the latest results from H$_2$/He-rich sub-Neptune evolution models \citep{tang_reassessing_2024}. The \texttt{smint} grid-based exploration reveals that GJ\,9827\,d's properties are at odds with predictions from such gas-dwarf atmosphere models (Fig. \ref{fig:tang_hhe_models}): the corresponding envelope mass fractions are so low that the envelope would be unstable to full thermally-driven loss, in tension with the non-detections of H and He escape (\citealp{kasper_non-detection_2020,carleo_multiwavelength_2021,krishnamurthy_absence_2023},\citetalias{piaulet-ghorayeb_jwstniriss_2024}). 

We further explore the stability of a hydrogen-dominated primary atmosphere on GJ\,9827\,d using bespoke evolutionary calculations, following \citet{louca_metallicity_2023} and \citet{louca_metal-enriched_2025}. We use the \texttt{MESA} code \citep{Paxton2011} to evolve the planetary interior and envelope, with stellar flux evolution modeled using the MIST database \citep{choi_mesa_2016}, and a hydrodynamic escape implementation following \citet{kubyshkina_grid_2018}. At each timestep, we compute the crossover mass to determine which species are dragged along with the hydrodynamic hydrogen outflow \citep{louca_metal-enriched_2025}. We explore a grid with initial planetary masses between 3.0-4.5\,$M_\oplus$ and an initial H/He envelope mass fractions between 1 and 30\%, evolving the models for up to 10 Gyr. During periods of strong hydrodynamic escape, the large hydrogen flux can also efficiently remove heavier species, whereas chemical fractionation becomes increasingly important as the escape rate decreases and heavy species are preferentially retained \citep{louca_metal-enriched_2025}.

We find that many of the low-mass evolutionary calculations terminate before reaching 10 Gyr as a result from strong atmospheric escape and eventual loss of their H/He envelopes. Among the tracks that retain an atmosphere to Gyr ages, we find enrichment factors of a few, reaching up to $\sim4$ times the initial heavy-element abundance, while the evolution of C/O remains small. This is considerably less enrichment than the factors of $\sim50$–$70$ found by \citet{louca_metal-enriched_2025} for a more massive planet. For GJ\,9827\,d, the stronger escape associated with its low mass instead removes heavy species together with hydrogen during the period of highest mass loss and, in the more extreme and the majority of the cases explored, ultimately strips the primordial envelope altogether before larger metal enrichment can develop. None of the models starting with these prescribed hydrogen-rich conditions surviving atmospheres simultaneously reproduce the observed mass and radius of GJ\,9827\,d: while some reproduce its present-day mass, their radii are substantially smaller than observed. Together with the strong atmospheric loss experienced by the lower-mass models or models with higher envelope fractions, this suggests that a primordial H/He-rich atmosphere would not have survived the evolution of GJ\,9827\,d, supporting a secondary origin for its present-day atmosphere.

Since carbon-bearing species are orders of magnitude less soluble than water in the silicate melt \citep{lichtenberg_redox_2021}, most of the volatile-derived oxygen is initially stored in the molten planetary interior ($>98$\% across the silicate and metal phases, including between 40\% and $>90$\% of the O budget by mole in the silicate phase across our GCE simulations). Meanwhile, carbon remains available for escape to space, plausibly entrained in the complete loss of a primordial hydrogen-dominated envelope. The loss of the primary atmosphere should lead to solidification of the surface, leading to the outgassing of a water-rich secondary atmosphere in line with both the low C/O ratio and high metallicity of GJ\,9827\,d \citep{kite_exoplanet_2020,postolec_atmospheric_2026}. 

While less studied on sub-Neptunes, impact devolatilization is another mechanism that could lead to the preferential loss of carbon via impacts, as long as water is shielded away in the magma ocean. This mechanism has been proposed as a major contributor to the evolution of the Earth's volatile budget \citep{tucker_evidence_2014,Schlichting2018,solomatova_genesis_2021}. The low mass of GJ 9827\,d safely places it within the regime where impact-driven atmospheric escape alone would lead to the complete loss of a primary H/He-dominated atmosphere \citep{zlimen_reconciling_2026}.

\subsection{Compositional evolution of an initially metal-rich atmosphere}\label{ssec:metalrich_formation}

\begin{figure}
    \centering
        \includegraphics[width=0.48\textwidth]{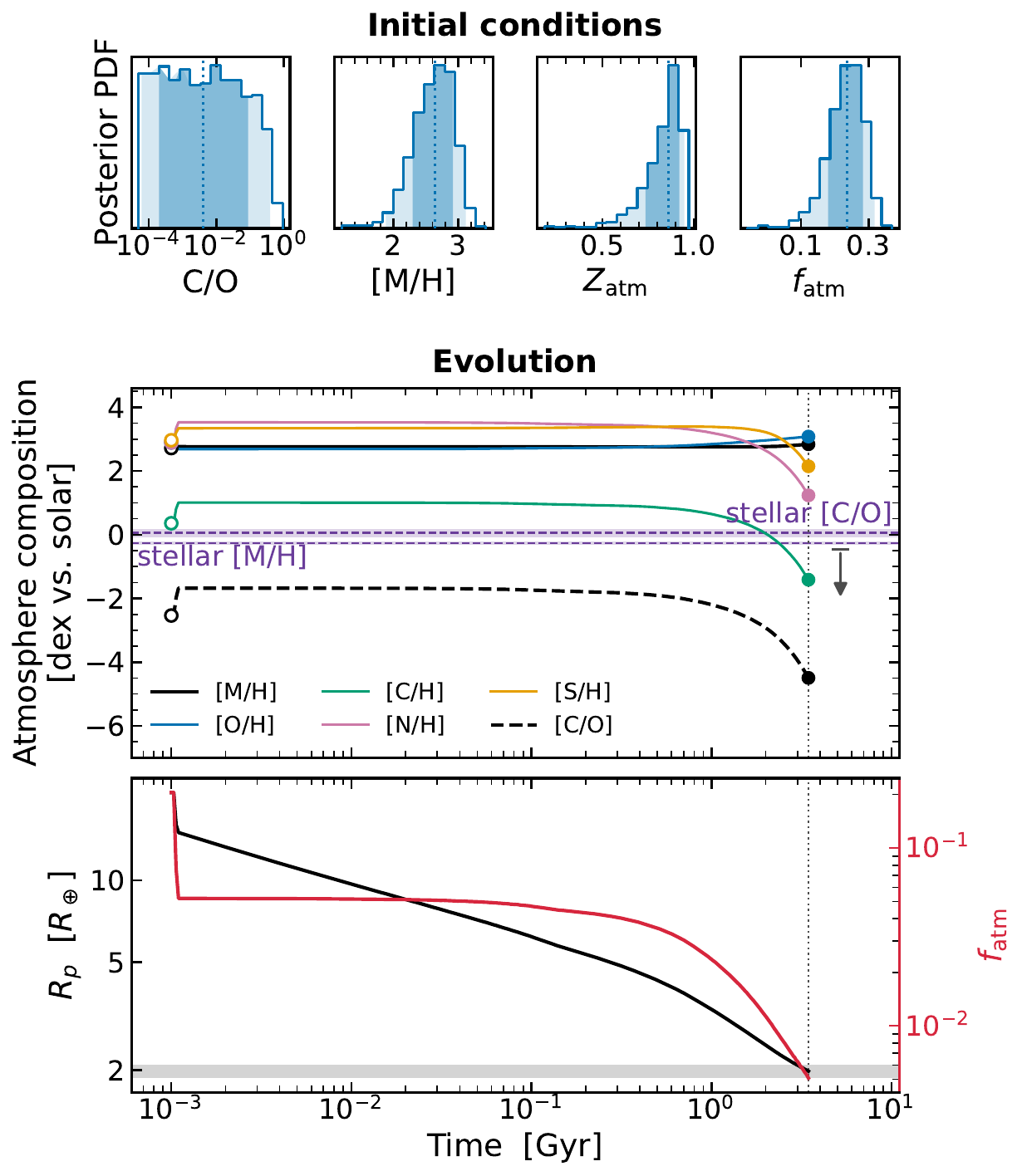}
    \vspace{5mm}
    \vspace{-5mm}\caption{
    Results from the \texttt{IsoFATE} retrieval of initial conditions that match GJ\,9827\,d's present atmospheric state. \textit{Top panels:} Posterior distributions on the atmosphere properties at the start of the simulation (C/O ratio, metallicity, atmosphere metal mass fraction and bulk atmosphere mass fraction). \textit{Middle panel:} For the best-fit model, time evolution of various elemental ratios relative to the solar values from \citet{asplund_chemical_2021} (metallicity and C/O ratio shown in black). The C/O ratio is shown as a dashed line with the upper limit inferred from the NIRSpec measurements is shown as a gray arrow. Stellar values for [M/H] and [C/O] are shown in purple. The maximum-likelihood model evolves over 3.2 Gyr, with an initial metallicity of 350$\times$ solar ([M/H]$=2.55$) and atmosphere mass fraction of 21\%. \textit{Lower panel:} Time evolution of the planet's radius (black) and atmosphere mass fraction (red, right-hand axis) for the best-fit model. The gray band indicates the present-day radius measurement for GJ\,9827\,d.}
    \label{fig:MAP_solution_IsoFATE}
\end{figure}

We explore the alternative scenario where the atmosphere of GJ 9827\,d would never have been fully stripped, and is instead the result of Gyr of compositional evolution through weakly fractionated escape and magma ocean-atmosphere volatile exchange. We find that the long-term survival of the atmosphere is contingent on a metal-rich composition already at formation, with the metallicity enhanced over time.

We use the version of the \texttt{IsoFATE} code \citep{cherubim_isofate_2025,cherubim_oxidation_2025} coupled with \texttt{Atmodeller} \citep{bower_diversity_2025} in order to model the radius and compositional evolution of GJ\,9827\,d accounting for magma ocean-atmosphere exchange, thermal evolution, and fractionated escape. In particular, we use a new tool leveraging the Markov Chain Monte Carlo module \texttt{emcee} \citep{foreman-mackey_emcee_2013} for parameter inference, to solve the inverse problem of mapping the present-day planetary radius (\citetalias{piaulet-ghorayeb_jwstniriss_2024}), planetary mass, upper limit on the atmospheric C/O ratio, and lower limit on the atmospheric H$_2$O abundance (Section \ref{sec:retrieval_comparison}) to the corresponding initial conditions (Cherubim et al. in prep.). The model itself follows \citet{cherubim_oxidation_2025}, now coupled with MCMC for parameter inference (see Appendix \ref{sec:isofate_extra}). One additional feature of the model presented is the choice of the binary pair in the escape fractionation calculation at each time step rather than keeping it fixed as H and He. Overall, we find that matching GJ\,9827\,d's atmosphere composition through compositional evolution of its primary atmosphere alone requires an initially metal-rich atmosphere ([M/H]$_\mathrm{ini} = 2.7 \pm 0.3$: $\sim 500\times$solar, $Z_\mathrm{ini}>0.6$; Fig. \ref{fig:MAP_solution_IsoFATE}). The radius evolution of the planet down to its present-day size is achieved partially through cooling of the envelope, but most substantially via atmospheric escape, with the envelope mass fraction decreasing from about 20\% initially to roughly 0.5\% at the end of the simulation.

In this scenario, the atmosphere begins water-rich ([O/H]$_\mathrm{ini}\sim 700 \times$solar), and accumulates even more water over time. Atmospheric water enrichment is driven by its magma solubility and gradual outgassing from the magma ocean, plus the strong escaping wind, which is mostly nonselective. Additionally, our model predicts the accumulation of O$_2$ in the atmosphere ($\sim 30$\% VMR), in line with predictions for the evolution of water-rich planets \citep{krissansen-totton_oxygen_2021}.  

For all elements besides O and H, their molecular species are orders of magnitude less soluble, providing no shield from escape. One exception is S$_2$, which can be more soluble than water. However, as H is preferentially lost and the atmosphere gradually oxidizes, and S is oxidized to SO$_2$, greatly decreasing its solubility. Even with this lower solubility in oxidizing conditions, the 68\% C.I. on the final fraction of S dissolved in the interior spans from 1\% to 40\% (5\% in the best-fit model).  Our model ignores sulfate solubility, which could potentially further enhance S retention in the interior. Overall, interior partitioning of sulfur in the case of a persistent magma ocean would lower atmospheric SO$_2$ abundances compared to a scenario where it was fully outgassed following solidification, suggesting SO$_2$ abundance measurement as a potential tie-breaker between the two scenarios.

Finally, our compositional constraints are best matched when the initial C/H ratio is substantially depleted (by about 2 orders of magnitude) relative to the highly super-solar O/H, M/H invoked to match the planet's composition (Fig. \ref{fig:MAP_solution_IsoFATE}). This is because the outflow is mostly non-fractionating, which results in carbon being removed at the same pace as other metals (but slower than O because of its dissolution). We discuss this tension in the following subsection.

\subsection{Formation from carbon-poor building blocks}\label{ssec:formation_lowC}

In both scenarios, the water budget required to match GJ 9827 d’s radius (tens of percent by mass) far exceeds the few percent that can be produced endogenically through hydrogen–magma ocean reactions \citep{kite_exoplanet_2020,rogers_most_2024, werlen_atmospheric_2025}, implying that most of this water was accreted at formation. Both scenarios therefore require formation from water-rich building blocks, and differ primarily in the evolutionary history of the atmosphere, and how it shapes the observed depletion in carbon relative to the elevated O/H ratio. 

The near-solar stellar C/O ratio we derive for GJ\,9827 (Section~\ref{sec:stellar_abundances}) indicates that any carbon depletion in the planet is not inherited from an anomalous host star composition. Atmospheric carbon has been proposed as a distinct tracer of formation location for sub-Neptunes, with the hypothesis of ``soot worlds'' or ``soot-water worlds'' proposed as alternatives to water worlds with Earth-like interiors, to explain the low densities of small sub-Neptunes \citep{Bergin2014,bergin_exoplanet_2023,li_soot_2026,lin_carbon-rich_2025}. 

Soot worlds are planets formed beyond the ``soot line'' in the protoplanetary disk (Fig. \ref{fig:scenarios_illustration}), the location where carbon-rich grains or polycyclic aromatic hydrocarbons are thermally destructed \citep{kress_soot_2010}. Such soot planets should have tens of percent by mass of hydrocarbon-rich material in their interior \citep{bergin_exoplanet_2023,li_soot_2026}. A key prediction of soot-planet models is the presence of abundant carbon-bearing species in their atmospheres, more abundant than water, which poses a challenge when compared to the water-rich, carbon-poor composition of GJ\,9827\,d (\citealp{bergin_exoplanet_2023,lin_carbon-rich_2025}; see also \citealp{dorn_atmospheric_2026} who employ the same GCE as in this work, with carbon partitioning caveat described in Appendix Section \ref{sec:carbon_partitioning}). This incompatibility may be lifted if carbon becomes so abundant that it is supersaturated and condenses out in a graphite layer (see Section \ref{ssec:condensation}), a modality that has yet to be explored by soot-world models. Independently, the high atmospheric metallicity of GJ\,9827\,d is also at odds with the $\leq 100\times$ solar metallicity expected from soot-planet models \citep{lin_carbon-rich_2025}. 

Similarly to the Earth, GJ\,9827\,d may therefore have formed from carbon-depleted solids. For Earth, such carbon depletion could have arisen from the irreversible sublimation of refractory carbon since volatile gases (CO, CO$_2$, CH$_4$) would not have condensed back into solids at nebular pressures in the inner disk as it cooled down \citep{Bergin2015,anderson_destruction_2017,grewal_delivery_2019,li_earths_2021}. Sub-Neptunes, however, are substantially more massive than Earth, and the formation of their cores may instead proceed through pebble accretion \citep{bitsch_growth_2015,Ormel2021,venturini_nature_2020,venturini_fading_2024}. Achieving a carbon-poor composition for the solids under a pebble accretion scenario would require forming the planet early, when the nebula is still hot, but stopping the migration of pebbles from colder disk regions (to avoid refractory carbon enrichment). Such barriers to migration would make it challenging to simultaneously explain the substantial water inventory required to achieve a high [M/H], essential for primary atmosphere survival (Section \ref{ssec:metalrich_formation}). This motivates further modeling of the range of outcomes of sub-Neptune formation in terms of not only water, but carbon inventories, to further assess the range of plausible starting conditions for sub-Neptune evolution.

\begin{figure*}
    \centering
    \includegraphics[width=0.99\textwidth]{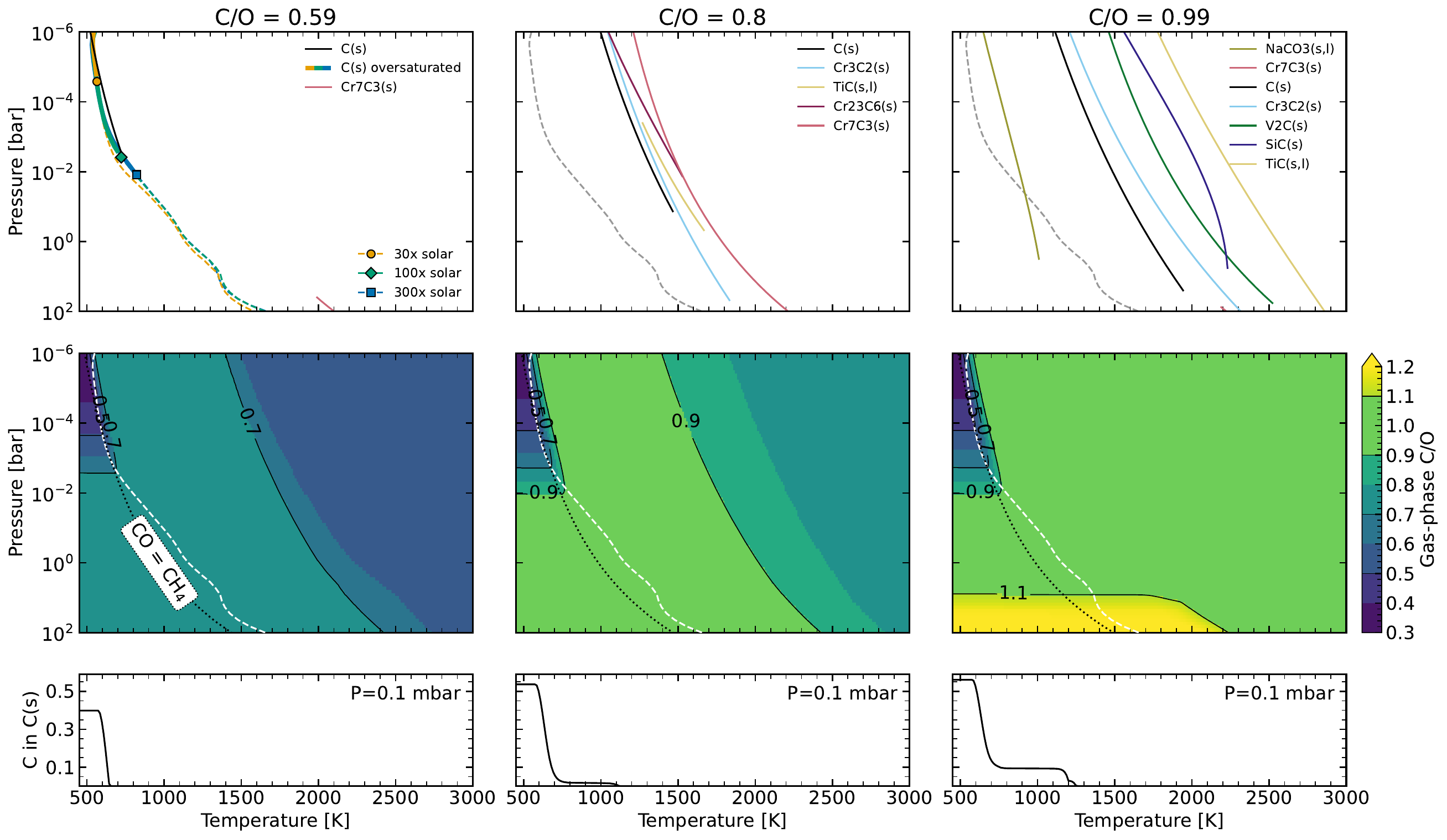}
    \vspace{5mm}
    \vspace{-5mm}\caption{Major carbon-bearing condensates in high-metallicity atmospheres with solar- to super-solar C/O ratios and their impact on the gas-phase C/O ratio, at temperatures and pressures relevant for evolutionary calculations. \textit{Top panels:} Fitted condensation curves for carbon-bearing species in high-metallicity (100$\times$solar) atmospheres (colored lines). Each column corresponds to a different C/O ratio (labeled at the top). Self-consistent temperature profiles calculated for GJ\,9827\,d with SCARLET for a 100$\times$ solar metallicity, solar C/O (C/O$=0.59$) atmosphere are overlaid as dashed lines. The left panel additionally displays temperature profiles calculated for 30$\times$ and 300$\times$ solar metallicity compositions. The range of pressures and temperatures where our model predicts the graphite activity $a_{gr}$ to be $>1$ (oversaturated) in the atmosphere, for a solar C/O ratio, is shown as a thicker overlay on each temperature profile, with a marker indicating the highest pressure where $a_{gr}>1$ in the atmosphere. \textit{Middle:} C/O ratio of the gas-phase species predicted by the isobaric rainout model once condensation has been accounted for, across temperatures representative of the thermal evolution and present-day state of GJ\,9827\,d's atmosphere. The dashed white curve is the 100$\times$ solar metallicity profile for GJ\,9827\,d, and the black dotted line tracks where CO$=$CH$_4$ in the gas-phase, which regulates the onset of graphite formation. \textit{Bottom:} Fraction of the total carbon locked in graphite as a function of temperature, at 10$^{-4}$ bar.}
    \label{fig:C_condensates_fastchem}
\end{figure*}

\subsection{Plausibility of condensation-driven compositional gradients}\label{ssec:condensation}

We study the particular case of graphite condensation as described by the CCO redox buffer \citep{h_keppler_graphite_2019}, and perform equilibrium-condensation modeling with \texttt{FastChem-cond} \citep{kitzmann_fastchem_2023} to evaluate the plausibility of reducing the gas-phase C/O ratio through the condensation of carbon-bearing species in a high-metallicity atmosphere. Our models predict that in a metal-rich atmosphere, graphite condensation could decrease the observable gas-phase C/O ratio to $\sim 0.4$ even if the elemental C/O ratio is $\gtrsim 0.59$, but that it may be challenged if the CO-to-CH$_4$ conversion pathway is inefficient at low pressures (e.g. due to substantial quenching), or by the difficulty of forming graphite in otherwise water-rich gas. We refer to \citet{peng_puffy_2024} and Weglarz et al. (in prep.) for the detailed modeling methodology, which we summarize below along with our results.

First, we perform a calculation of the graphite activity, $a_{gr}$, relative to the gas-phase composition, temperature and pressure conditions predicted by self-consistent radiative-convective SCARLET models at three metallicities (30, 100, and 300$\times$ solar metallicity) and two C/O ratios (0.3 or the solar value of 0.59; \citealp{asplund_chemical_2021}), both assuming $T_\mathrm{int}=100$\,K (following the methods of \citealp{piaulet-ghorayeb_window_2025}). In each atmosphere layer, we follow \citet{peng_puffy_2024} to estimate whether $a_{gr}>1$, where the atmosphere is oversaturated with graphite (C(s)) that can precipitate out. This prescription allows for estimates of the pressure range over which graphite grains are allowed to condense, even at high metallicities \citep{peng_puffy_2024}. We find that graphite precipitation is not predicted for the C/O$=0.3$ case, but that for C/O$=0.59$ it can occur at pressures lower than about 10 mbar for metallicities $\gtrsim 100\times$ solar (Fig. \ref{fig:C_condensates_fastchem}; deeper than the photosphere observed with JWST; Table \ref{tab:retrieval_results}).

Second, we use \texttt{FastChem 4} \citep{kitzmann_fastchem_2026} to track the condensation of a broader range of species from the gas-phase (Weglarz et al., in prep). We explore isobaric conditions ranging from pressures of 10$^{-6}$ bar to 10$^{3}$ bar. At each pressure, we evolve compositions by varying the temperature from 6000\,K down to 100\,K, tracking the condensation point and number density of each species as a function of temperature. We perform this exercise with a metallicity fixed to the highest computationally-stable value of 100$\times$solar ([Fe/H]=+2.0) and for C/O ratios ranging from 0.1 to 1.6. We fit a pressure-temperature relationship following an empirical variation of the Clausius-Clapeyron law to the grid points in pressure-temperature space where the number density of each condensate exceeds 10$^4$ cm$^{-3}$ (determined to result in the minimal cloud opacity visible in JWST observations; Weglarz et al., in prep).

While condensates include TiC, SiC, vanadium carbide, and a variety of chromium carbides at high temperatures, the condensation of graphite at lower temperatures has the largest effect on reducing the gas-phase C/O ratio (Fig. \ref{fig:C_condensates_fastchem}). Ti, Cr and V condensates have a minor effect due to their lower abundances. The larger amount of Si available lets SiC absorb at most to 1 to 2\% of the atmospheric carbon in these high-metallicity conditions. Our isobaric rainout calculation predicts that graphite locks about 40 to 50\% of the atmospheric carbon at 600\,K (Fig. \ref{fig:C_condensates_fastchem}). In the solar-C/O (C/O$=0.59$) and the C/O$=0.8$ case, the sharp transitions in the gas-phase C/O ratio near 1500\,K are due to the condensation of the oxygen-bearing species, specifically forsterite (MgSiO$_4$) and SiO/SiO$_2$, that together remove about 20\% of the gas-phase oxygen. While graphite starts forming at higher temperatures in the high C/O ratio models, it initially only locks a few percent of the carbon budget and instead forms more readily at low pressures, around the CO to CH$_4$ conversion region (Fig. \ref{fig:C_condensates_fastchem}). 

These predictions match the graphite activity-based estimates from the \citet{peng_puffy_2024} model, and overall suggest that C/O ratios lower than about 0.4 could not be explained by graphite condensation alone. Deep-atmosphere quenching, which would prevent efficient CO to CH$_4$ conversion at low pressures, may limit even further the formation of graphite and its potential impact on the gas-phase C/O ratio, motivating future modeling work. 
Finally, we note that while predicted by equilibrium-condensation models \citep{moses_compositional_2013,mbarek_clouds_2016}, laboratory experiments have struggled to form graphite, especially in water-rich conditions \citep{li_impact_2025,wang_formation_2025,Helling2017}, motivating validation with further laboratory work covering the broader range of temperatures and pressures relevant for sub-Neptune evolution models.

\section{Summary and Conclusions} \label{sec:conclusion}

In this work, we analyzed new JWST/NIRSpec transit observations of GJ\,9827\,d and combined them with existing HST/WFC3 and JWST/NIRISS measurements of the planet's transmission spectrum, as well as CFHT/SPIRou observations of its host star. We performed extensive modeling of interior-atmosphere chemistry, atmospheric escape and evolution to constrain the atmospheric composition of this small 2--$R_\oplus$ sub-Neptune in the context of its formation and history. Our findings and conclusions are summarized below:

\begin{itemize}
    \item The two additional JWST/NIRSpec transits support transit-timing variations for GJ\,9827\,d, with peak-to-peak amplitudes of 10 to 12 minutes, suggesting a nonzero eccentricity of $0.07$--$0.09$, motivating further dynamical modeling of the GJ\,9827 system.
    \item We derive new elemental abundances for GJ\,9827 using high-resolution CFHT/SPIRou spectra, which confirm a highly subsolar metallicity of $-0.26\pm0.05$, consistent across volatile and refractory species. 
    \item The JWST/NIRSpec G395H spectrum does not exhibit prominent absorption features. Atmosphere retrievals performed on the combination of all near-infrared observations of the planet, including HST/WFC3 and JWST/NIRISS SOSS transmission measurements, indicate a high metallicity, but carbon-poor atmosphere ([M/H]$>2.4$ to $2.9$, and C/O$<0.1$--$0.5$ at 2$\sigma$ across model assumptions). With this abundance pattern, GJ\,9827\,d adds to the compositional diversity of the sub-Neptune population, and deviates from the detection of prominent absorption from carbon-bearing species on other sub-Neptunes. 
    \item We model the partitioning of volatile species across GJ\,9827\,d's metal, silicate, and gas phase, and find that the high atmospheric metallicity is best reproduced when starting from hydrogen-poor building blocks, while the low C/O ratio can only be reconciled with a low bulk C inventory, with virtually no solid-derived carbon. 
    \item We propose that GJ\,9827\,d's composition could be the result of a primary-to-secondary atmosphere transition, supported by simulations of fractionated atmospheric escape that predict the complete loss of a primary hydrogen-dominated atmosphere. In this scenario, water would initially have been shielded from loss through dissolution in the magma ocean more efficiently than carbon-bearing species, entrained alongside hydrogen. The present-day atmosphere of GJ\,9827\,d would therefore be the product of outgassing as the magma ocean solidified.
    \item Alternatively, we find that GJ\,9827\,d could have retained both its atmosphere as well as a magma ocean throughout its entire lifetime, if its atmosphere formed with a high enough initial metallicity [M/H]$_\mathrm{ini}\sim2.7$, $Z_\mathrm{ini}>0.6$). Matching observational limits on the planet's water abundance and C/O ratio requires carbon-poor formation (by about two orders of magnitude compared to O and other metals). Reconciling the low initial carbon abundance with the ice-rich formation required for such a high metallicity is challenging and requires further modeling exploration. 
    \item A high (more than percent-level) abundance of sulfur-bearing molecules such as SO$_2$ would support a secondary origin for the atmosphere, as a magma ocean can hold up to about 40\% of the total sulfur budget in our simulations, lowering the amount of S$_2$ and SO$_2$ gas in the atmosphere.
    \item We model the conditions for condensation of carbon-bearing species in high-metallicity atmospheres in order to evaluate its potential impact on reducing the gas-phase C/O ratio, and find that only graphite condensation could plausibly lead to a substantial reduction of GJ\,9827\,d's C/O down to $\sim 0.4$ at photospheric pressures, locking up to $\sim 50$\% of the gas-phase carbon at low pressures. Quenching from deeper, CO-dominated regions, or kinetic inhibition of graphite formation due to the high activation energy required by its formation pathways would however limit the impact of this process on our observations, and require further experimental tests. 
    \item Rather than equilibrium condensates, organic haze, or soot, could form on GJ\,9827\,d at the high end of the range of C/O ratios allowed by its transmission spectrum, and could further contribute to muting absorption features. A deeper observational campaign of GJ\,9827\,d in transmission is required to further constrain the relative abundances of H$_2$O, CO$_2$, and CH$_4$ in its atmosphere, as well as search for SO$_2$, since a measurement of the SO$_2$ abundance could inform whether GJ\,9827\,d's atmosphere is primary or outgassed in nature. 
\end{itemize}

\section{Data Availability} \label{sec:data}

The NIRSpec/G395H transmission spectrum of GJ\,9827\,d is provided as a machine-readable table on Zenodo\footnote{Zenodo link to be updated upon paper acceptance.}. All other models and data products will be provided upon reasonable request.

\section{Author Contributions} \label{sec:contributions}

C. P.-G. designed the study, performed the light curve analysis, atmosphere modeling with SCARLET, GCE modeling, dynamical analysis, interpretation, wrote the manuscript and created all the figures in the main text and Appendices A through G. N. G. performed the stellar abundance analysis under the supervision of R.C.. J.W. performed the \texttt{FastChem-cond} modeling under the supervision of E.M.-R.K. Y.M. performed evolutionary calculations for GJ\,9827\,d with the help of A.L. B.P. performed the graphite precipitation modeling. T.F. and E.A. provided the \texttt{exoTEDRF} and \textit{Eureka!} data reductions. E.R. performed a suite of POSEIDON retrievals for robustness checks, with help from R.M. E.M. provided the \texttt{smint} composition inference results for GJ\,9827\,d under the supervision of C.P.-G. M.C.N. ran one Aurora retrieval as a robustness check and probing potential CS$_2$, with collaboration from L.W.. J. Y. provided context for the haze modeling discussion. C.Ch. ran the \texttt{IsoFATE} models presented in this study. H.S., M.-L.S., A.W., C.D. and E.Y. reviewed the GCE modeling methodology and discussion. P.B. informed the range of saturated wavelengths in the NIRISS observations. All other authors were co-investigators the JWST proposal GO 4098, reviewed the manuscript and provided comments and feedback.

\vspace{10mm}
\begin{acknowledgments}
C. P.-G. thanks D. Fabrycky, V. Nagpal, R. Arora, C. Wirth, M. Brady, S. Moran, D. Samra, and E. Bergin for helpful discussions. C. P.-G. also thanks S. Grimm for making the full \citet{grimm_new_2026} sub-Neptune global chemical equilibrium code publicly available.
This work is based on observations with the NASA/ESA/CSA James Webb Space Telescope, obtained at the Space Telescope Science Institute (STScI) operated by AURA, Inc. All of the JWST data presented in this paper were obtained from the Mikulski Archive for Space Telescopes (MAST) at the Space Telescope Science Institute. 
The data used in this paper can be found in MAST at the following DOI: \dataset[10.17909/4t9h-kg73]{http://10.17909/4t9h-kg73}. 
The authors wish to recognize and acknowledge the very significant cultural role and reverence that the summit of Maunakea has always had within the indigenous Hawaiian community. We are most fortunate to have the opportunity to conduct observations from this mountain. 

C.P.-G. acknowledges support from the E. Margaret Burbidge Prize Postdoctoral Fellowship from the Brinson Foundation, from the Suzuki Postdoctoral Fellowship, and from the Heising-Simons Foundation through the 51 Pegasi b Fellowship program. C.Ca. acknowledges the support from the Swiss National Science Foundation (SNSF) under the grant SPECTRE (No 200021\_215200). C.D. acknowledges support from the SNSF under grant TMSGI2\_211313. C.G. acknowledges funding from the ERC Horizon Europe funding programme in support of the Synergy Grant Geoastronomy (grant agreement number 101166936).
T.R.F. acknowledges usage of Delta at the University of Illinois Urbana-Champaign (National Science Foundation grant 2005572) through allocation PHY260183 the NSF ACCESS program (supported by NSF grants 2138259, 2138286, 2138307, 2137603, and 2138296), and the University of Chicago’s Research Computing Center. Y.M acknowledges funding from the European Research Council (ERC) under the European Union’s Horizon 2020 research and innovation programme (grant agreement no. 101088557, N-GINE). 
C.Ch. was supported by NASA through the NASA Hubble Fellowship grant HST-HF2-51601.001-A, awarded by the Space Telescope Science Institute, which is operated by the Association of Universities for Research in Astronomy, Inc., for NASA, under contract NAS5-26555.
We acknowledge the use of A.I. tools (Claude, GitHub Copilot) for assistance with coding, and all analyses have been thoroughly vetted by the authors. All the text in this manuscript was written by the authors themselves, and they take full responsibility for its contents.

\end{acknowledgments}

\clearpage

\appendix
\onecolumngrid                 
\let\twocolumngrid\relax       
\setcounter{figure}{0}
\renewcommand{\thefigure}{A\arabic{figure}}
\setcounter{table}{0}
\renewcommand{\thetable}{A\arabic{table}}
\setcounter{equation}{0}
\renewcommand{\theHfigure}{A\arabic{figure}}
\renewcommand{\theHtable}{A\arabic{table}}

\section{ExoTEP light-curve fitting}
Here we provide additional details on the astrophysical and systematics model assumptions and sensitivity tests performed when fitting the light curves extracted from the \texttt{exoTEDRF} and \textit{Eureka!} data reductions. 

\subsection{Broadband light-curve fits}\label{sec:appendix_whitelc}

We perform broadband light curve fitting independently on the time series extracted for each detector and visit, because of detector-to-detector discrepancies in the measured transit times further discussed below. The best-fitting orbital parameter values and transit times from this light curve fitting step are recorded and fixed for the spectroscopic light curve fits (Table \ref{table:wlc_param}). Given the higher signal to noise of the broadband light curves, we also use them to constrain some of the instrumental and astrophysical systematics that impact the time series and use that information when fitting the chromatic light curves. 

As a first step, we perform outlier rejection on the time series of stellar spectra. We reject 3.5$\sigma$ time outliers in the white light curve flux over 10 iterations, each time identifying the outliers relative to the local running median of the light curve computed with a window size of 20 integrations. We use cross-correlation on the stellar spectra to record the wavelength shift, specifically the wavelength shift of the peak of the cross-correlation function (CCF) relative to the first spectrum in the time series, and the CCF peak width. We discard points that are outliers at greater than 3$\sigma$ in the CCF peak position, or in the recorded spatial position of the trace on the detector (y position) and the width of the Gaussian fitted to the flux distribution summed over the spectral dimension to determine the y position. The same criteria are used to detect and reject outliers in each of the spectroscopic light curves described in the following subsection. Following standard practice, we record the time-dependent eigenvalues associated with a principal component analysis (PCA) of the time series of detector images, for each detector and visit separately. We discard points that are $>5\sigma$ outliers in the eigenvalues of the 2-5th PCA components (the first one contains the transit).

The light curves are fitted using both astrophysical and systematics models which are multiplied to obtain the final light curve model. The parameters of the astrophysical and systematics model are fitted jointly using the Markov Chain Monte Carlo (MCMC) sampler \texttt{emcee} \citep{foreman-mackey_emcee_2013}. 

The astrophysical model consists of a transit model calculated using the \texttt{batman} module \citep{kreidberg_batman_2015}. Because of GJ 9827\,d's high impact parameter (Table \ref{table:wlc_param}, \citealp{prieto-arranz_mass_2018}), we fix the limb darkening coefficients to precomputed values obtained with \texttt{ExoTiC-LD} \citep{david_grant_2022_7437681} using PHOENIX models \citep{husser_new_2013} integrated over the wavelengths covered by our observations. We adopt quadratic limb-darkening coefficients for full consistency with the NIRISS spectrum \citepalias{piaulet-ghorayeb_jwstniriss_2024}. We find that fitting the limb-darkening coefficients negatively impacts our ability to fit the light curves when allowing the orbital parameters to vary (which is important in a system with potential transit-timing variations), especially in the presence of instrumental and/or astrophysical systematics.

In order to assess our ability to fit jointly the light curves measured by each detector and in each visit, we performed initial light curve fits where the systematics model was a simple linear trend with time in order to capture the downward slope observed in the NRS1 detector. We found systematic differences between the measured transit times for the same astrophysical event but different NIRSpec detectors that range from less than 25 seconds for the first visit to about 30 seconds in the second visit, with the NRS2 detector systematically recording an earlier transit time. This difference is of the same order of magnitude as the $\sim 30$\,second NRS1-NRS2 time difference reported previously in NIRSpec/G395H observations of TOI-836\,c \citep{wallack_jwst_2024}. Such a mismatch between detector timestamps has been found to be widespread for observations obtained prior to March 2025 and can lead to a varying time offset between the NRS1 and NRS2 detectors across time series observations\footnote{\url{https://jwst-docs.stsci.edu/known-issues/nirspec-known-issues/nirspec-bots-known-issues\#NIRSpecBOTSKnownIssues-Detectortimestampdifferences}}. Since the uncertainty on our measured transit times is vastly shorter than these time offsets, we elect to fit each detector time series separately for each visit.

In terms of the planetary orbital parameters (scaled orbital distance $a/R_\star$ and impact parameter $b$), we sought to use the information contained in the NIRISS transits in order to prevent any offsets in the joint NIRISS+NIRSpec spectrum arising from the use of slightly different parameters (see e.g. \citealp{carter_benchmark_2024}). For the NIRISS analysis \citepalias{piaulet-ghorayeb_jwstniriss_2024}, both parameters were fitted to the cleaner light curve of the second visit, and their joint posterior was used as a correlated Gaussian prior to fit the first visit. We tried using the same prior to fit each of the NIRSpec light curves individually, but found that the transit shape could not be reproduced, even when fitting freely the limb-darkening coefficients rather than fixing them to model values. 
When letting both $a/R_\star$ and $b$ vary as free parameters, we found that the measured transit durations were 2 to 3 minutes larger than the median value predicted based on the NIRISS data obtained $\sim 1$\,year prior, regardless of whether limb-darkening coefficients were fixed or fitted. While small mutual inclinations associated with nodal precession would only produce TTVs of a few seconds on circular orbits, they may lead to transit duration variations (TDVs). Therefore, in all our fits, we hold $a/R_\star$ fixed to the median value retrieved from the first NIRISS visit (which used the first visit constraints as a prior), and only fit the impact parameter $b$. Since the NRS2 light curves have a higher scatter but observed the same astrophysical event probed by the NRS1 time series, we use the constraints on $b$ from each visit's NRS1 broadband light curve fit as a Gaussian prior for the NRS2 visit fit. 

Besides a visit-long slope, the light curves exhibit both short-timescale variability on the timescale of about 100 integrations, as well as longer-timescale smooth trends potentially due to stellar variability, as was observed in the NIRISS time series. We added terms to the systematics model that account for linear detrending with PCA eigenvalues for the top 4 components (i.e. up to the 5th component explaining the most variance, since the component encoding the planetary transit was not included in the detrending) and included in the final systematics model only the components where the corresponding slope parameter was nonzero within its uncertainties. For both the NRS1 and NRS2 light curves, our final model detrends against the second, fourth, and fifth PCA component from the NRS1 detector image decomposition of the relevant visit. The time series of NRS2 detector eigenvalues showed more scatter and were less efficient for detrending.
The PCA detrending alone already captures scatter on the order of 20 to 80 ppm in each broadband light curve (Fig. \ref{fig:wlc_fit}). 

We additionally fitted the a nonparametric Gaussian process (GP) model such that our final uncertainties accurately capture any residual correlated noise in the observations, which can arise from stellar surface features or granulation (e.g. \citealp{coulombe_highly_2025}). We use the Mat\'{e}rn 3/2 kernel implemented in the \texttt{celerite} module \citep{celerite} in order to encode correlation between residuals close in time (see \citealp{piaulet-ghorayeb_strict_2025} for a discussion). We fit two parameters $\log_{10} \sigma_\mathrm{GP}$ and $\log_{10} \rho_\mathrm{GP}$ where $\sigma_\mathrm{GP}$ is the correlation amplitude, and $\rho_\mathrm{GP}$ the correlation timescale. We bin our lightcurves by a factor of 20 prior to each fit which aids with the speed of GP model computation, and helps illuminate correlations between residuals and PCA eigenvalue variations. Finally, we noticed two in-transit features that when unaccounted for bias the transit depth to unphysically larger values in the visit 2 light curve for the first transit, which could be due to the transit chord crossing small regions of the photosphere that are brighter, such as plages or faculae, or stellar variability on shorter timescales than what the GP models. Since these features are not captured by the GP model because of their short timescale, we model them as Gaussians with fitted central time, amplitude, and width. 

Overall, our full model includes three fitted astrophysical parameters ($R_p/R_\star$, $b$, $T_c$), 5 systematics model parameters: a scale $c$, a slope with time, and 3 parameters for the PCA eigenvalues detrending. Three additional parameters describe the noise and covariance structure of the residuals: a white noise scatter term $\sigma_\mathrm{phot}$, as well as $\log_{10} \sigma_\mathrm{GP}$ and $\log_{10} \rho_\mathrm{GP}$. Where applicable, 6 additional parameters describe the bright in-transit crossing events as described above. The MCMC chains are run for a total of 10,500 steps, with 4 times as many walkers as fitted parameters, and we discard 60\% of the samples as burn-in to obtain the constraints on astrophysical parameters and light curve systematics (Table \ref{table:wlc_param}, Fig. \ref{fig:wlc_fit}).

This fitting procedure results in consistent broadband transit depths measured across all transits and detectors, and no residual red noise is apparent in the residuals when examining the binned-RMS with increasing bin size compared to white-noise expectations.


\begin{table*}
\caption{Fitted white-light-curve parameters for both JWST NIRSpec/G395H transits of GJ 9827\,d, reduced independently with \texttt{exoTEDRF} and \textit{Eureka!} and fitted using ExoTEP. The semi-major axis listed in the final block was held fixed at the value quoted, common to all fits. }
\label{table:wlc_param}
\centering
\begin{tabular}{lcccc}
\hline
\hline
Fitted parameter & \multicolumn{2}{c}{Visit 1} & \multicolumn{2}{c}{Visit 2} \\
\cline{2-3} \cline{4-5}
 & NRS1 & NRS2 & NRS1 & NRS2 \\
\hline
\multicolumn{5}{l}{\textbf{\texttt{exoTEDRF} reduction}} \\
Mid-transit time $T_c - 2460000$ [BJD$_\mathrm{TDB}$] & $469.76054 \pm 0.00014$ & $469.76026 \pm 0.00015$ & $643.40819 \pm 0.00012$ & $643.40786_{-0.00018}^{+0.00017}$ \\
Planet-to-star radius ratio $R_p/R_*$ & $0.03099_{-0.00056}^{+0.00044}$ & $0.03031_{-0.00046}^{+0.00037}$ & $0.03057_{-0.00053}^{+0.00059}$ & $0.03089_{-0.00054}^{+0.00058}$ \\
Impact parameter $b$ & $0.8834 \pm 0.0014$ & $0.8827 \pm 0.0011$ & $0.8830 \pm 0.0014$ & $0.8838 \pm 0.0010$ \\
Transit depth $(R_p/R_*)^2$ [ppm] & $960.2_{-34.1}^{+27.3}$ & $918.7_{-27.4}^{+22.8}$ & $934.7_{-32.3}^{+36.6}$ & $954.0_{-32.8}^{+35.9}$ \\
\hline
\multicolumn{5}{l}{\textbf{\textit{Eureka!} reduction}} \\
Mid-transit time $T_c - 2460000$ [BJD$_\mathrm{TDB}$] & $469.76040_{-0.00014}^{+0.00013}$ & $469.76037 \pm 0.00013$ & $643.40817 \pm 0.00014$ & $643.40781_{-0.00021}^{+0.00020}$ \\
Planet-to-star radius ratio $R_p/R_*$ & $0.03118_{-0.00044}^{+0.00039}$ & $0.03002_{-0.00041}^{+0.00036}$ & $0.03067_{-0.00056}^{+0.00064}$ & $0.03086_{-0.00065}^{+0.00072}$ \\
Impact parameter $b$ & $0.8842_{-0.0013}^{+0.0014}$ & $0.8828 \pm 0.0010$ & $0.8834 \pm 0.0014$ & $0.8838 \pm 0.0011$ \\
Transit depth $(R_p/R_*)^2$ [ppm] & $972.3_{-27.4}^{+24.3}$ & $901.1_{-24.5}^{+21.7}$ & $940.5_{-34.3}^{+39.4}$ & $952.2_{-39.9}^{+44.8}$ \\
\hline
Scaled semi-major axis $a/R_*$ & \multicolumn{4}{c}{$\equiv 19.56$} \\
\hline
\end{tabular}
\end{table*}

\subsection{Spectroscopic light-curve fitting}\label{sec:appendix_speclc}

\begin{figure*}
    \centering
    \includegraphics[width=0.85\textwidth]{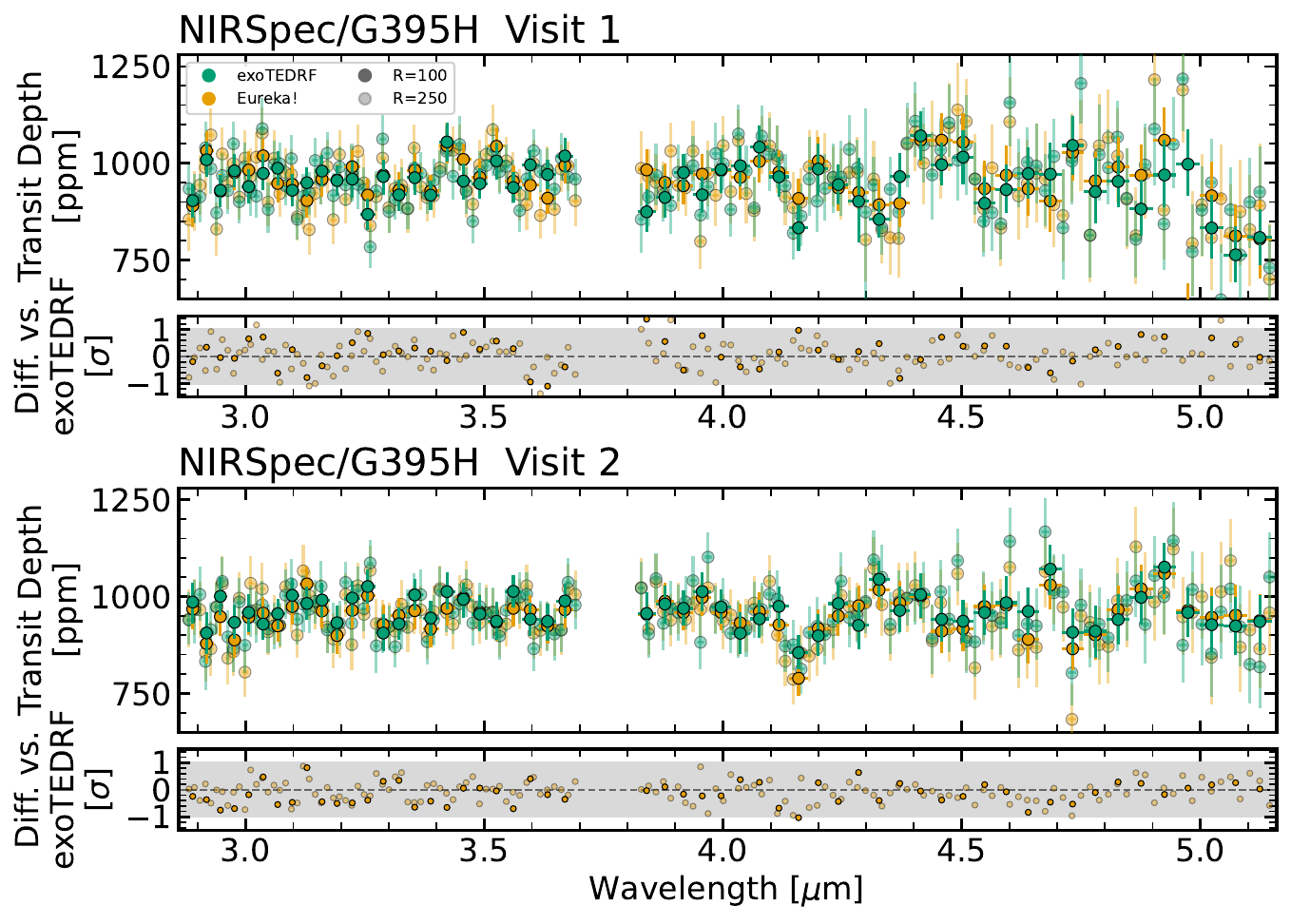}
    \vspace{5mm}
    \vspace{-5mm}\caption{Transmission spectra of GJ 9827\,d measured with JWST/NIRSpec G395H for the first (top panels) and second visit (bottom panels). The \texttt{exoTEDRF} and Eureka! spectra are shown in green and orange, with R=100 spectra shown in the foreground, and R=250 spectra displayed in semi-transparent color in the background. The residual panels show the difference between the Eureka! and \texttt{exoTEDRF} points in units of $\sigma$ ($\pm1\sigma$ region shaded in gray), demonstrating the excellent agreement between the two data reductions.}
    \label{fig:compare_reductions}
\end{figure*}

The light curves are binned on a wavelength grid shared across visits and for the two data reductions, and the astrophysical model we adopt has the planet-to-star radius ratio $R_p/R_\star$ as the only fitted astrophysical parameter for each spectroscopic light curve - with all other parameters set to their white light curve best-fit values. The limb-darkening coefficients in each wavelength bin are computed using \texttt{ExoTiC-LD}.

We test a few different approaches to account for the chromaticity in the spectroscopic light curve systematics. The light curve S/N in individual bins was insufficient for the Mat\'{e}rn 3/2 GP kernel lengthscale parameter to be reliably constrained. We therefore tried either keeping the lengthscale fixed to the white light curve best-fit value and only fitting a multiplicative factor to the $\sigma_\mathrm{GP}$ value constrained by the broadband light curve fit (`lenGP' approach), or simply fitting a wavelength-bin-specific scaling factor to the mean GP model predicted for the best-fit GP parameters in the white light curve fit stage (`scaleGP' hereafter). These two choices were motivated by  previous approaches used to account for chromaticity in the underlying correlation structure of the residuals \citep{piaulet-ghorayeb_strict_2025}, with the first method being the most flexible of the two since the GP kernel is applied to the specific residual structure of each wavelength bin. For the PCA component, we tried either fitting the individual eigenvalue-scaling coefficients to each light curve (`fitPCA'), or simply recording the PCA term in the white light curve best-fitting systematics model, and scaling this term by a fitted factor (`scalePCA'; assumes achromatic time variability). For the fitted Gaussians, we fixed the central-time and width of each Gaussian to the broadband light curve fit results and only fitted the amplitude independently in individual bins. While the amplitudes are fitted, their posterior distributions in individual spectroscopic bins span the entire prior range and we cannot constrain the spectrum of any potential stellar heterogeneities intersecting the transit chord. 

We selected the best combination of GP and PCA treatment based on visual inspection of the binned-RMS behaviour with increasing bin size, and of the residuals after best-fit model subtraction in each spectroscopic light curve. We found that the most successful prescription was the combination of `lenGP' and `scalePCA' for \texttt{exoTEDRF} (see Fig. \ref{fig:wlc_fit}), and `fitPCA' for \textit{Eureka!}, with no noticeable impact on the final spectrum (Fig. \ref{fig:compare_reductions}).

\section{Exploration of system dynamics}\label{sec:dynamical_analysis}

We perform dynamical simulations of the GJ\,9827 system where we explore the impact of planetary eccentricities and mutual inclinations on the amplitude and variability of GJ\,9827\,d's TTVs. For this analysis, we adopt the planetary parameters from \citet{passegger_compact_2024} for all planets and perform forward calculations with the intent of diagnosing the likely origin of the TTVs rather than performing a TTV fit, deferring system-wide TTV and photodynamical fits to future work (Livingston et al., in prep.).

\begin{figure*}
    \centering
    \includegraphics[width=0.9\textwidth]{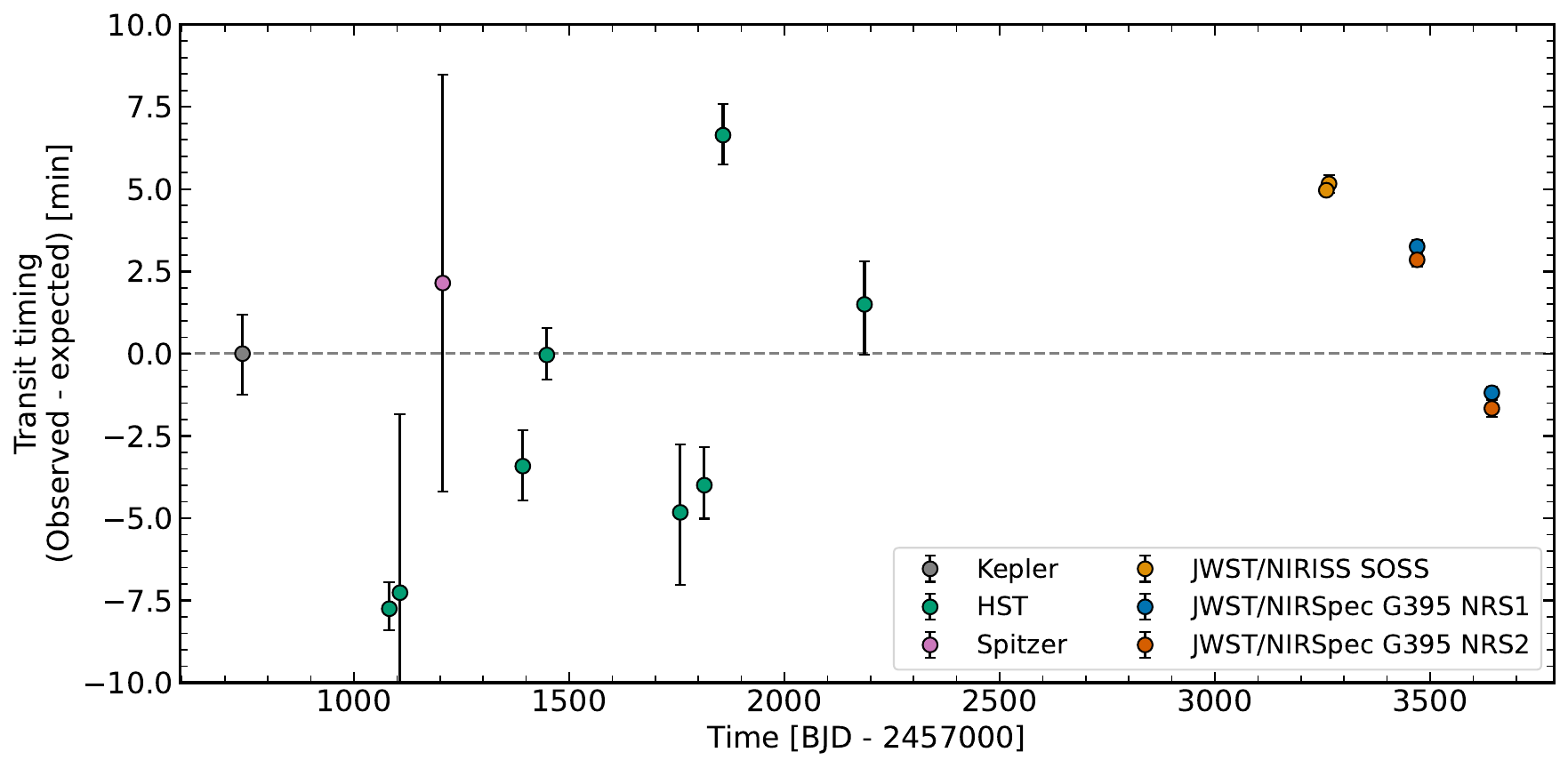}
    \vspace{5mm}
    \vspace{-5mm}\caption{Time series of the transit timing variations (TTVs) of GJ\,9827\,d measured with the \textit{Kepler}, \textit{HST}, \textit{Spitzer}, and \textit{JWST}, obtained after subtracting the \citet{kosiarek_physical_2021} ephemeris (dashed line, from \textit{Kepler} and \textit{Spitzer} data). We elect to use this ephemeris for this illustration rather than a recalculated ephemeris including all the JWST data, because the small JWST uncertainties dominate the least-squares fit and result in all pre-JWST datasets having negative TTVs. Two \textit{HST} measurements with uncertainties of $>15$ minutes on the measured transit time are not shown (see \citetalias{roy_water_2023}). }
    \label{fig:ttv_data}
\end{figure*}

We report both the distance to mean motion resonance $\Delta$ and the associated super-period $P_\mathrm{super}$ for each of the planet pairs in Table \ref{table:mmr_summary}, illustrating that none of the near-commensurabilities in the GJ\,9827 system are first-order. We calculate the quantities as follows:

\begin{equation}
    \Delta \equiv \frac{P_{\rm out}}{P_{\rm in}}\frac{j-k}{j} - 1
\end{equation}
\noindent where $P_{\rm out}$ and $P_{\rm in}$ are the orbital periods of the outer and inner planets, and the outer planet performs approximately $j$ orbits for every $k$ orbits of the inner planet. For the super-period, we report:

\begin{equation}
    P_{j} = \left| \frac{j}{P_{\rm out}} - \frac{j-k}{P_{\rm in}} \right|^{-1}.
\end{equation}

\begin{deluxetable*}{lccccc}
\tablecaption{Nearest mean-motion commensurabilities in the GJ 9827 system. We quote the fractional distance from resonance $\Delta$ and the superperiod, with uncertainties propagated from \citet{passegger_compact_2024}.
\label{table:mmr_summary}}
\tablewidth{0pt}
\tablehead{
  \colhead{Pair} &
  \colhead{$P_{\rm out}/P_{\rm in}$} &
  \colhead{Nearest MMR} &
  \colhead{Order} &
  \colhead{$\Delta$} &
  \colhead{$P_{\rm super}$ (days)}
}
\startdata
b--c & 3.01752 & 3:1 & 2 & $+0.00584$ & $208.26 \pm 0.10$ \\
c--d & 1.70001 & 5:3 & 2 & $+0.02001$ & $62.00 \pm 0.01$ \\
b--d & 5.12981 & 5:1 & 4 & $+0.02596$ & $47.77 \pm 0.01$ \\
\enddata
\end{deluxetable*}

While near-first-order MMR systems can lead to large TTVs even on circular orbits \citep{lithwick_extracting_2012}, the near-resonant TTV signal appears only for eccentric orbits in configurations involving second- and higher-order MMRs \citep{deck_transit_2016,Hadden2016}. We test four model setups (described below) to assess the impact of the planets' eccentricities and inclinations on the peak-to-peak amplitudes and RMS of the TTVs around a linear ephemeris.

We use \texttt{REBOUND} v5.1.1 \citep{rein_rebound_2012} together with the \texttt{WHFast} integrator \citep{rein_whfast_2015}. We adopt $dt=P_b/50$ in our forward integrations, which we check against the 15th-order adaptive-step-size integrator IAS15 \citep{rein_IAS15_2015}. Our setup includes the star and three planets, with baselines of 1,200 to 3,000 days comparable to the time difference between the \textit{K2} and \textit{JWST} observations (Fig. \ref{fig:ttv_data}). We find transit times by bisecting on the sky-plane crossing, with the observer along the $+x$ direction, flagging when the star-planet $y$ separation intersects zero from below, with $x>0$. 

We noticed that while the quoted eccentricity values from \citet{passegger_compact_2024} are $e_b = 0.03^{+0.03}_{-0.02}$, $e_c = 0.05^{+0.06}_{-0.03}$, $e_d = 0.13^{+0.12}_{-0.09}$, they are subject to the positive bias from marginalizing a 2D posterior on $\left(\sqrt{e} \cos \omega, \sqrt{e} \sin \omega \right)$ centered on zero. Therefore, we also test using instead the eccentricity values from the quoted medians on $\left(\sqrt{e} \cos \omega, \sqrt{e} \sin \omega \right)$ directly (see M4, M5 below).

\begin{deluxetable*}{cccccc}
\tablecaption{Results from the M4 model, illustrating the TTV rms and peak-to-peak (`pk-pk') amplitude for GJ\,9827\,d as a function of the $f$ factor multiplying the median $(\sqrt{e}\cos\omega,\,\sqrt{e}\sin\omega)$ components of
\citet{passegger_compact_2024}. The nominal median values are bolded. \label{tab:eccscan}}
\tablewidth{0pt}
\tablehead{
  \colhead{$f$} &
  \colhead{$e_{\rm b}$} &
  \colhead{$e_{\rm c}$} &
  \colhead{$e_{\rm d}$} &
  \colhead{rms} &
  \colhead{pk--pk}  \\
  \colhead{} & \colhead{} & \colhead{} & \colhead{} &
  \colhead{(min)} & \colhead{(min)} 
}
\startdata
0.30 & 0.0026 & 0.0013 & 0.0253 & 0.14 & 0.55  \\
0.50 & 0.0043 & 0.0022 & 0.0422 & 0.25 & 0.89  \\
0.70 & 0.0060 & 0.0031 & 0.0591 & 0.42 & 1.94  \\
0.85 & 0.0072 & 0.0038 & 0.0718 & 1.09 & 5.43  \\
\textbf{1.00} & \textbf{0.0085} & \textbf{0.0045} & \textbf{0.0845} & \textbf{3.73} & \textbf{15.47}  \\
1.30 & 0.0111 & 0.0059 & 0.1098 & 11.60 & 43.21  \\
\enddata

\end{deluxetable*}

For each simulation, we use a least-squares fit to the transit times at each of the modeled epochs to obtain the best-fit linear ephemeris and subtract it to obtain the time series of TTVs. Our four model setups are as follows:
\begin{itemize}
    \item M1: all orbits are circular and coplanar. This baseline case probes any synodic ``chopping'' terms.
    \item M2: coplanar orbits, with the eccentricities of all three planets fixed to $\{0, 0.03, 0.05, 0.1\}$. For each choice of $e$, we perform 12 random realizations for the pericenter phases of the three planets. This model serves as a test of the scaling of the TTV signal with eccentricity, and a probe of the uncertainty introduced by the unknown orientation of the pericenter.
    \item M3: circular orbits with nonzero mutual inclinations. The node and inclination $i_\mathrm{sky}=87.41^\circ$ is fixed for planet d, and we test three common node offsets for planets b and c (shared between the two planets): $\Delta \Omega \in \{20^\circ, 45^\circ,90^\circ \}$, which correspond to mutual inclinations of $i_\mathrm{mut} \in \{0.90^\circ, 1.98^\circ, 3.66^\circ\}$ respectively. For circular orbits, we track the transit duration as:

    \begin{equation}
        T_{14} \simeq \frac{P}{\pi}\frac{R_\star}{a}\sqrt{1-b^2}, \qquad b = \frac{a}{R_\star}\cos i.
    \end{equation}
    \item M4: coplanar orbits, with eccentricity values set to either (1) the marginalized median values quoted in \citet{passegger_compact_2024}, or (2) the vector medians $(e_b=0.0085, e_c=0.0045, e_d=0.085)$. We also run a series of models with the eccentricities scaled relative to case (2) by a constant factor $f \in \{0.3, 0.4, 0.5, 0.7, 0.85, 1.0, 1.3\}$. 
    
\end{itemize}

As expected, M1 leads to much smaller predicted TTVs than the $\sim 5$-$10$ minute peak-to-peak signal measured for GJ\,9827\,d (Fig. \ref{fig:ttv_data}) -- with a predicted amplitude of $<20$ seconds when no eccentricity is introduced. The M2 realizations reveal the breadth of the spread in expected TTV amplitudes when accounting for random pericenter phases. The min-max TTV peak-to-peak amplitude values range from 0.2 to 0.9 minutes for $e_b=e_c=e_d=0.03$, to 0.2-10 minutes for $e_b=e_c=e_d=0.05$, all the way to a 0.5 minutes - 188 minutes spread (median TTV full amplitude of 32 minutes) for $e_b=e_c=e_d=0.1$. In what follows, we report only median values, but this uncertainty should be kept in mind as it stresses the importance of system-specific fits rather than forward models for precise and accurate eccentricity estimates. In terms of the precession signal probed by M3, our models set an upper limit of $\lesssim 2^\circ$ on the mutual inclination, since the $\Delta \Omega=45^\circ$ model predicts TDVs of 23 seconds between the two NIRSpec transits, in $\sim 2.2\sigma$ tension with the measurements.

Finally, the M4 results (Table \ref{tab:eccscan}) suggest a low but non-zero eccentricity for GJ\,9827\,d. Matching a $5$ to $10$\,min TTV peak-to-peak signal requires $e_d$ of $0.07$-$0.085$. Meanwhile, the higher $e_d=0.13$ reported by \citet{passegger_compact_2024} from their marginalized estimate (in conjunction with their quoted values for the two other planets) would lead to predicted peak-to-peak amplitudes of 63 minutes, incompatible with the observations as they currently stand. We also note that the $e_d=0.13$ estimate would lead to a likely unstable dynamical configuration for the system, with $e_d\leq 0.1$ for dynamical stability (established using \texttt{spock}; \citealp{Tamayo2020} \citealp{kosiarek_physical_2021}).

\section{Atmosphere retrievals}\label{sec:appendix_retrievals}

We detail the sets of retrievals performed with SCARLET, \textit{Aurora}, and \texttt{POSEIDON} in the following subsections. 

\subsection{SCARLET} \label{sec:scarlet}

We use the SCARLET atmosphere modeling and Bayesian retrieval framework \citep{benneke_atmospheric_2012,benneke_how_2013,benneke_strict_2015,benneke_sub-neptune_2019, benneke_water_2019, pelletier_where_2021, piaulet_evidence_2023,piaulet-ghorayeb_jwstniriss_2024,piaulet-ghorayeb_strict_2025} to constrain the atmospheric composition of GJ 9827 d under three sets of assumptions regarding the chemistry. First, we perform free chemistry retrievals, where the abundances of each species are fitted as pressure-independent free parameters. Second, we conduct chemical equilibrium retrievals, where the abundances of each species throughout the atmosphere are dictated by thermochemical equilibrium given the atmospheric metallicity, C/O ratio, and local pressure and temperature. As an intermediate case between chemical equilibrium and free chemistry approaches, we additionally performed quenched-chemistry retrievals, where the upper atmosphere composition reflects chemical equilibrium at a deeper pressure $P_\mathrm{quench}$ and a characteristic temperature $T_\mathrm{quench}$, while the spectrum is calculated for a lower photosphere temperature $T_\mathrm{phot}$ \citep{benneke_jwst_2024}. 

The atmospheric forward model solves the radiative transfer equation for a plane-parallel atmosphere, where the abundances of the species present are dictated by the sampled composition parameters and hydrostatic equilibrium is solved accounting for the variation in the planet mass over the pressure grid \citep{piaulet_evidence_2023}. The considered opacity sources include H$_2$-H$_2$ and H$_2$-He collision-induced absorption (CIA), as well as H$_2$O \citep{ExoMol_H2O}, CH$_4$ \citep{Yurchenko2024}, CO \citep{Li2015}, CO$_2$ \citep{Yurchenko2026}, NH$_3$ \citep{ExoMol_NH3}, HCN \citep{Harris_HCN_2006,Barber2014}, SO$_2$ \citep{Exomol_SO2}, CS$_2$ \citep{sharpe_gas-phase_2004,gordon_hitran2020_2022}. N$_2$ is included as a potential component of the atmosphere but does not contribute to the absorption, except through N$_2$-N$_2$ CIA in N$_2$-dominated atmospheres \citep{gordon_hitran2020_2022}. We similarly include CO$_2$-CO$_2$ CIA \citep{Terragni2025}. In free chemistry retrievals, we test different assumptions in terms of the background gas motivated by the expected composition outcomes for $\sim 2\,R_\oplus$ sub-Neptunes \citep{cherubim_oxidation_2025}: H$_2$/He (with an assumed Jupiter-like He/H$_2$ = 0.157), N$_2$, or H$_2$O. We note, however, that just like H$_2$/He-dominated atmospheres, N$_2$-dominated atmospheres may also not be stable on GJ\,9827\,d unless mixed with a substantial fraction of metals such as CO$_2$ \citep{van_looveren_hazard_2025}, and treat these tests as probing different plausible mean molecular weight regimes. These retrieval variations enable us to assess the impact of the retrieval prior bias towards a large abundance of the ``filler'' gas on our inferences \citep{benneke_atmospheric_2012}. The forward model adjusts the planetary 10-mbar radius in each model so as to obtain the best match to the observed spectrum, iterating through the hydrostatic equilibrium and radiative transfer steps for consistency. The models are computed using line-by-line opacities at a resolving power of 15,625 (or 31,250 for the higher-resolution-model test) in the retrieval, and integrated within each observed spectral bandpass for the likelihood evaluation step. 

We marginalize over the impact of both potential gray clouds and hazes via two parameters: $p_\mathrm{cloud}$ which describes the cutoff pressure at which clouds become opaque, and $c_\mathrm{haze}$ representing the slope enhancement factor relative to Rayleigh scattering. For the temperature, we fit a single parameter $T_\mathrm{phot}$ representing the photosphere temperature which is probed over the limited pressure range observed in transmission. Finally, we also marginalize over any instrument offsets relative to NIRISS/SOSS order 1 for NIRISS/SOSS order 2, \textit{HST}/WFC3, NIRSpec/G395H NRS1, and NRS2. 

For sampling, we choose SCARLET's implementation of the \texttt{nestle} module \citep{skilling_nested_2004, skilling_nested_2006} for Nested Sampling, with at least 1,000 and at most 10,000 live points, and report the posterior distribution obtained from equally-weighted samples. 

We test the sensitivity of our results to the resolving power of the data ($R=100$ or $R=250$), of our forward models ($R=15,625$ vs. $R=31,250$), and to the transit-light-source effect (TLS) using the same methodology and set of stellar models for GJ\,9827 as in \citetalias{piaulet-ghorayeb_jwstniriss_2024} (from the \texttt{pymsg} package; \citealp{husser_new_2013,piaulet-ghorayeb_stctm_2025}). We find that our results are insensitive to these choices (Appendix Fig. \ref{fig:retrieval_variations}; quantitative results from the retrieval with TLS quoted in Appendix Table \ref{tab:retrieval_results}). Finally, we note that wavelengths 0.98 to 1.45 $\mu$m in the NIRISS spectrum suffered from undetected saturation. The JWST pipeline version available in 2024 lacked saturation flagging for datasets with $N_{\rm group}=2$, leading to this previous oversight. We perform H$_2$/He-background free chemistry and quenched-chemistry retrievals excluding this spectral region, and confirm that this partial saturation does not change our conclusions (Appendix Fig. \ref{fig:retrieval_variations}).

\begin{figure*}
    \centering
    \includegraphics[width=0.99\textwidth]{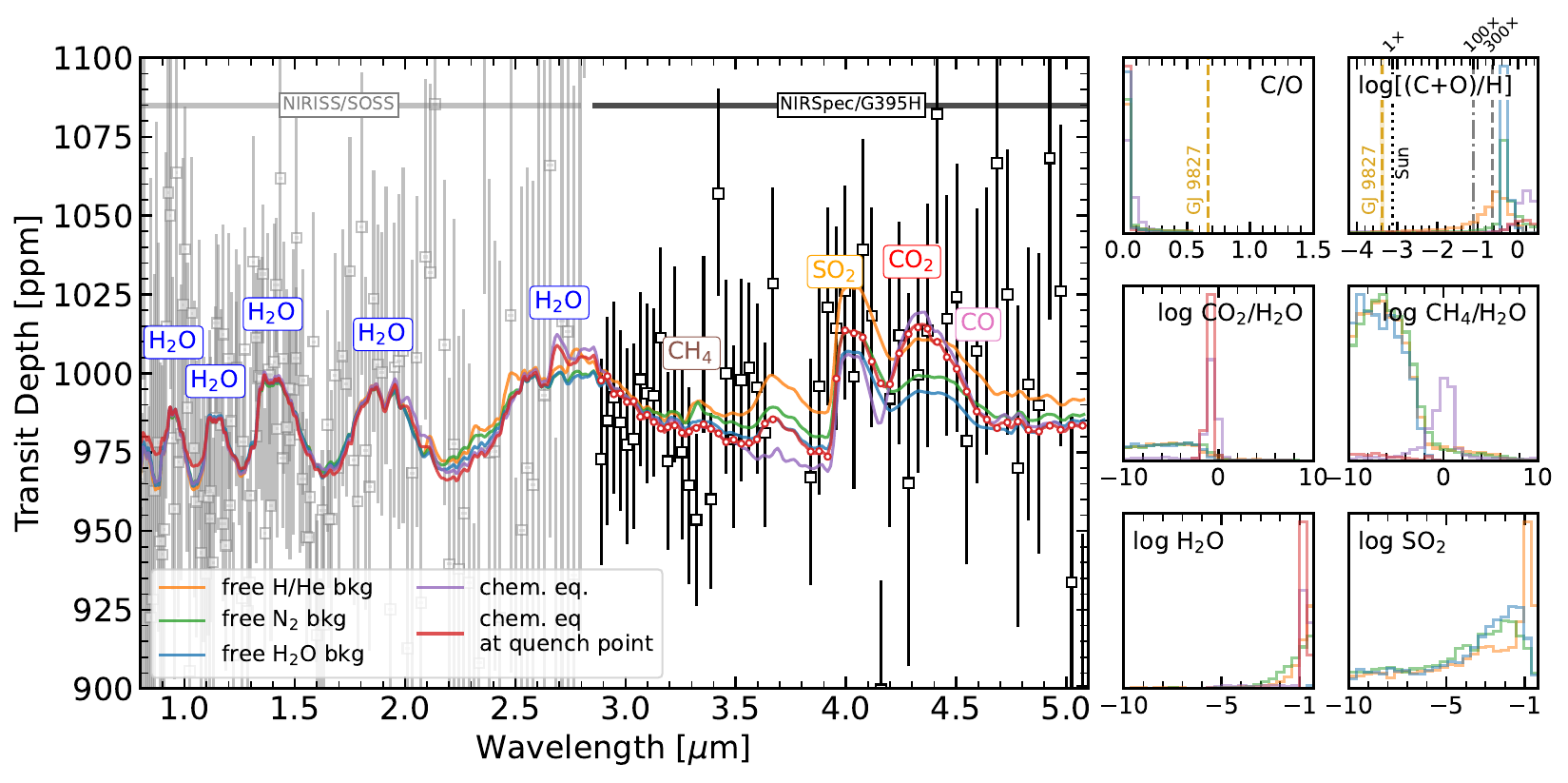}
    \vspace{5mm}
    \vspace{-5mm}\caption{Detailed breakdown of the atmospheric abundance constraints and modeling results from the nominal set of free-chemistry and chemically-consistent SCARLET retrievals. \textit{Left panel:} Measured HST/WFC3 and NIRISS/SOSS (gray squares) and NIRSpec/G395H (black squares) $R=100$ transmission spectrum reduced with \texttt{exoTEDRF} and fitted with ExoTEP, with prominent molecular features labeled and representative SCARLET models overlaid. We show the best-fit models from the three baseline free chemistry retrievals, each with a different background gas (see legend), as well as the chemical equilibrium retrieval (purple), and the quenched-chemistry retrieval (red). Bandpass-integrated predicted transit depths from the quenched-chemistry model are shown as red circles. \textit{Right:} Posterior and derived probability distributions on the molecular abundances, abundance ratios, C/O ratio and metallicity of GJ\,9827\,d, with colors matching the models shown on the left. The metallicity constraint is shown as (C+O)/H, with stellar and solar values annotated. We highlight metallicities corresponding to 1, 100, and 300$\times$ solar for reference. We extract from the chemically-consistent retrievals the photospheric abundances of CO$_2$, H$_2$O, CH$_4$, and SO$_2$ which are superimposed with the free-chemistry constraints in the corresponding panels. }
    \label{fig:nominal_retrievals_detailed_constraints}
\end{figure*}

We report the results from several variations from the baseline SCARLET atmosphere retrieval setup in Figs. \ref{fig:nominal_retrievals_detailed_constraints} and \ref{fig:retrieval_variations}, and in Table \ref{tab:retrieval_results}.

\begin{figure*}
    \centering
    \includegraphics[width=0.99\textwidth]{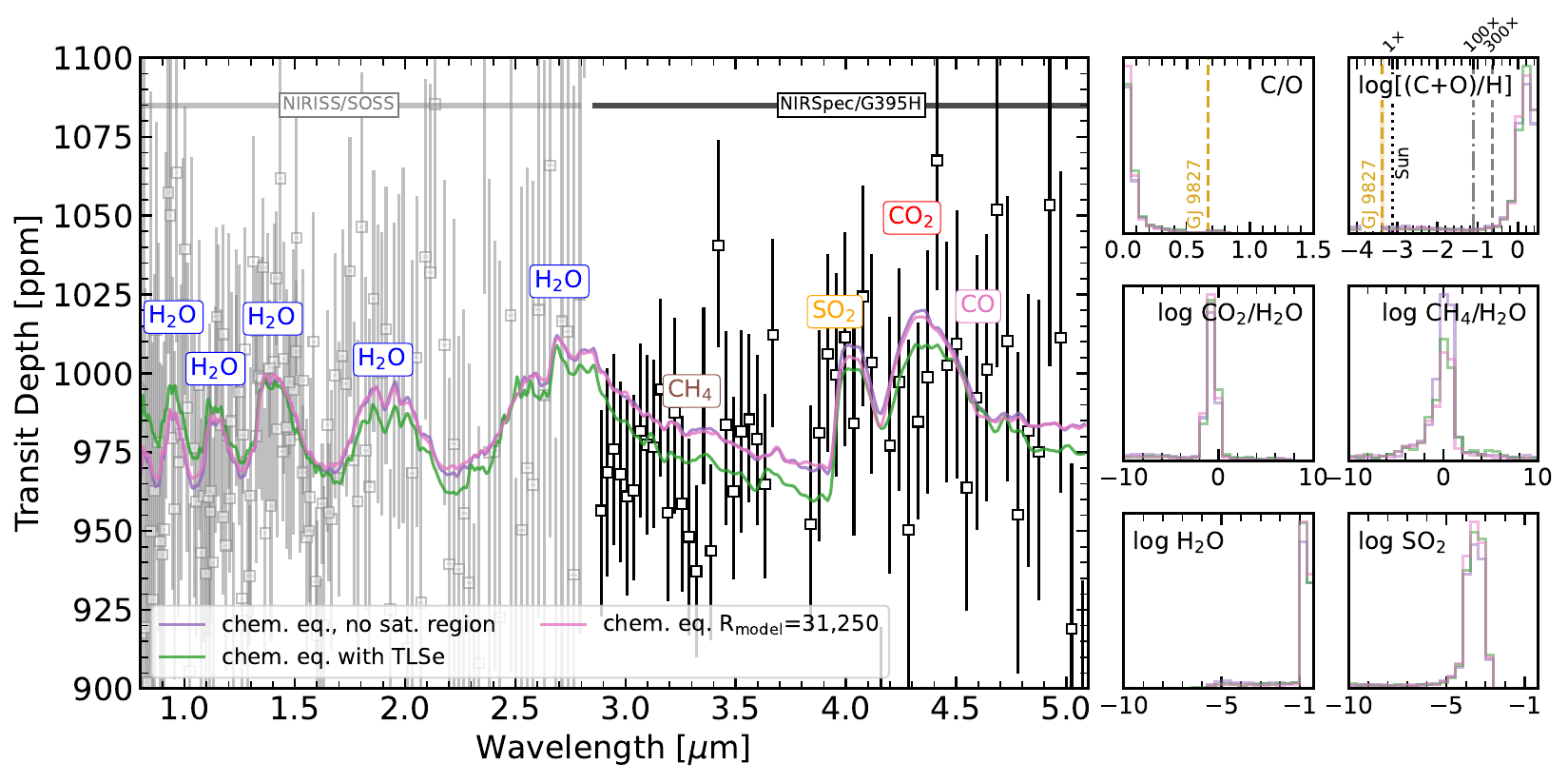}
    \vspace{5mm}
    \vspace{-5mm}\caption{Same as Fig. \ref{fig:nominal_retrievals_detailed_constraints} but for three variations from the baseline SCARLET chemical-equilibrium retrieval: excluding any saturated pixels from the NIRISS spectrum (purple), accounting for the TLSe (green), and increasing the model resolution by a factor of two (pink). The conclusion of a high metallicity, low-C/O atmosphere holds against each of these sensitivity tests.}
    \label{fig:retrieval_variations}
\end{figure*}

\begin{table*}
    \caption{\label{tab:retrieval_results} {Constraints on the atmospheric composition of GJ\,9827\,d from the suite of SCARLET retrievals performed on its combined HST/WFC3, JWST/NIRISS SOSS and JWST/NIRSpec G395H transmission spectrum. Parameters where the posterior spans the entire range of the prior across all runs are omitted, or reported as ``N/A'' for the runs where no constraints were obtained. ``with TLSe'' refers to the retrieval where the potential contribution of unocculted starspots was accounted for; in that case, the temperature difference between spots and the photosphere $\Delta T_\mathrm{spot}$ is assumed to be $<-50$\,K in the retrieval prior, while the spot covering fraction can range from 0 to 0.5, explaining the nonzero-retrieved $\Delta T_\mathrm{spot}$, while $f_\mathrm{spot}$ consistent with zero is in line with previous results from \citetalias{piaulet-ghorayeb_jwstniriss_2024}. Mean molecular weight (MMW) and metal mass fraction ($Z$) values are derived from the posterior samples, with $Z$ computed from elemental abundances. For free chemistry retrievals, the metallicity reported corresponds to the elemental (C$+$O)/H value, relative to solar. For chemically-consistent runs, the SO$_2$ abundance is not fitted independently, and the value quoted corresponds to the 10-mbar abundance of SO$_2$ in the models. Instrument offsets are defined relative to the NIRISS order 1 spectrum. For bounded constraints, we report the 68\% C.I., while lower or upper limits are reported at 2$\sigma$.}}
    \centering
    \begin{tabular}{lcccccc}
    \hline
    \hline
    Parameter & \multicolumn{3}{c}{Free Chemistry} & \multicolumn{3}{c}{Chemically-Consistent} \\
    & H/He bkg & N$_2$ bkg & H$_2$O bkg & Chem. Eq. & Quenched & With TLSe \\
    \hline
    \multicolumn{1}{l}{\textbf{Atmosphere composition}}\\
    $\log_{10}$ CH$_4$/H$_2$O & $<0.22$ & $<-0.42$ & $<-3.24$ & -- & -- & -- \\ 
    $\log_{10}$ CO$_2$/H$_2$O & $<1.07$ & $<0.41$ & $<-1.76$ & -- & -- & -- \\ 
    $\log_{10}$ NH$_3$/H$_2$O & $<0.05$ & $<-0.80$ & $<-4.00$ & -- & -- & -- \\ 
    $\log_{10}$ SO$_2^{(a)}$ & N/A & $-1.97^{+1.53}_{-3.59}$ & $-1.79^{+1.21}_{-3.38}$ & $-3.61^{+0.87}_{-1.68}$ & $-1.71^{+0.08}_{-0.04}$ & $-3.62^{+0.88}_{-1.65}$ \\ 
    C/O & $<0.33$ & $<0.31$ & $<0.10$ & $<0.40$ & $<0.04$ & $<0.46$ \\ 
    $\log_{10}$ M/H [$\times$solar] & $2.63^{+0.39}_{-0.57}$ & $2.83^{+0.48}_{-0.12}$ & $2.82^{+0.03}_{-0.01}$ & $>2.5$ & $>2.97$ & $>2.4$ \\ 
    MMW [amu] & $10.19^{+7.93}_{-6.54}$ & $>21.04$ & $18.13^{+1.29}_{-0.26}$ & $19.22^{+4.67}_{-2.11}$ & $19.13^{+0.09}_{-0.07}$ & $19.37^{+4.78}_{-2.09}$ \\ 
    Metal mass fraction $Z$ & $0.89^{+0.07}_{-0.23}$ & $>0.96$ & $0.889^{+0.013}_{-0.002}$ & $>0.80$ & $>0.93$ & $>0.8$ \\ 
    $\log_{10} p_\mathrm{cloud}$ [bar] & $-3.74^{+3.85}_{-0.86}$ & $>-4.61$ & $>-2.99$ & N/A & $>-2.92$ & N/A \\ 
    \multicolumn{4}{l}{\textbf{Instrument offsets}}\\
    $\delta_\mathrm{HST}$ [ppm]& $15.02^{+4.75}_{-5.14}$ & $14.76^{+4.04}_{-5.72}$ & $14.39^{+5.19}_{-4.54}$ & $13.88^{+5.42}_{-4.53}$ & $15.07^{+4.75}_{-5.28}$ & $15.21^{+3.72}_{-6.35}$ \\ 
    $\delta_\mathrm{NIRISS~o2}$ [ppm]& $-22.93^{+21.69}_{-13.37}$ & $-15.52^{+16.50}_{-16.08}$ & $-22.75^{+20.64}_{-13.35}$ & $-15.89^{+16.34}_{-19.28}$ & $-21.04^{+17.56}_{-17.29}$ & $-5.12^{+21.94}_{-18.88}$ \\ 
    $\delta_\mathrm{NIRSpec~NRS1}$ [ppm]& $78.66^{+7.64}_{-6.89}$ & $76.90^{+6.36}_{-5.27}$ & $77.29^{+6.93}_{-4.86}$ & $76.28^{+5.68}_{-6.35}$ & $78.69^{+6.23}_{-5.50}$ & $68.78^{+8.72}_{-7.06}$ \\ 
    $\delta_\mathrm{NIRSpec~NRS2}$ [ppm]& $84.43^{+21.82}_{-9.31}$ & $89.22^{+8.72}_{-10.19}$ & $87.07^{+8.75}_{-10.61}$ & $98.33^{+8.10}_{-11.19}$ & $102.76^{+7.70}_{-7.73}$ & $89.18^{+10.58}_{-14.05}$ \\ 
    \multicolumn{4}{l}{\textbf{Unocculted stellar heterogeneity}}\\
    $f_\mathrm{spot}$& -- & -- & -- & -- & -- & $0.06^{+0.14}_{-0.06}$ \\ 
    $\Delta T_\mathrm{spot}$ [K]& -- & -- & -- & -- & -- & $-126.27^{+76.21}_{-343.28}$ \\ 
    \hline
    \hline
    \\
    \end{tabular}
    
\end{table*}

\subsection{POSEIDON} \label{sec:poseidon}

We additionally use \texttt{POSEIDON} \citep{macdonald_hd_209458b_2017, macdonald_poseidon_2023} to conduct further independent atmospheric retrievals on GJ~9827~d. These retrievals jointly fit the combined JWST/NIRSpec G395H data presented here, the JWST/NIRISS SOSS spectrum from \citet{piaulet-ghorayeb_jwstniriss_2024} and HST/WFC3 spectrum from \citetalias{roy_water_2023}. Through \texttt{POSEIDON}, we use PyMultiNest (a nested sampling algorithm; \citealp{feroz_multinest_2009, buchner_x-ray_2014}), with 2000 live points per retrieval for Bayesian parameter inference and model comparison. Radiative transfer calculations are conducted using \texttt{TRIDENT}. Further details regarding the forward models and opacity database are described in \citet{macdonald_trident_2022}.

For each retrieval, we calculate model transmission spectra from 0.58 to 5.3\,$\mu$m on a wavelength grid with a resolving power of R = 20,000 using opacity sampling. The resulting spectra are convolved and integrated over the corresponding instrumental wavelength bins before comparison with the NIRISS SOSS, WFC3, and NIRSpec G395H observations. All our retrievals fit for three relative instrumental offsets: one for HST/WFC3, one for NIRSpec/G395H NRS1, and one for NIRSpec/G395H NRS2, while the NIRISS/SOSS spectra remain unshifted.

We consider three different types of planetary models. First, our primary atmospheric model (the results of which are reported in Section~\ref{sec:retrieval_comparison}) is a free-chemistry multi-gas atmosphere containing H$_2$, He, N$_2$, HCN, H$_2$O, CO, CO$_2$, CH$_4$, NH$_3$, H$_2$S, and SO$_2$. Second, to quantify the Bayesian evidence for H$_2$O, we additionally repeat the multi-gas retrieval with H$_2$O removed while otherwise retaining the same atmosphere as a nested model. Lastly, we fit an atmosphere-free model in which the planetary atmosphere is disabled, providing a flat-spectrum model for comparison. Multi-gas retrievals including H$_2$O are favored at 2.5$\sigma$ relative to retrievals without H$_2$O, consistent with previous results, with the slightly lower significance relative to our previous results (\citetalias{piaulet-ghorayeb_jwstniriss_2024}) likely driven by the addition of the NIRSpec/G395H data that does not contain water absorption features, but requires two additional parameters for the NRS1 and NRS2 offsets.

For models containing a planetary atmosphere, we fit a reference radius, R$_{\mathrm{p,ref}}$, at a reference pressure of 10 mbar and adopt an isothermal atmospheric temperature profile. We marginalize over aerosols using the \citet{macdonald_hd_209458b_2017} cloud and haze prescription, consisting of an optically thick gray cloud deck parameterized by the cloud-top pressure P$_\mathrm{cloud}$ and a two-parameter power-law haze described by log a$_\mathrm{haze}$ and $\gamma_\mathrm{haze}$. Our atmospheric models use the \texttt{POSEIDON} v1.3 opacity database, through which we adopt the line list from \citet{ExoMol_H2O} for H$_2$O, \citet{Yurchenko2024} for CH$_4$, \citet{Yurchenko2020} for CO$_2$, \citet{Li2015} for CO, \citet{Coles2019} for NH$_3$, \citet{Harris_HCN_2006} and \citet{Barber2014} for HCN, and \citet{Azzam2016} for H$_2$S. We also include all applicable CIA opacity pairs available in \texttt{POSEIDON} \citep{macdonald_trident_2022,Karman2019}. 
The retrieved molecular mixing ratios are assumed to be constant with pressure and follow a permutation-invariant centered-log-ratio (CLR) prior \citep{benneke_atmospheric_2012}. H$_2$ and He comprise the bulk atmosphere, with the remaining abundance after accounting for the retrieved molecular species assigned to H$_2$/He. 

Our retrieval priors span a wide range to encompass potential atmospheres, aerosol properties, and instrumental offsets. Our multi-gas retrievals use the following priors: R$_{\mathrm{p,ref}}$ = $\mathcal{U}$ (0.60R$_\mathrm{p}$, 1.15R$_\mathrm{p}$), T$_\star$ (K) = $\mathcal{U}$ (100, 1000), log a$_\mathrm{haze}$ = $\mathcal{U}$ (-4, 8), $\gamma_\mathrm{haze}$ = $\mathcal{U}$ (-20, 2), log P$_\mathrm{cloud}$ (bar) = $\mathcal{U}$ (-7, 2), and log X = $\mathcal{CLR}$ (-12, 0). The offset priors are $\mathcal{U}$(-500, 500) ppm for WFC3 and $\mathcal{U}$(-200, 200) ppm for both NRS1 and NRS2.

Lastly, we also include stellar contamination in our retrievals. For the flat, full multi-gas, and H$_2$O-excluded multi-gas models, we perform retrievals assuming no stellar heterogeneities, unocculted spots or faculae only, and both spots and faculae. \texttt{POSEIDON} incorporates these effects through a wavelength-dependent stellar contamination factor applied to the planetary transmission spectrum, with the photospheric and heterogeneous-region spectra generated by interpolating the PHOENIX stellar-atmosphere grid \citep{husser_new_2013} using \texttt{PyMSG} \citep{townsend_msg_2023}.
We adopt T$_\star$ = 4236 $\pm$ 12 K. For the single-heterogeneity models, we use f$_\mathrm{het}$ = $\mathcal{U}$ (0.0, 0.5), T$_\mathrm{het}$ (K) = $\mathcal{U}$ (2300, 1.2T$_\star$), and T$_\mathrm{phot}$ (K) = $\mathcal{N}$ (T$_\star$, 12). For models containing both spots and faculae, we instead retrieve five stellar parameters with priors f$_\mathrm{spot}$ = $\mathcal{U}$ (0.0, 0.5), f$_\mathrm{fac}$ = $\mathcal{U}$ (0.0, 0.5), T$_\mathrm{spot}$ (K) = $\mathcal{U}$ (2300, T$_\star$+36), T$_\mathrm{fac}$ (K) = $\mathcal{U}$ (T$_\star$-36, 1.2T$_\star$), and T$_\mathrm{phot}$ (K) = $\mathcal{N}$ (T$_\star$, 12). The stellar surface gravity is fixed to log g = 4.719 (cgs) \citep{passegger_compact_2024}. However, as the stellar-contamination models are disfavored relative to the corresponding uncontaminated models in our Bayesian model comparisons, we do not report their retrieved posteriors further. 

\subsection{Aurora} \label{sec:aurora}

We additionally model the transmission spectrum of GJ\,9827\,d using the \textit{Aurora} framework \citep{Welbanks2021}, which combines an atmospheric forward model with Bayesian parameter estimation. The version of \textit{Aurora} used here includes developments from a number of studies \citep{pinhas_retrieval_2018,Welbanks2019a,Nixon2020,Nixon2022,Nixon2024}. We adopt the stellar radius $R_*=0.58\,R_{\odot}$ and the planet mass $M_p=3.02\,M_{\oplus}$ from \citet{passegger_compact_2024}. We retrieve both the planet radius $R_p$ and the reference pressure $P_{\rm ref}$.

We assume an isothermal pressure-temperature (P-T) profile at temperature $T_{\rm iso}$, with 100 layers evenly spaced in log-pressure from 10$^{-9}$ to 100 bar. We assume that H\textsubscript{2} and He, in solar proportions (He/H\textsubscript{2}\,=\,0.17; \citealt{Asplund2009}), make up the remainder of the atmosphere after the species listed below. We include absorption from the following species: H\textsubscript{2}O \citep{Rothman2010}, CH\textsubscript{4} \citep{Yurchenko2014}, NH\textsubscript{3} \citep{Yurchenko2011}, HCN \citep{Barber2014}, CO \citep{Rothman2010}, CO\textsubscript{2} \citep{Rothman2010}, SO\textsubscript{2} \citep{Underwood2016}, H\textsubscript{2}S \citep{Azzam2016} and CS\textsubscript{2} \citep{gordon_hitran2020_2022}, as well as H\textsubscript{2}/H\textsubscript{2} and H\textsubscript{2}/He collision-induced absorption \citep{Richard2012}. We account for aerosol opacity following an inhomogeneous cloud and haze model \citep{line_influence_2016,macdonald_hd_209458b_2017}, in which a fraction $\phi$ of the terminator is covered by an optically thick grey cloud deck with cloud-top pressure $P_c$ and hazes, while the remaining $1-\phi$ is clear. The haze opacity is parametrised as $\sigma_{\rm haze}=a\,\sigma_0(\lambda/\lambda_0)^{\gamma}$, where $\sigma_0=5.31\times10^{-31}$\,m$^2$ is the H\textsubscript{2} Rayleigh scattering cross section at $\lambda_0=350$\,nm, $a$ is the Rayleigh enhancement factor and $\gamma$ is the scattering slope \citep{lecavelier_des_etangs_rayleigh_2008,macdonald_hd_209458b_2017}. We do not include the effect of unocculted stellar heterogeneities in these retrievals. We fit for transit depth offsets of the HST/WFC3, NIRSpec/G395H NRS1 and NRS2 spectra relative to the NIRISS/SOSS spectrum, with no offset between the two SOSS orders. The 19 free parameters have priors $\log_{10}X_i=\mathcal{U}(-12,\,-0.3)$ for each of the nine species, $T_{\rm iso}=\mathcal{U}(100,\,1000)$\,K, $\log_{10}(P_{\rm ref}/{\rm bar})=\mathcal{U}(-9,\,2)$, $R_p=\mathcal{N}(0.177,\,0.01)\,R_{\rm J}$, $\log_{10}a=\mathcal{U}(-4,\,10)$, $\gamma=\mathcal{U}(-20,\,2)$, $\log_{10}(P_c/{\rm bar})=\mathcal{U}(-9,\,2)$ and $\phi=\mathcal{U}(0,\,1)$, with offset priors of $\mathcal{U}(-500,\,500)$\,ppm for WFC3 and $\mathcal{U}(-200,\,200)$\,ppm for NRS1 and NRS2.

The above inputs are used to generate a high-resolution spectrum (0.55--5.3\,$\mu$m at $R=25{,}000$) using opacity sampling, which is binned to the resolution of the observations. We fit the NIRISS/SOSS spectrum of \citetalias{piaulet-ghorayeb_jwstniriss_2024} at $R=100$ (order 2: 0.60--0.85\,$\mu$m; order 1: 0.85--2.82\,$\mu$m), the HST/WFC3 G141 spectrum of \citetalias{roy_water_2023} (1.10--1.67\,$\mu$m) and the visit-averaged NIRSpec/G395H NRS1 (2.88--3.70\,$\mu$m) and NRS2 (3.82--5.16\,$\mu$m) spectra at $R=250$, giving 313 data points in total. The model parameter space is explored using the Nested Sampling algorithm \citep{skilling_nested_2006}, specifically PyMultiNest \citep{buchner_x-ray_2014}, a Python interface for MultiNest \citep{feroz_multinest_2009}, with 1000 live points.

\section{Stellar abundances}\label{sec:appendix_stellar_abundances}

Our custom modeling pipeline \citep[see][]{Gromek2026} constrains stellar elemental abundances by applying the spectral synthesis method to individual stellar lines. We use MARCS stellar atmosphere models \citep{gustafsson_grid_2008} as well as the Turbospectrum radiative transfer code \citep{alvarez_near-infrared_1998,plez_turbospectrum_2012}, implemented via modified iSpec routines \citep{blanco-cuaresma_determining_2014,blanco-cuaresma_modern_2019}. 

Our first step consists of a global continuum normalization and radial velocity shift. We then identify candidate spectral lines using the VALD3 line list \citep{Ryabchikova_2015}, followed by a visual inspection step to discard heavily contaminated or blended lines. 

For each spectral line, we generate a grid of synthetic spectra by varying the elemental abundance [X/H] between –0.75 and +0.75 dex in steps of 0.25 dex, followed by an interpolation of the model spectra to a finer grid with 0.015 dex resolution. The stellar parameters ($T_{\rm eff}$, [M/H],
$\log g$) used to synthesize the model spectra are adopted from \citet{passegger_compact_2024}. While more recent stellar parameters exist \citep{Howard2025}, they are derived from a reanalysis of a subset of the data presented in \citet{passegger_compact_2024}, and consistent with the ones we adopt (with a larger uncertainty on $T_\mathrm{eff}$). In order to address the impact of macro- and micro-turbulent velocities ($v_\mathrm{mac}$ and $v_\mathrm{mic}$). We treat them as nuisance parameters, derived by minimizing the scatter in the derived abundances of the CO lines, which are particularly sensitive to these effects, while varying $v_\mathrm{mac}$ and $v_\mathrm{mic}$ across steps of 0.25 km/s. Once the model spectra are synthesized and interpolated, we perform an additional pseudo-continuum normalization in the spectral region surrounding each line to match their continua to the observed spectrum, reducing biases in the final results. We then perform a $\chi^2$ minimization between the models and the observed spectrum within each line region in order to determine the best-fit abundances on a line-by-line basis. We report the abundance of each element as the weighted average of the abundances obtained for each spectral line, weighted by the line-depth divided by the line-specific root-mean-square error between the best-fit model and the observed spectrum. We assess the uncertainties due to the distribution of line-by-line abundance measurements $\sigma_\mathrm{ran}$, reporting the standard deviation of all the abundances measured for a single element divided by $\sqrt{N}$ where $N$ is the number of lines considered. The [M/H] value we report is calculated by converting each of the log-elemental abundances to absolute abundances, summing the contributions from all metals before converting back to [M/H] \citep{hinkel_concise_2022}, and repeating the process with 100,000 samples drawn from the posterior distributions on each measured [X/H] in order to obtain the quoted random uncertainty. We further estimate the systematic uncertainties associated with  $T_\mathrm{eff}$, [M/H], $\log~g$, $v_\mathrm{mac}$, and $v_\mathrm{mic}$ by independently resampling 15 times each stellar parameter from its Gaussian uncertainty distribution and repeating the full analysis \citep{Gromek2026}. The total uncertainty for each elemental abundance was computed by summing the random and systematic components in quadrature (see Fig. \ref{fig:stellar_abundances}, Table \ref{tab:stellar_abundances}). We find that the star appears marginally enriched in Mg and Si relative to Fe, which is in line with population-level trends for planet-hosting stars \citep{sharma_chemical_2024}.

\begin{deluxetable*}{lccc}
\tablecaption{Elemental abundances of GJ\,9827 derived from the 23 co-added SPIRou spectra. Abundances are reported relative to solar  \citep{asplund_chemical_2021}. The quoted uncertainties include both the line-by-line random error as well as the systematic contributions from the uncertainties on $T_{\rm eff}$, [M/H],
$\log g$, $v_{\rm mic}$, and $v_{\rm mac}$. \label{tab:stellar_abundances}}
\tablewidth{0pt}
\tablehead{
\colhead{Element} &  \colhead{$N_{\rm lines}$} &
\colhead{[X/H]} & \colhead{[X/Fe]}  \\
\colhead{} &  \colhead{} &
\colhead{(dex)} & \colhead{(dex)} 
}
\startdata
C  &  176 & $-0.23 \pm 0.12$ & $+0.09 \pm 0.13$  \\
O  &    35 & $-0.29 \pm 0.04$ & $+0.03 \pm 0.05$ \\
Na &   6 & $-0.23 \pm 0.03$ & $+0.10 \pm 0.05$ \\
Mg &    7 & $-0.17 \pm 0.18$ & $+0.15 \pm 0.18$ \\
Al &   11 & $-0.27 \pm 0.03$ & $+0.05 \pm 0.05$ \\
Si &   32 & $-0.26 \pm 0.04$ & $+0.06 \pm 0.06$ \\
K  &    4 & $-0.32 \pm 0.04$ & $+0.003 \pm 0.052$ \\
Ca &   17 & $-0.40 \pm 0.03$ & $-0.08 \pm 0.05$ \\
Sc &    2 & $-0.28 \pm 0.07$ & $+0.04 \pm 0.08$ \\
Ti &   57 & $-0.41 \pm 0.05$ & $-0.09 \pm 0.07$ \\
Cr &   28 & $-0.39 \pm 0.05$ & $-0.07 \pm 0.06$ \\
Mn &   9 & $-0.24 \pm 0.05$ & $+0.08 \pm 0.06$ \\
Fe &   80 & $-0.32 \pm 0.04$ & --           \\
Ni &    6 & $-0.31 \pm 0.08$ & $+0.02 \pm 0.09$ \\
\hline
\multicolumn{4}{c}{Derived quantities} \\
\hline
$[M/H]$  & \multicolumn{3}{c}{$-0.26 \pm 0.05$}  \\
$[\alpha/\mathrm{Fe}]$ & \multicolumn{3}{c}{$0.03 \pm 0.05$}  \\
C/O                       & \multicolumn{3}{c}{$0.67^{+0.22}_{-0.17}$}  \\
Mg/Si                     & \multicolumn{3}{c}{$1.35^{+0.71}_{-0.46}$}  \\
Fe/Si                     & \multicolumn{3}{c}{$0.78^{+0.11}_{-0.10}$}  \\ 
\\
\enddata
\tablecomments{C and O abundances are obtained from CO and OH molecular lines (see Section \ref{sec:appendix_stellar_abundances}).}
\end{deluxetable*}

\section{Bulk composition inference with SMINT}\label{sec:smint}

We use a new version of \texttt{smint} (Mirkin \& Piaulet-Ghorayeb in prep.) incorporating the evolutionary models of \citet{aguichine_evolution_2024} and \citet{tang_reassessing_2024} to infer the range of hydrogen and water mass fractions that are compatible with GJ\,9827\,d's properties, within the context of layered structure models with Earth-like interior compositions. 

\begin{figure*}
    \centering
    \includegraphics[width=0.7\textwidth]{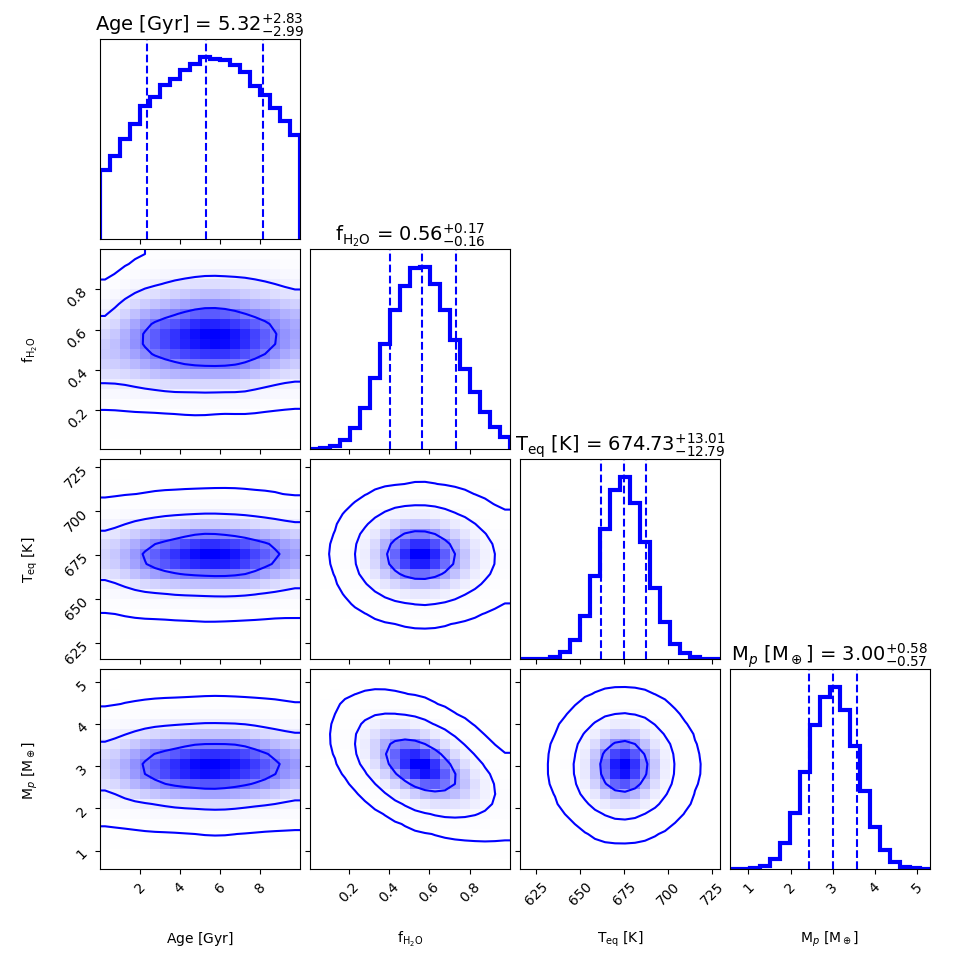}
    \vspace{5mm}
    \vspace{-5mm}\caption{Results from the water mass fraction inference for GJ\,9827\,d under the ``water world'' hypothesis using \texttt{smint}. All models assume an Earth-like interior composition, and we present the joint and marginalized posterior distributions on the water mass fraction and system age. The contours indicate 1-, 2-, and 3-$\sigma$ confidence intervals.}
    \label{fig:water_world_smint}
\end{figure*}

For the \citet{aguichine_evolution_2024} models, we adopt the M-star evolution model grid rather than the G dwarf grid, since GJ\,9827 is a K7-type star. Within the Bayesian inference framework of \texttt{smint}, we adopt the following literature-based priors: a Gaussian prior on the planetary mass, stellar age, and equilibrium temperature (the latter from Monte Carlo sampling of the orbital parameter constraints, assuming a zero Bond albedo), following \citet{passegger_compact_2024}. The Gaussian likelihood calculation matches the measured planet radius \citepalias{piaulet-ghorayeb_jwstniriss_2024} to the 20-mbar radius predicted by our interpolated \citet{aguichine_evolution_2024} grid. We sample the parameter space of \{age; bulk water mass fraction $f_\mathrm{H_2O}$; $T_\mathrm{eq}$; $M_p$\} using the \texttt{emcee} sampler \citep{foreman-mackey_emcee_2013}, with four times as many walkers as fitted parameters. The chains run for 10,000 steps, which is $>250\times$ the mean autocorrelation time \citep{goodman_ensemble_2010}; the first 60\% of each chain are discarded as burn-in to produce the posterior distributions (Fig. \ref{fig:water_world_smint}).

\begin{figure*}
    \centering
    \includegraphics[width=0.9\textwidth]{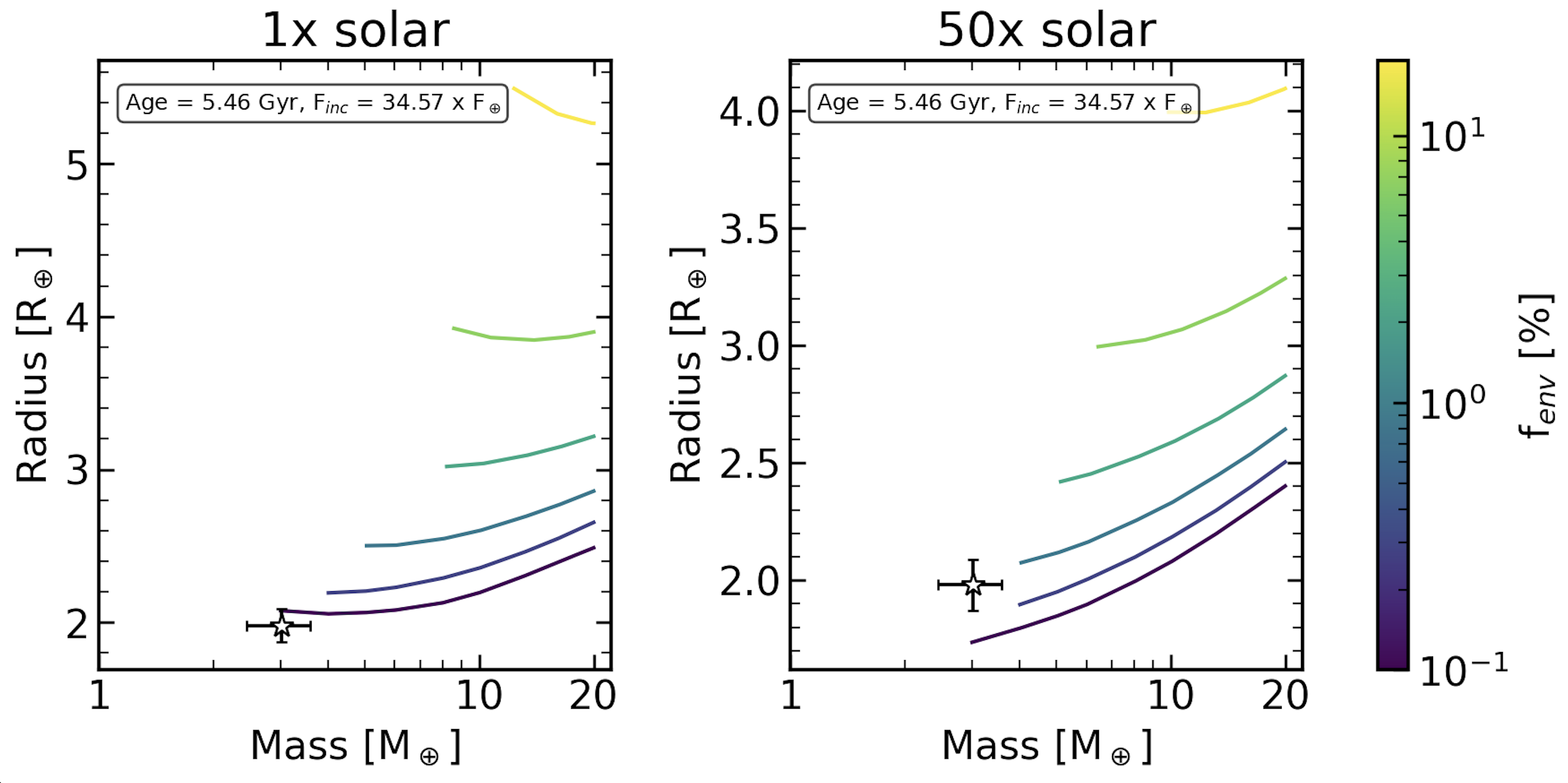}
    \vspace{5mm}
    \vspace{-5mm}\caption{Interpolated mass-radius curves from the \citet{tang_reassessing_2024} model grids using \texttt{smint}, for 1$\times$solar (left) and 50$\times$solar metallicity (right) envelopes and at the median age and insolation flux drawn from our prior distribution for GJ\,9827\,d. The colors of the mass-radius curves encode different few values of the envelope mass fraction (see colorbar). The lower bound on planet mass for each curve is set by the validity range of the models presented in \citet{tang_reassessing_2024}. The small size and low density of GJ\,9827\,d place it outside of the predictions of such ``gas dwarf'' models.}
    \label{fig:tang_hhe_models}
\end{figure*}

Since GJ 9827\,d's parameters lie outside of the range of masses and radii predicted by the \citet{tang_reassessing_2024} models regardless of metallicity, we simply show the position of the planet relative to constant-composition curves in Figure \ref{fig:tang_hhe_models}.

\section{Global Chemical Equilibrium Modeling}\label{sec:appendix_modeling_GCE}

Our adopted model is described in detail in \citet{grimm_new_2026}, and we summarize our specific setup here and in Fig. \ref{fig:stellar_abundances}. In what follows, we refer to the ``core'' as the solid/molten interior of the planet underlying the atmosphere.

The total mass of the planet is set to the median value from the global transit+RV fit from \citet{passegger_compact_2024}, and the pressure at the  AMOI is calculated self-consistently from the equilibrated gas-phase (atmosphere) mass. 
The initial elemental budget is set in three steps. First, the baseline elemental abundances of Si and Na are fixed to a bulk Earth-like preset (see \citealp{young_earth_2023,werlen_atmospheric_2025,grimm_new_2026}). For carbon, we introduce 0.5 wt\% C in the composition of the solids (6.5\% in terms of C/Si molar ratio; \citealp{werlen_atmospheric_2025}), which is similar to estimates of the Earth's core carbon content (see e.g. \citealp{Wood2013}). The bulk Mg and Fe abundances of the solids are set by the user-imposed Mg/Si and Fe/Si ratios, which we nominally fix to median values derived from stellar abundances, although this assumption is relaxed as a sensitivity test (see Appendix Section \ref{sec:appendix_modeling_GCE}). Water is added to the total as H and O such that its final mass fraction matches the prescribed value for $w_\mathrm{H2O}$, then the addition of primary gas (with mass fraction $w_\mathrm{H2}$ of the total) further impacts the H, O, and C budget (98 wt\% H, and 2 wt\% C and O). We modify the preset C/O ratio of the primordial gas to match the value inferred through stellar characterization (Table \ref{tab:stellar_abundances}). Since He is generally unreactive and is not included in our model \citep{schlichting_chemical_2022,grimm_new_2026}, it is not accounted for in the GCE modeling. However, we introduce He in solar abundance when computing atmosphere models with SCARLET to interpret the GCE outputs in terms of photosphere composition. Although magma-atmosphere interactions can impact both nitrogen and sulfur-bearing species, our model does not include N or S chemistry. Since no prominent N or S-bearing species are meaningfully constrained by our observations of GJ\,9827\,d, we focus our interpretation on C- and O-bearing species.

The final partitioning of species across reservoirs is then solved for given the 19 reactions considered \citep{young_earth_2023,werlen_atmospheric_2025}, with the equilibrium conditions set by the Gibbs free energies evaluated at the temperature at the AMOI and the temperature $T_\mathrm{s-m}$ at the boundary between the silicate and metallic phases. Mass balance is tracked for each of the seven elements (H, C, O, Na, Mg, Si, Fe), and we enforce the closure conditions of mole fractions summing to unity within each phase. We set $T_\mathrm{s-m}=T_\mathrm{AMOI}$, supported by the single-phase behaviour of the MgSiO$_3$-H$_2$-Fe system for planets formed with $\gtrsim 1$\% H$_2$ by mass \citep{young_differentiation_2025}. We note that since $T_\mathrm{s-m}$ is the average temperature for metal-silicate equilibration, it is likely to be slightly higher than the AMOI temperature, but the validity range of the available Gibbs free energy data does not extend to temperatures $\gtrsim 4,000$\,K \citep{grimm_new_2026}. Coupled interior-atmosphere modeling of GJ\,9827\,d suggested values of $T_\mathrm{AMOI}$ ranging from $\sim 5,000$\,K for cloud-free atmospheres, to $\sim 6,500$\,K when the radiative feedback of clouds is accounted for. These modeled structures matched both the mass and radius of GJ\,9827\,d as well as its metal-enriched composition, but did not account for thermal evolution or the constraint on $T_\mathrm{AMOI}$ imposed by the binodal curve representing two-phase equilibrium between the silicate and gas-dominated phases, which restricts it to $\lesssim 4,500$\,K \citep{young_differentiation_2025}. We therefore elect to choose a nominal value of $T_\mathrm{AMOI}=3,500$\,K, which is in agreement with typical temperatures along the binodal curve \citep{young_differentiation_2025}. We performed a sensitivity test on $T_\mathrm{AMOI}$ (see Appendix Section \ref{sec:appendix_modeling_GCE}, Fig. \ref{fig:sensitivity_Tamoi}) demonstrating negligible impact of a higher $T_\mathrm{s-m}$ on our conclusions. 

Non-ideal behaviour is accounted for in the metal phase for silicon, oxygen, and carbon \citep{young_earth_2023, werlen_atmospheric_2025}, which is a departure from \citet{schlichting_chemical_2022}. We ignore the non-ideality in the gas and silicate phase as described in \citet{werlen_effects_2026}, who found that including activity coefficients for only a subset of species in these phases can lead to artificial trends. We explore different prescriptions to describe the activity coefficient of carbon specific to its partitioning into the metal phase, and elect to turn off this process altogether (Appendix \ref{sec:carbon_partitioning}, Fig. \ref{fig:lngc_atmosphere_comparison}). The baseline implementation of C partitioning 
resulted in virtually no carbon in the atmosphere of planets formed water-poor \citep{steinmeyer_coupled_2026,grimm_new_2026}, which cannot reproduce the hydrogen-rich compositions of e.g. K2-18\,b, TOI-270\,d, or TOI-1130\,b where C species have been detected \citep{madhusudhan_carbon-bearing_2023,shorttle_distinguishing_2024,wogan_jwst_2024,benneke_jwst_2024,nixon_magma_2025,barat_jwst_2026}. 

\begin{figure*}
    \centering
    \includegraphics[width=0.9\textwidth]{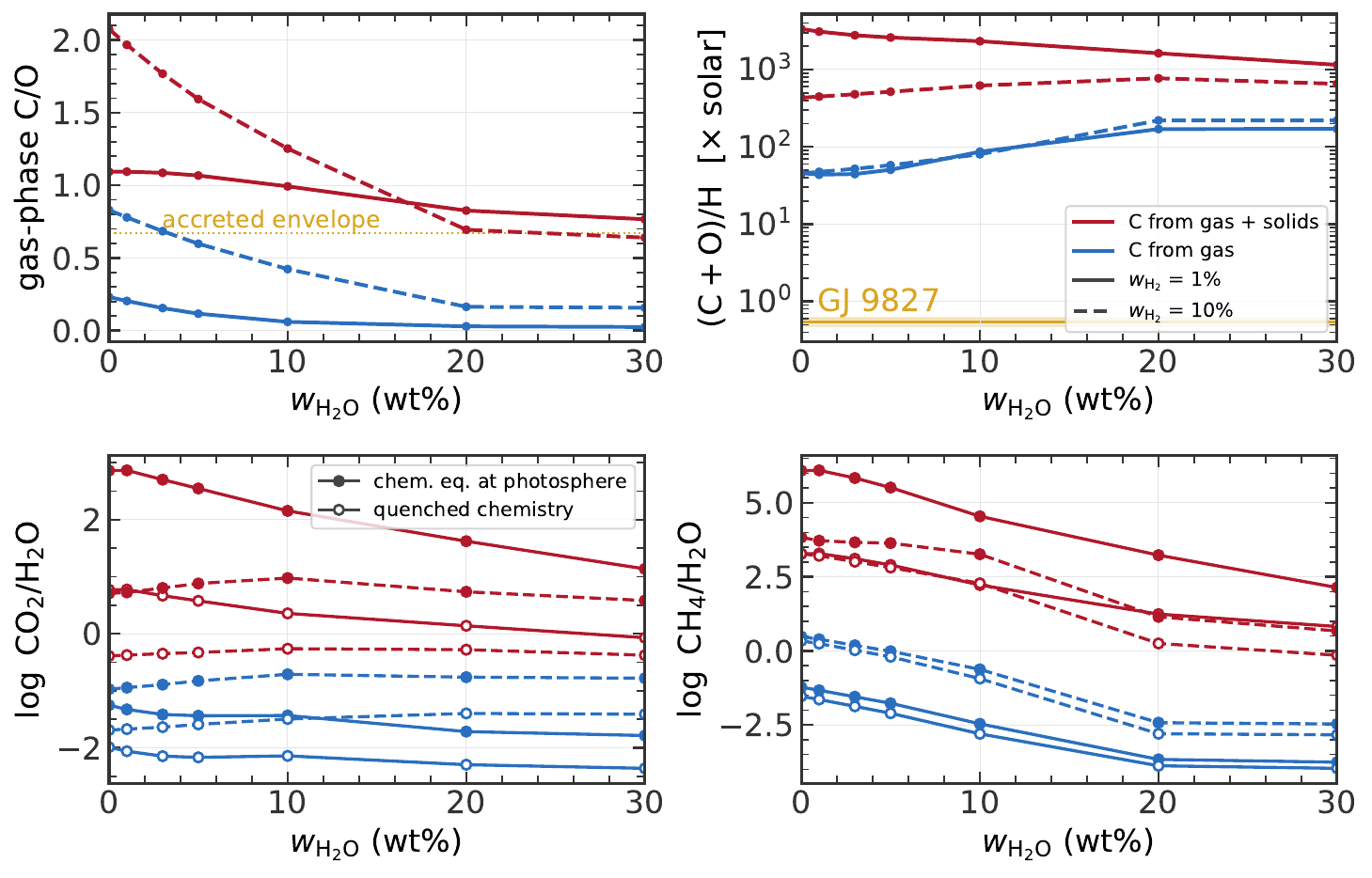}
    \vspace{5mm}
    \vspace{-5mm}\caption{Summary of the magma ocean chemistry outcomes for GJ\,9827\,d. Two values are tested for the initial C inventory (either from both the accreted solids and the primordial gas - in red, or only from the gas - in blue), and the mass fraction of primordial gas, $w_{\rm H2}$ (solid vs. dashed lines). \textit{Top left panel:} Comparison between the accreted envelope C/O ratio (set to the stellar value shown as the dotted line) and the C/O ratio of the equilibrated gas-phase composition. Introducing carbon into the composition of the solids is enough to push the C/O ratio to super-solar (and super-stellar) values, suggesting either a carbon-poor formation environment for GJ\,9827\,d, or an atmospheric carbon depletion mechanism. \textit{Top right panel:} Predicted atmospheric metallicities (from relative abundances of C, O, and H) for each of the model setups, with the corresponding stellar value shown as the yellow band. The high metallicity retrieved for GJ\,9827\,d is consistent with models where abundant additional oxygen is provided through accretion of water-rich bodies (high $w_{\rm H_2O}$). \textit{Bottom panels:} Molecular abundance ratios predicted for GJ\,9827\,d from each of the sets of elemental abundances represented in the GCE model outcomes, either in a case where chemical equilibrium holds all the way to the photosphere (filled circles) or to 1 bar, in a quenched chemistry calculation (open circles). The CH$_4$/H$_2$O ratio appears to be a stronger independent tracer of C content, with a smaller dependence on the chemical quenching assumptions.}
    \label{fig:GCE_2x2panel}
\end{figure*}

\begin{figure*}
    \centering
    \includegraphics[width=0.9\textwidth]{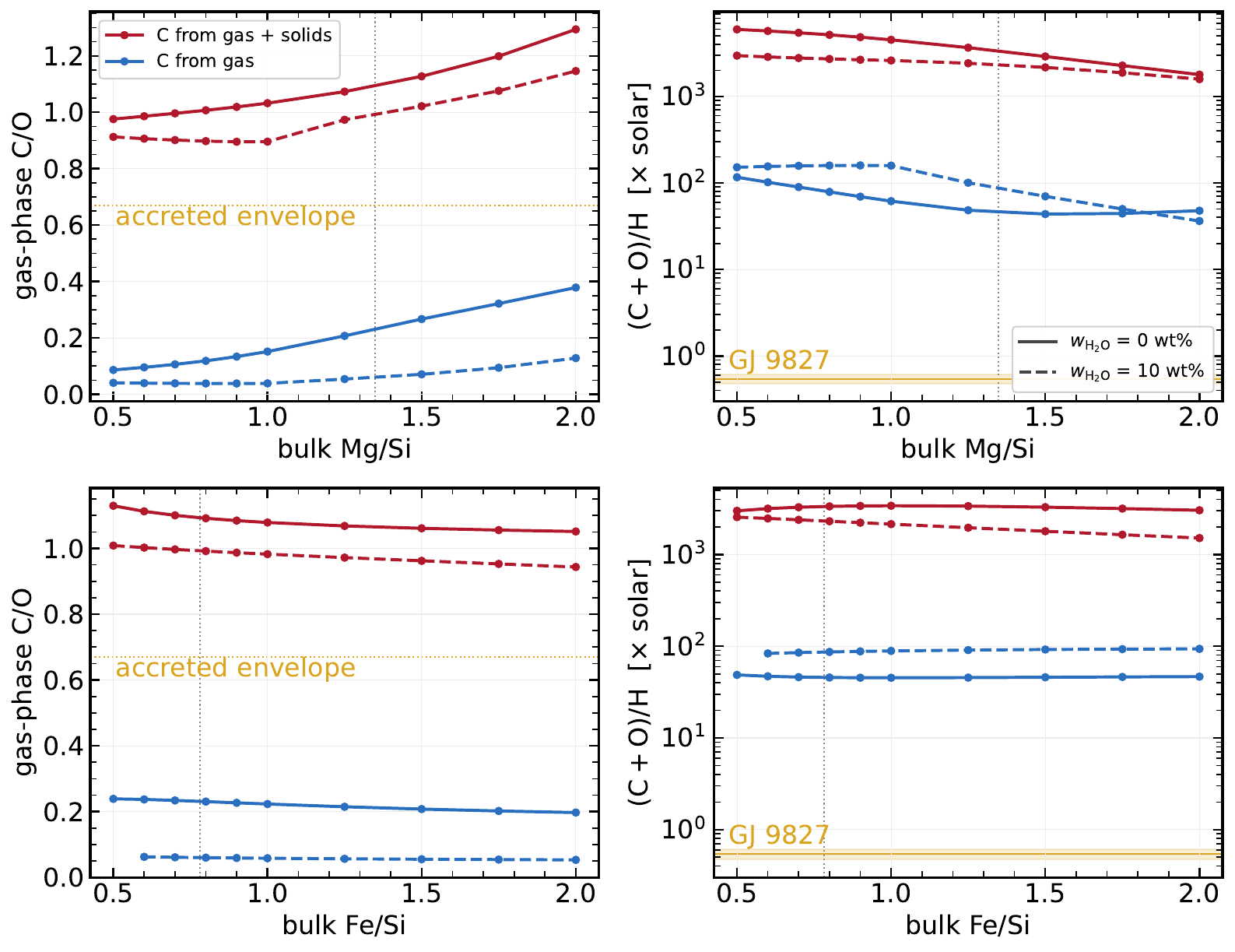}
    \vspace{5mm}
    \vspace{-5mm}\caption{Exploration of the sensitivity of magma ocean-atmosphere equilibration outcomes to the assumed Mg/Si (top panels) and Fe/Si (bottom panels). For two initial carbon inventories (envelope C only in blue, and additional C from the solids in red), and two values of $w_{\rm H2O}$ (solid and dashed lines), we show how the gas-phase C/O ratio (left panels) and metallicity (right panels) evolves based on bulk elemental ratios. The simulations assume $w_{\rm H2}=1$\% and $T_{\rm AMOI}=3500$\,K. The stellar C/O and metallicity are highlighted in orange. Overall, sub-Neptune compositional outcomes depend very weakly on the bulk compositions, except for the C/O ratio which can increase by 0.2 to $\sim 0.4$ for high Mg/Si ratios. These findings contrast starkly with those of \citet{grimm_new_2026} which found a high dependence of both the C/O ratio and metallicity on the relative amounts of Mg, Si, and Fe -- highlighting the impact of the adopted C partitioning treatment (see Section \ref{sec:carbon_partitioning}).}
    \label{fig:Mg_Fe_Si_tests}
\end{figure*}

We present the full results of our baseline GCE calculations, detailing the dependence of the gas-phase C/O ratio, metallicity, and abundance ratios at each $w_{\rm H2O}$ point in Fig. \ref{fig:GCE_2x2panel}. Our sensitivity tests on the impact of varying $T_\mathrm{AMOI}$ and the bulk elemental ratios Fe/Si, Mg/Si are presented in Figs. \ref{fig:sensitivity_Tamoi}
 and \ref{fig:Mg_Fe_Si_tests}.
 
\begin{figure*}
    \centering
    \includegraphics[width=0.9\textwidth]{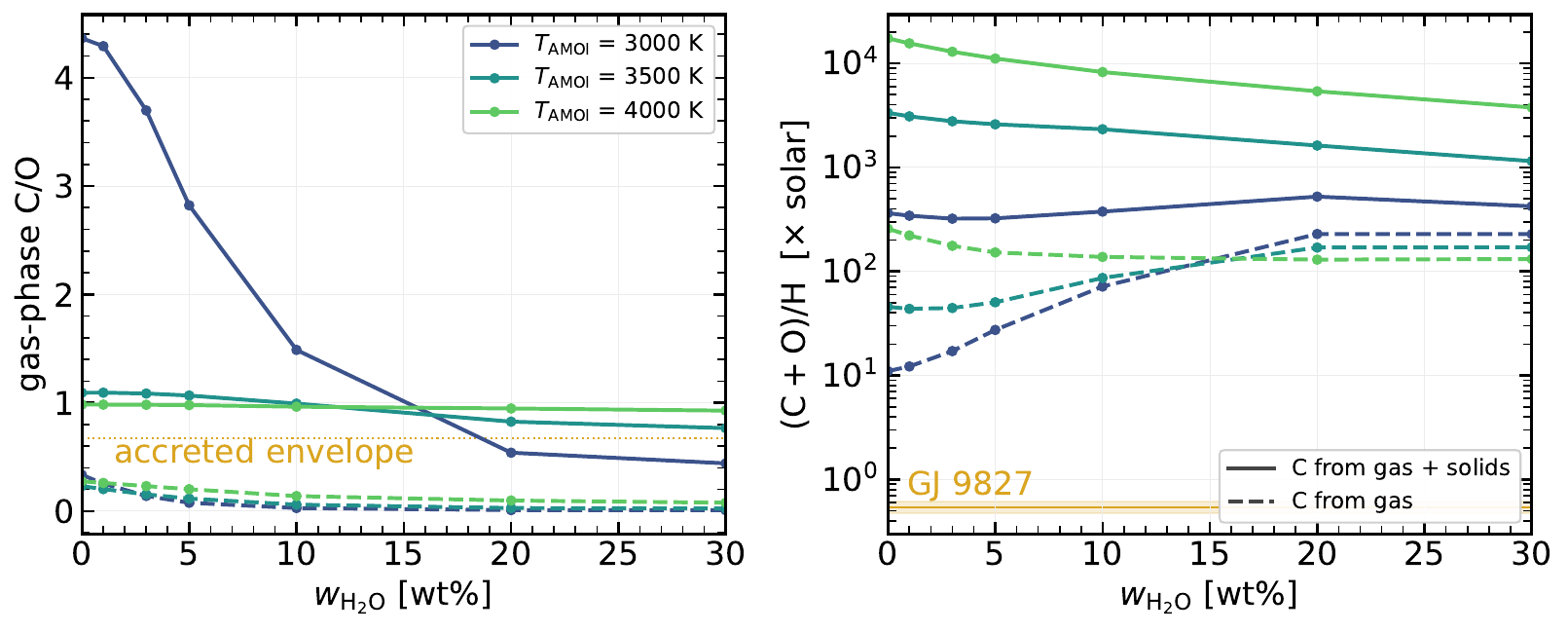}
    \vspace{5mm}
    \vspace{-5mm}\caption{Impact of the value of $T_{\rm AMOI}$ on the predicted atmospheric compositions from the GCE model. We show the gas-phase equilibrated C/O ratio (left panel) and metallicity (right panel) as a function of $w_{\rm H2O}$, for $w_{\rm H2}=1$\% and three values of the atmosphere-magma ocean interface temperature (shades of green). Stellar C/O and metallicity values are shown in orange. The values tested encompass the range of plausible temperatures from the theory of silicate-hydrogen phase equilibration. We find that compositions are most sensitive to the $T_{\rm AMOI}$ assumption for water-poor conditions -- but do not change to the extent that introducing an additional C source form the solids (solid lines) leads to high predicted C/O ratios compared to the observations.}
    \label{fig:sensitivity_Tamoi}
\end{figure*}

\section{Carbon Partitioning Treatment}\label{sec:carbon_partitioning}
\setcounter{equation}{0}

Carbon partitioning between the metal and silicate phases was introduced into the global equilibrium framework of \citet{schlichting_chemical_2022} by \citet{werlen_atmospheric_2025}, through the reaction
${\rm C}_{\rm metal} + {\rm O}_{\rm metal} \rightleftharpoons {\rm CO}_{\rm silicate}$. This extension enables the framework to address the metal-phase carbon budget, which the tests describe below build up on. In adapting the implementation to our grid of initial water contents, we found that the governing factor for the high variation in C/O ratio with initial ice content was the prescription adopted for the activity coefficient of carbon in the metal, $\gamma_{C_{\rm metal}}$. We document this sensitivity here insofar as it bears directly on the interpretation of our results. 

The treatment draws on two distinct experimental studies: the standard-state Gibbs free energy of the reaction, $\Delta \hat{G}^{\circ}_{\rm rxn}$, is calibrated on the metal-silicate carbon
partitioning measurements reported in \citet{blanchard_metal-silicate_2022}, which
\citet{werlen_atmospheric_2025} adopt in preference to
\citet{grewal_delivery_2019} and \citet{fischer_carbon_2020} on the grounds of the $^{13}$C methodology and the exclusion of quenched metal blebs. The reaction free energy is set by the associated $\log K_D(T)$ fits. However, the compositional non-ideality of carbon in the metal is accounted for through an activity coefficient of the form
\begin{equation}
\ln \gamma_{\mathrm{C}}^{\mathrm{metal}}
  = -\,2.303 \times 19.5 \times
    \ln\!\left(1 - X_{\mathrm{O}}^{\mathrm{metal}}\right),
\end{equation}
with the exponent 19.5 originating in Eq.~1 of \citet{fischer_carbon_2020} where it appears in the oxygen-dependent term of an empirical fit to the carbon partition coefficient. Fundamentally, this means that the oxygen dependence of carbon's siderophility is represented twice in the calculation: once explicitly, through the $X_{\mathrm{O}}^{\mathrm{metal}}$ stoichiometry of the reaction, and once implicitly, through the compositional terms already folded into the empirical partitioning fit from which $\gamma_{\mathrm{C}}^{\mathrm{metal}}$ is derived.

\begin{figure*}
    \centering
    \includegraphics[width=0.9\textwidth]{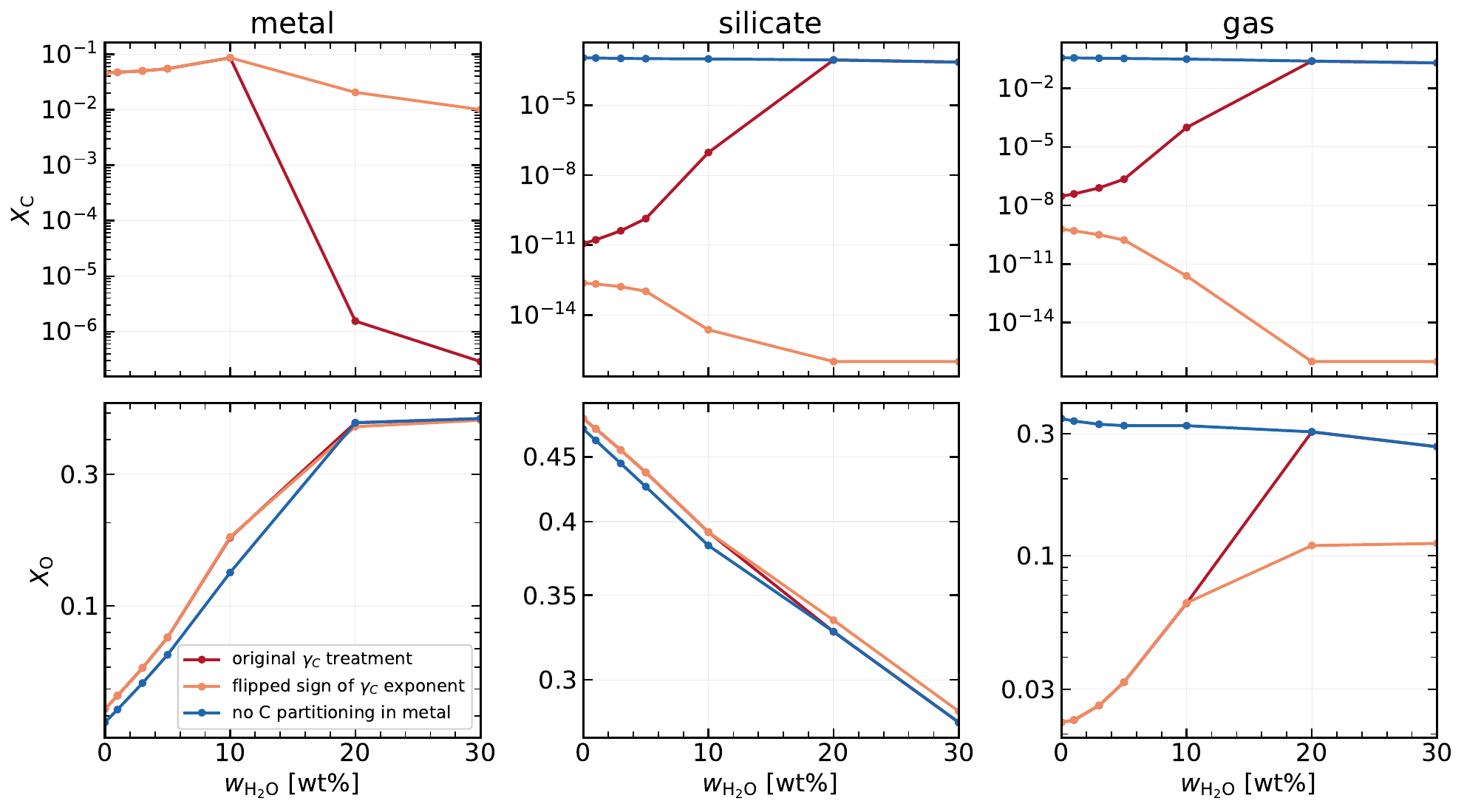}
    \vspace{5mm}
    \vspace{-5mm}\caption{Concentration of carbon (top panels) and oxygen (bottom panels) across planetary reservoirs under different assumptions for carbon partitioning. We test three prescriptions: one where the activity coefficient regulating the storing of carbon in the metal phase follows the implementation illustrated in \citet{werlen_atmospheric_2025} and \citet{grimm_new_2026} (O pushes out C from the metal phase; red), one where C partitions more readily in the metal in the presence of O (the intended behavior reported by \citet{fischer_carbon_2020}; ``flipped sign'', in orange), and one where C partitioning in the metal phase is turned off altogether (blue, adopted in this work). We show a breakdown of the C and O content of the metal (left), silicate (middle), and gas (right) phases after chemical equilibration for various values of $w_{\rm H2O}$. In all the simulations, we adopt $w_{\rm H2}=1$\%, $T_{\rm AMOI}=T_{\rm s-m}=3500$\,K, and both Fe/Si and Mg/Si are set to the stellar values. }
    \label{fig:lngc_Xi_comparison}
\end{figure*}

\begin{figure*}
    \centering
    \includegraphics[width=0.9\textwidth]{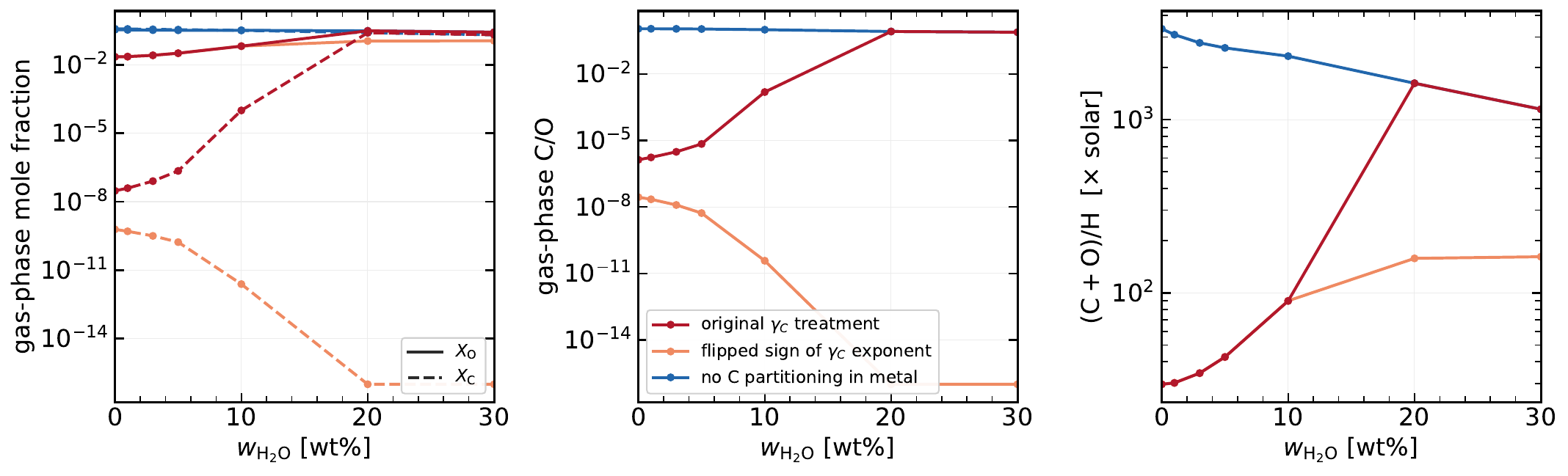}
    \vspace{5mm}
    \vspace{-5mm}\caption{Impact of the carbon partitioning prescription on the atmospheric observables. The simulation setup is the one illustrated in Fig. \ref{fig:lngc_Xi_comparison}, and the color scheme is matched. The mole fractions of O and C in the gas phase as a function of $w_{\rm H2O}$ are shown on the left. We show the corresponding C/O ratio (middle panel) and metallicity (right panel) in the planetary atmosphere. Interpreting the \citet{fischer_carbon_2020} coefficient to describe the $C_{\rm metal} + O_{\rm metal} \rightleftharpoons CO_{\rm silicate}$ reaction leads to extremely low atmospheric C/O ratios (orange), and the opposite sign adopted in the ``original $\gamma_c$'' treatment \citep{werlen_atmospheric_2025,grimm_new_2026} leads to an artificial transition from C fully stored in the metal phase at low $w_{\rm H2O}$ to extremely high gas-phase C abundances as soon as enough O is introduced in the metal phase (red). }
    \label{fig:lngc_atmosphere_comparison}
\end{figure*}

The consequence in our application is that $\gamma_{\mathrm{C}}^{\mathrm{metal}}$ rises steeply with $X_{\mathrm{O}}^{\mathrm{metal}}$, which itself increases with $w_{\rm H_2O}$, effectively expelling carbon from the metal as oxygen is added. We note that this is opposite in sense to the behavior of the partition coefficient reported by \citet{fischer_carbon_2020}, which scales as $\left(1 - X_{\mathrm{O}}^{\mathrm{metal}}\right)^{-19.5}$. As a sensitivity test, we reversed the sign of the exponent (Fig.~\ref{fig:lngc_Xi_comparison}), which results in essentially all the carbon remaining sequestered in the metal irrespective of $w_{\rm H_2O}$ -- yielding low atmospheric C/O across the entire grid
(Fig.~\ref{fig:lngc_atmosphere_comparison}) rather than only at low $w_{\rm H_2O}$ as in the original implementation \citep{werlen_atmospheric_2025,grimm_new_2026}. Neither approach is satisfactory at this stage, as the sign-reversed case would predict carbon-poor atmospheres for all warm sub-Neptunes, in tension with the growing set of carbon-bearing detections in this population
\citep{madhusudhan_carbon-bearing_2023,benneke_jwst_2024,barat_first_2024,
schlawin_possible_2024}. 

Therefore, we elect to set $\ln \gamma_{\mathrm{C}}^{\mathrm{metal}}$ to a large fixed value that suppresses carbon partitioning into the metal (see Figs. \ref{fig:lngc_Xi_comparison} and \ref{fig:lngc_atmosphere_comparison}), and treat this as the lower bound on the interior carbon sink. We discuss this caveat in Section \ref{ssec:deepinteriorcarbon}, and leave to future work the development of a self-consistent prescription in which the reaction free energy and the metal-phase activity model are derived from a single internally consistent dataset.

\section{Additional supporting material for the \texttt{IsoFATE} analysis}\label{sec:isofate_extra}

\begin{figure*}
    \centering
    \includegraphics[width=0.9\textwidth]{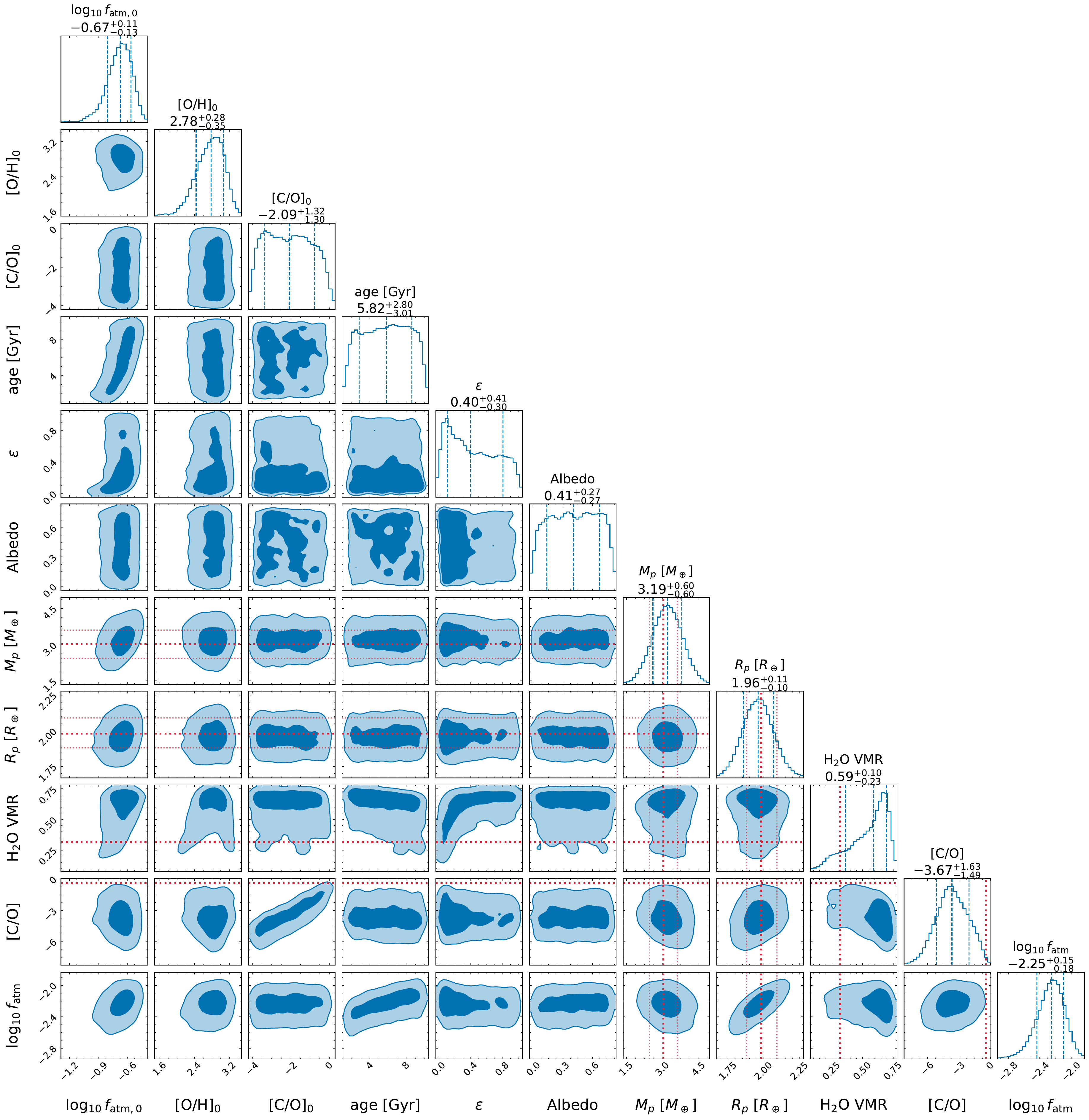}
    \vspace{5mm}
    \vspace{-5mm}\caption{Joint and marginalized posterior distributions on the planet formation conditions and present-day atmosphere properties of GJ\,9827\,d predicted by the \texttt{IsoFATE} Bayesian inference. Shaded regions indicate 1 and 2-$\sigma$ confidence intervals. The inference framework matches the lower limit on H$_2$O VMR, and upper limit on C/O ratio (red dotted vertical lines) constrained by the atmosphere analysis, shown in the last two columns.}
    \label{fig:compare_isofate_corner}
\end{figure*}

The \texttt{IsoFATE} MCMC analysis presented in the main text explores a multi-dimensional parameter space to match the present-day radius, mass, water abundance, and C/O ratio of GJ\,9827\,d. The modeling uses \texttt{IsoFATE}'s new \texttt{dynamic\_phi} feature for the time-dependent selection of the pair of species considered in the fractionation calculation. The fitted parameters include the initial bulk atmosphere mass fraction (fitted as $\log_{10} f_\mathrm{atm,0}$), the initial atmosphere metal mass fraction [M/H]$_0$ and C/O ratio ($\log_{10}$ C/O$_0$), as well as the planet mass using a Gaussian prior informed by the literature value for GJ\,9827\,d \citep{passegger_compact_2024}. We also fit the planetary age, with a uniform prior from 1 to 10 Gyr, the efficiency term $\epsilon$ modulating photoevaporative mass-loss \citep{watson_dynamics_1981,yelle_aeronomy_2004,murray-clay_atmospheric_2009,owen_planetary_2012}, and the Bond albedo of the planet's atmosphere (assumed to be constant over time). The retrieval method will be described in more detail in Cherubim et al. (in prep). We show the joint and marginalized posterior distributions on the fitted parameters in Fig. \ref{fig:compare_isofate_corner}. 

\clearpage
\bibliographystyle{aasjournal}

\bibliography{references_clean}

\begin{thebibliography}{}
\expandafter\ifx\csname natexlab\endcsname\relax\def\natexlab#1{#1}\fi
\providecommand{\url}[1]{\href{#1}{#1}}
\providecommand{\dodoi}[1]{doi:~\href{http://doi.org/#1}{\nolinkurl{#1}}}
\providecommand{\doeprint}[1]{\href{http://ascl.net/#1}{\nolinkurl{http://ascl.net/#1}}}
\providecommand{\doarXiv}[1]{\href{https://arxiv.org/abs/#1}{\nolinkurl{https://arxiv.org/abs/#1}}}

\bibitem[{V. {Adibekyan} {et~al.}(2024){Adibekyan}, {Deal}, {Dorn}, {Dittrich}, {Soares}, {Sousa}, {Santos}, {Bitsch}, {Mordasini}, {Barros}, {Bossini}, {Campante}, {Delgado Mena}, {Demangeon}, {Figueira}, {Moedas}, {Martirosyan}, {Israelian}, \& {Hakobyan}}]{Adibekyan2024}
{Adibekyan}, V., {Deal}, M., {Dorn}, C., {et~al.} 2024, \bibinfo{title}{{Linking the primordial composition of planet building disks to the present-day composition of rocky exoplanets},} \aap, 692, A67, \dodoi{10.1051/0004-6361/202452193}

\bibitem[{A. Aguichine {et~al.}(2024)Aguichine, Batalha, Fortney, Nettelmann, Owen, \& Kempton}]{aguichine_evolution_2024}
Aguichine, A., Batalha, N., Fortney, J.~J., {et~al.} 2024, \bibinfo{title}{Evolution of steam worlds: energetic aspects,} \dodoi{10.48550/arXiv.2412.17945}

\bibitem[{E.-M. Ahrer {et~al.}(2025)Ahrer, Radica, Piaulet-Ghorayeb, Raul, Wiser, Welbanks, Acuna, Allart, Coulombe, Louca, MacDonald, Saidel, Evans-Soma, Benneke, Christie, Beatty, Cadieux, Cloutier, Doyon, Fortney, Gagnebin, Gapp, Innes, Knutson, Komacek, Krissansen-Totton, Miguel, Pierrehumbert, Roy, \& Schlichting}]{ahrer_escaping_2025}
Ahrer, E.-M., Radica, M., Piaulet-Ghorayeb, C., {et~al.} 2025, \bibinfo{title}{Escaping {Helium} and a {Highly} {Muted} {Spectrum} {Suggest} a {Metal}-{Enriched} {Atmosphere} on {Sub}-{Neptune} {GJ3090b} from {JWST} {Transit} {Spectroscopy},} \dodoi{10.3847/2041-8213/add010}

\bibitem[{R. Alvarez \& B. Plez(1998)Alvarez \& Plez}]{alvarez_near-infrared_1998}
Alvarez, R., \& Plez, B. 1998, Near-infrared narrow-band photometry of {M}-giant and {Mira} stars: models meet observations, arXiv, \dodoi{10.48550/arXiv.astro-ph/9710157}

\bibitem[{D.~E. Anderson {et~al.}(2017)Anderson, Bergin, Blake, Ciesla, Visser, \& Lee}]{anderson_destruction_2017}
Anderson, D.~E., Bergin, E.~A., Blake, G.~A., {et~al.} 2017, \bibinfo{title}{Destruction of {Refractory} {Carbon} in {Protoplanetary} {Disks},} The Astrophysical Journal, 845, 13, \dodoi{10.3847/1538-4357/aa7da1}

\bibitem[{M. Asplund {et~al.}(2021)Asplund, Amarsi, \& Grevesse}]{asplund_chemical_2021}
Asplund, M., Amarsi, A.~M., \& Grevesse, N. 2021, \bibinfo{title}{The chemical make-up of the {Sun}: {A} 2020 vision,} Astronomy and Astrophysics, 653, A141, \dodoi{10.1051/0004-6361/202140445}

\bibitem[{M. {Asplund} {et~al.}(2009){Asplund}, {Grevesse}, {Sauval}, \& {Scott}}]{Asplund2009}
{Asplund}, M., {Grevesse}, N., {Sauval}, A.~J., \& {Scott}, P. 2009, \bibinfo{title}{{The Chemical Composition of the Sun},} \araa, 47, 481, \dodoi{10.1146/annurev.astro.46.060407.145222}

\bibitem[{A.~A.~A. {Azzam} {et~al.}(2016){Azzam}, {Tennyson}, {Yurchenko}, \& {Naumenko}}]{Azzam2016}
{Azzam}, A. A.~A., {Tennyson}, J., {Yurchenko}, S.~N., \& {Naumenko}, O.~V. 2016, \bibinfo{title}{{ExoMol molecular line lists - XVI. The rotation-vibration spectrum of hot H$_{2}$S},} \mnras, 460, 4063, \dodoi{10.1093/mnras/stw1133}

\bibitem[{S. Barat {et~al.}(2024)Barat, Désert, Goyal, Vazan, Kawashima, Fortney, Bean, Line, Panwar, Jacobs, Shivkumar, Sikora, Baeyens, Oklopcić, David, \& Livingston}]{barat_first_2024}
Barat, S., Désert, J.-M., Goyal, J.~M., {et~al.} 2024, \bibinfo{title}{First {Comparative} {Exoplanetology} {Within} a {Transiting} {Multi}-planet {System}: {Comparing} the atmospheres of {V1298} {Tau} b and c,} \url{http://arxiv.org/abs/2407.14995}

\bibitem[{S. Barat {et~al.}(2026)Barat, Fairnington, Courreges, Huang, Vanderburg, Morley, Korth, Parviainen, Brandeker, Zhou, Evans-Soma, Sha, Lin, Wright, Morrissey, Nabbie, Evans, Guillot, Horne, Radford, Schwarz, Shporer, Srdoc, \& Suarez}]{barat_jwst_2026}
Barat, S., Fairnington, T., Courreges, S., {et~al.} 2026, \bibinfo{title}{{JWST} unveils a high mean molecular weight atmosphere for mini-{Neptune} {TOI}-1130b: {Evidence} for formation beyond the water ice line,} The Astrophysical Journal Letters, 1002, L32, \dodoi{10.3847/2041-8213/ae5f8b}

\bibitem[{R.~J. {Barber} {et~al.}(2014){Barber}, {Strange}, {Hill}, {Polyansky}, {Mellau}, {Yurchenko}, \& {Tennyson}}]{Barber2014}
{Barber}, R.~J., {Strange}, J.~K., {Hill}, C., {et~al.} 2014, \bibinfo{title}{{ExoMol line lists - III. An improved hot rotation-vibration line list for HCN and HNC},} \mnras, 437, 1828, \dodoi{10.1093/mnras/stt2011}

\bibitem[{T.~J. Bell {et~al.}(2022)Bell, Ahrer, Brande, Carter, Feinstein, Caloca, Mansfield, Zieba, Piaulet, Benneke, Filippazzo, May, Roy, Kreidberg, \& Stevenson}]{bell_eureka_2022}
Bell, T.~J., Ahrer, E.-M., Brande, J., {et~al.} 2022, \bibinfo{title}{Eureka!: {An} {End}-to-{End} {Pipeline} for {JWST} {Time}-{Series} {Observations},} Journal of Open Source Software, 7, 4503, \dodoi{10.21105/joss.04503}

\bibitem[{B. Benneke(2015)Benneke}]{benneke_strict_2015}
Benneke, B. 2015, \bibinfo{title}{Strict {Upper} {Limits} on the {Carbon}-to-{Oxygen} {Ratios} of {Eight} {Hot} {Jupiters} from {Self}-{Consistent} {Atmospheric} {Retrieval},} arXiv e-prints, arXiv:1504.07655.
\newblock \url{https://ui.adsabs.harvard.edu/#abs/2015arXiv150407655B/abstract}

\bibitem[{B. Benneke \& S. Seager(2012)Benneke \& Seager}]{benneke_atmospheric_2012}
Benneke, B., \& Seager, S. 2012, \bibinfo{title}{Atmospheric {Retrieval} for {Super}-{Earths}: {Uniquely} {Constraining} the {Atmospheric} {Composition} with {Transmission} {Spectroscopy},} The Astrophysical Journal, 753, 100, \dodoi{10.1088/0004-637X/753/2/100}

\bibitem[{B. Benneke \& S. Seager(2013)Benneke \& Seager}]{benneke_how_2013}
Benneke, B., \& Seager, S. 2013, \bibinfo{title}{How to {Distinguish} between {Cloudy} {Mini}-{Neptunes} and {Water}/{Volatile}-{Dominated} {Super}-{Earths},} The Astrophysical Journal, 778, 153, \dodoi{10.1088/0004-637X/778/2/153}

\bibitem[{B. Benneke {et~al.}(2017)Benneke, Werner, Petigura, Knutson, Dressing, Crossfield, Schlieder, Livingston, Beichman, Christiansen, {Jessica Krick}, Gorjian, Howard, Sinukoff, Ciardi, \& Akeson}]{benneke_spitzer_2017}
Benneke, B., Werner, M., Petigura, E., {et~al.} 2017, \bibinfo{title}{Spitzer {Observations} {Confirm} and {Rescue} the {Habitable}-zone {Super}-{Earth} {K2}-18b for {Future} {Characterization},} The Astrophysical Journal, 834, 187, \dodoi{10.3847/1538-4357/834/2/187}

\bibitem[{B. Benneke {et~al.}(2019{\natexlab{a}})Benneke, Knutson, Lothringer, Crossfield, Moses, Morley, Kreidberg, Fulton, Dragomir, Howard, Wong, Désert, McCullough, Kempton, Fortney, Gilliland, Deming, \& Kammer}]{benneke_sub-neptune_2019}
Benneke, B., Knutson, H.~A., Lothringer, J., {et~al.} 2019{\natexlab{a}}, \bibinfo{title}{A sub-{Neptune} exoplanet with a low-metallicity methane-depleted atmosphere and {Mie}-scattering clouds,} Nature Astronomy, 3, 813, \dodoi{10.1038/s41550-019-0800-5}

\bibitem[{B. Benneke {et~al.}(2019{\natexlab{b}})Benneke, Wong, Piaulet, Knutson, Lothringer, Morley, Crossfield, Gao, Greene, Dressing, Dragomir, Howard, McCullough, Kempton, Fortney, \& Fraine}]{benneke_water_2019}
Benneke, B., Wong, I., Piaulet, C., {et~al.} 2019{\natexlab{b}}, \bibinfo{title}{Water {Vapor} and {Clouds} on the {Habitable}-zone {Sub}-{Neptune} {Exoplanet} {K2}-18b,} The Astrophysical Journal Letters, 887, L14, \dodoi{10.3847/2041-8213/ab59dc}

\bibitem[{B. Benneke {et~al.}(2024)Benneke, Roy, Coulombe, Radica, Piaulet, Ahrer, Pierrehumbert, Krissansen-Totton, Schlichting, Hu, Yang, Christie, Thorngren, Young, Pelletier, Knutson, Miguel, Evans-Soma, Dorn, Gagnebin, Fortney, Komacek, MacDonald, Raul, Cloutier, Acuna, Lafrenière, Cadieux, Doyon, Welbanks, \& Allart}]{benneke_jwst_2024}
Benneke, B., Roy, P.-A., Coulombe, L.-P., {et~al.} 2024, {JWST} {Reveals} {CH}\$\_4\$, {CO}\$\_2\$, and {H}\$\_2\${O} in a {Metal}-rich {Miscible} {Atmosphere} on a {Two}-{Earth}-{Radius} {Exoplanet}, arXiv.
\newblock \url{http://arxiv.org/abs/2403.03325}

\bibitem[{E.~A. {Bergin} {et~al.}(2015){Bergin}, {Blake}, {Ciesla}, {Hirschmann}, \& {Li}}]{Bergin2015}
{Bergin}, E.~A., {Blake}, G.~A., {Ciesla}, F., {Hirschmann}, M.~M., \& {Li}, J. 2015, \bibinfo{title}{{Tracing the ingredients for a habitable earth from interstellar space through planet formation},} Proceedings of the National Academy of Science, 112, 8965, \dodoi{10.1073/pnas.1500954112}

\bibitem[{E.~A. {Bergin} \& L.~I. {Cleeves}(2018){Bergin} \& {Cleeves}}]{Bergin2018}
{Bergin}, E.~A., \& {Cleeves}, L.~I. 2018, \bibinfo{title}{{Chemistry During the Gas-Rich Stage of Planet Formation},} in Handbook of Exoplanets, ed. H.~J. {Deeg} \& J.~A. {Belmonte}, 137, \dodoi{10.1007/978-3-319-55333-7_137}

\bibitem[{E.~A. {Bergin} {et~al.}(2014){Bergin}, {Cleeves}, {Crockett}, \& {Blake}}]{Bergin2014}
{Bergin}, E.~A., {Cleeves}, L.~I., {Crockett}, N., \& {Blake}, G.~A. 2014, \bibinfo{title}{{Exploring the Origins of Carbon in Terrestrial Worlds},} Faraday Discussions, 168, 61, \dodoi{10.1039/C4FD00003J}

\bibitem[{E.~A. Bergin {et~al.}(2023)Bergin, Kempton, Hirschmann, Bastelberger, Teal, Blake, Ciesla, \& Li}]{bergin_exoplanet_2023}
Bergin, E.~A., Kempton, E. M.-R., Hirschmann, M., {et~al.} 2023, \bibinfo{title}{Exoplanet {Volatile} {Carbon} {Content} as a {Natural} {Pathway} for {Haze} {Formation},} The Astrophysical Journal Letters, 949, L17, \dodoi{10.3847/2041-8213/acd377}

\bibitem[{B. Bitsch {et~al.}(2015)Bitsch, Lambrechts, \& Johansen}]{bitsch_growth_2015}
Bitsch, B., Lambrechts, M., \& Johansen, A. 2015, \bibinfo{title}{The growth of planets by pebble accretion in evolving protoplanetary discs,} Astronomy and Astrophysics, 582, A112, \dodoi{10.1051/0004-6361/201526463}

\bibitem[{I. Blanchard {et~al.}(2022)Blanchard, Rubie, Jennings, Franchi, Zhao, Petitgirard, Miyajima, Jacobson, \& Morbidelli}]{blanchard_metal-silicate_2022}
Blanchard, I., Rubie, D.~C., Jennings, E.~S., {et~al.} 2022, \bibinfo{title}{The {Metal}-{Silicate} {Partitioning} of {Carbon} {During} {Earth}'s {Accretion} and its {Distribution} in the {Early} {Solar} {System},} Earth and Planetary Science Letters, 580, 117374, \dodoi{10.1016/j.epsl.2022.117374}

\bibitem[{S. Blanco-Cuaresma(2019)Blanco-Cuaresma}]{blanco-cuaresma_modern_2019}
Blanco-Cuaresma, S. 2019, \bibinfo{title}{Modern stellar spectroscopy caveats,} Monthly Notices of the Royal Astronomical Society, 486, 2075, \dodoi{10.1093/mnras/stz549}

\bibitem[{S. Blanco-Cuaresma {et~al.}(2014)Blanco-Cuaresma, Soubiran, Heiter, \& Jofré}]{blanco-cuaresma_determining_2014}
Blanco-Cuaresma, S., Soubiran, C., Heiter, U., \& Jofré, P. 2014, \bibinfo{title}{Determining stellar atmospheric parameters and chemical abundances of {FGK} stars with {iSpec},} Astronomy and Astrophysics, 569, A111, \dodoi{10.1051/0004-6361/201423945}

\bibitem[{D.~J. Bower {et~al.}(2025)Bower, Thompson, Hakim, Tian, \& Sossi}]{bower_diversity_2025}
Bower, D.~J., Thompson, M.~A., Hakim, K., Tian, M., \& Sossi, P.~A. 2025, \bibinfo{title}{Diversity of rocky planet atmospheres in the {C}-{H}-{O}-{N}-{S}-{Cl} system with interior dissolution, non-ideality, and condensation: {Application} to {TRAPPIST}-1e and sub-{Neptunes},} \dodoi{10.48550/arXiv.2507.00499}

\bibitem[{B. Breza {et~al.}(2025)Breza, Nixon, \& Kempton}]{breza_not_2025}
Breza, B., Nixon, M.~C., \& Kempton, E. M.-R. 2025, \bibinfo{title}{Not {All} {Sub}-{Neptune} {Exoplanets} {Have} {Magma} {Oceans},} \dodoi{10.48550/arXiv.2509.20429}

\bibitem[{C.~L. Brinkman {et~al.}(2024)Brinkman, Polanski, Huber, Weiss, Valencia, \& Plotnykov}]{brinkman_revisiting_2024}
Brinkman, C.~L., Polanski, A.~S., Huber, D., {et~al.} 2024, Revisiting the {Relationship} {Between} {Rocky} {Exoplanet} and {Stellar} {Compositions}: {Reduced} {Evidence} for a {Super}-{Mercury} {Population}, arXiv, \dodoi{10.48550/arXiv.2409.08361}

\bibitem[{C.~L. Brinkman {et~al.}(2025)Brinkman, Weiss, Huber, Lee, Kolecki, Tenn, Zhang, Narayanan, Polanski, Dai, Bean, Beard, Brady, Brodheim, Brown, Chontos, Deich, Edelstein, Fulton, Giacalone, Gibson, Gilbert, Halverson, Handley, Hill, Holcomb, Holden, Householder, Howard, Isaacson, Kaye, Laher, Lanclos, Ong, Payne, Petigura, Pidhorodetska, Poppett, Roy, Rubenzahl, Saunders, Schwab, Seifahrt, Shaum, Sirk, Smith, Smith, Stefánsson, Stürmer, Thorne, Turtelboom, Tyler, Valliant, Van~Zandt, Walawender, Yee, Yeh, \& Zink}]{brinkman_compositions_2025}
Brinkman, C.~L., Weiss, L.~M., Huber, D., {et~al.} 2025, \bibinfo{title}{The {Compositions} of {Rocky} {Planets} in {Close}-in {Orbits} {Tend} to {Be} {Earth}-like,} The Astronomical Journal, 170, 109, \dodoi{10.3847/1538-3881/ade677}

\bibitem[{J. Buchner {et~al.}(2014)Buchner, Georgakakis, Nandra, Hsu, Rangel, Brightman, Merloni, Salvato, Donley, \& Kocevski}]{buchner_x-ray_2014}
Buchner, J., Georgakakis, A., Nandra, K., {et~al.} 2014, \bibinfo{title}{X-ray spectral modelling of the {AGN} obscuring region in the {CDFS}: {Bayesian} model selection and catalogue,} Astronomy and Astrophysics, 564, A125, \dodoi{10.1051/0004-6361/201322971}

\bibitem[{R. Burn {et~al.}(2024)Burn, Mordasini, Mishra, Haldemann, Venturini, Emsenhuber, \& Henning}]{burn_radius_2024}
Burn, R., Mordasini, C., Mishra, L., {et~al.} 2024, A radius valley between migrated steam worlds and evaporated rocky cores, arXiv.
\newblock \url{http://arxiv.org/abs/2401.04380}

\bibitem[{H. Bushouse {et~al.}(2023)Bushouse, Eisenhamer, Dencheva, Davies, Greenfield, Morrison, Hodge, Simon, Grumm, Droettboom, Slavich, Sosey, Pauly, Miller, Jedrzejewski, Hack, Davis, Crawford, Law, Gordon, Regan, Cara, MacDonald, Bradley, Shanahan, Jamieson, Teodoro, Williams, \& Pena-Guerrero}]{bushouse_2023}
Bushouse, H., Eisenhamer, J., Dencheva, N., {et~al.} 2023, JWST Calibration Pipeline, 1.12.3 Zenodo, \dodoi{10.5281/zenodo.8404029}

\bibitem[{R. Calder {et~al.}(2025)Calder, Shorttle, Nicholls, Lichtenberg, \& Guimond}]{calder_most_2025}
Calder, R., Shorttle, O., Nicholls, H., Lichtenberg, T., \& Guimond, C.-M. 2025, \bibinfo{title}{Most {Rocky} {Sub}-{Neptunes} are {Molten}: {Mapping} the {Solidification} {Shoreline} for {Gas} {Dwarf} {Exoplanets},} \dodoi{10.48550/arXiv.2512.05816}

\bibitem[{I. Carleo {et~al.}(2021)Carleo, Youngblood, Redfield, Casasayas~Barris, Ayres, Vannier, Fossati, Palle, Livingston, Lanza, Niraula, Alvarado-Gómez, Chen, Gandolfi, Guenther, Linsky, Nagel, Narita, Nortmann, Shkolnik, \& Stangret}]{carleo_multiwavelength_2021}
Carleo, I., Youngblood, A., Redfield, S., {et~al.} 2021, \bibinfo{title}{A {Multiwavelength} {Look} at the {GJ} 9827 {System}: {No} {Evidence} of {Extended} {Atmospheres} in {GJ} 9827b and d from {HST} and {CARMENES} {Data},} The Astronomical Journal, 161, 136, \dodoi{10.3847/1538-3881/abdb2f}

\bibitem[{A.~L. Carter {et~al.}(2024)Carter, May, Espinoza, Welbanks, Ahrer, Alderson, Brahm, Feinstein, Grant, Line, Morello, O'Steen, Radica, Rustamkulov, Stevenson, Turner, Alam, Anderson, Batalha, Battley, Bayliss, Bean, Benneke, Berta-Thompson, Brande, Bryant, Burleigh, Coulombe, Crossfield, Damiano, Désert, Flagg, Gill, Inglis, Kirk, Knutson, Kreidberg, Morales, Mansfield, Moran, Murray, Nixon, de~la Roche, Rackham, Schlawin, Sing, Wakeford, Wallack, Wheatley, Zieba, Aggarwal, Barstow, Bell, Blecic, Caceres, Crouzet, Cubillos, Daylan, de~Val-Borro, Decin, Fortney, Gibson, Heng, Hu, Kempton, Lagage, Lothringer, Lustig-Yaeger, Mancini, Mayne, Mayorga, Molaverdikhani, Nasedkin, Ohno, Parmentier, Powell, Redfield, Roy, Taylor, \& Zhang}]{carter_benchmark_2024}
Carter, A.~L., May, E.~M., Espinoza, N., {et~al.} 2024, \bibinfo{title}{A {Benchmark} {JWST} {Near}-{Infrared} {Spectrum} for the {Exoplanet} {WASP}-39b,} Nature Astronomy, \dodoi{10.1038/s41550-024-02292-x}

\bibitem[{C. Cherubim(2025)Cherubim}]{cherubim_isofate_2025}
Cherubim, C. 2025, \bibinfo{title}{{IsoFATE}: {Isotopic} {Fractionation} via {ATmospheric} {Escape},} Astrophysics Source Code Library, ascl:2503.018.
\newblock \url{https://ui.adsabs.harvard.edu/abs/2025ascl.soft03018C}

\bibitem[{C. Cherubim {et~al.}(2025)Cherubim, Wordsworth, Bower, Sossi, Adams, \& Hu}]{cherubim_oxidation_2025}
Cherubim, C., Wordsworth, R., Bower, D., {et~al.} 2025, \bibinfo{title}{An {Oxidation} {Gradient} {Straddling} the {Small} {Planet} {Radius} {Valley},} \dodoi{10.48550/arXiv.2503.05055}

\bibitem[{H. Chi {et~al.}(2014)Chi, Dasgupta, Duncan, \& Shimizu}]{chi_partitioning_2014}
Chi, H., Dasgupta, R., Duncan, M.~S., \& Shimizu, N. 2014, \bibinfo{title}{Partitioning of carbon between {Fe}-rich alloy melt and silicate melt in a magma ocean – {Implications} for the abundance and origin of volatiles in {Earth}, {Mars}, and the {Moon},} Geochimica et Cosmochimica Acta, 139, 447, \dodoi{10.1016/j.gca.2014.04.046}

\bibitem[{J. Choi {et~al.}(2016)Choi, Dotter, Conroy, Cantiello, Paxton, \& Johnson}]{choi_mesa_2016}
Choi, J., Dotter, A., Conroy, C., {et~al.} 2016, \bibinfo{title}{Mesa {Isochrones} and {Stellar} {Tracks} ({MIST}). {I}. {Solar}-scaled {Models},} The Astrophysical Journal, 823, 102, \dodoi{10.3847/0004-637X/823/2/102}

\bibitem[{P.~A. Coles {et~al.}(2019)Coles, , Yurchenko, \& Tennyson}]{ExoMol_NH3}
Coles, P.~A., , Yurchenko, S.~N., \& Tennyson, J. 2019, \bibinfo{title}{{ExoMol molecular line lists XXXV: a rotation-vibration line list for hot ammonia},} Mon. Not. R. Astron. Soc., 490, 4638, \dodoi{10.1093/mnras/stz2778}

\bibitem[{P.~A. {Coles} {et~al.}(2019){Coles}, {Yurchenko}, \& {Tennyson}}]{Coles2019}
{Coles}, P.~A., {Yurchenko}, S.~N., \& {Tennyson}, J. 2019, \bibinfo{title}{{ExoMol molecular line lists - XXXV. A rotation-vibration line list for hot ammonia},} \mnras, 490, 4638, \dodoi{10.1093/mnras/stz2778}

\bibitem[{L.-P. Coulombe {et~al.}(2025{\natexlab{a}})Coulombe, Benneke, Krissansen-Totton, L'Heureux, Piaulet-Ghorayeb, Radica, Roy, Ahrer, Cadieux, Miguel, Schlichting, Delgado-Mena, Monaghan, Adamski, Raul, Cloutier, Komacek, Taylor, Gapp, Allart, Bouchy, Martins, Cook, Doyon, Evans-Soma, Larue, Mascareño, \& Wardenier}]{coulombe_possible_2025}
Coulombe, L.-P., Benneke, B., Krissansen-Totton, J., {et~al.} 2025{\natexlab{a}}, \bibinfo{title}{Possible {Evidence} for the {Presence} of {Volatiles} on the {Warm} {Super}-{Earth} {TOI}-270 b,} The Astronomical Journal, 170, 226, \dodoi{10.3847/1538-3881/adfc6a}

\bibitem[{L.-P. Coulombe {et~al.}(2025{\natexlab{b}})Coulombe, Radica, Benneke, D'Aoust, Dang, Cowan, Parmentier, Albert, Lafrenière, Taylor, Roy, Pelletier, Allart, Artigau, Doyon, Jayawardhana, Johnstone, Kaltenegger, Langeveld, MacDonald, Rowe, \& Turner}]{coulombe_highly_2025}
Coulombe, L.-P., Radica, M., Benneke, B., {et~al.} 2025{\natexlab{b}}, Highly reflective white clouds on the western dayside of an exo-{Neptune}, arXiv, \dodoi{10.48550/arXiv.2501.14016}

\bibitem[{I.~J.~M. Crossfield {et~al.}(2025)Crossfield, Ahrer, Brande, Kreidberg, Lothringer, Piaulet-Ghorayeb, Polman, Welbanks, Kirk, Powell, \& Khorshid}]{crossfield_mapping_2025}
Crossfield, I. J.~M., Ahrer, E.-M., Brande, J., {et~al.} 2025, Mapping the {SO2} {Shoreline} in {Gas} {Giant} {Exoplanets}, arXiv, \dodoi{10.48550/arXiv.2509.14318}

\bibitem[{R. Dasgupta {et~al.}(2013)Dasgupta, Chi, Shimizu, Buono, \& Walker}]{dasgupta_carbon_2013}
Dasgupta, R., Chi, H., Shimizu, N., Buono, A.~S., \& Walker, D. 2013, \bibinfo{title}{Carbon solution and partitioning between metallic and silicate melts in a shallow magma ocean: {Implications} for the origin and distribution of terrestrial carbon,} Geochimica et Cosmochimica Acta, 102, 191, \dodoi{10.1016/j.gca.2012.10.011}

\bibitem[{R. Dasgupta \& D. Walker(2008)Dasgupta \& Walker}]{dasgupta_carbon_2008}
Dasgupta, R., \& Walker, D. 2008, \bibinfo{title}{Carbon solubility in core melts in a shallow magma ocean environment and distribution of carbon between the {Earth}’s core and the mantle,} Geochimica et Cosmochimica Acta, 72, 4627, \dodoi{10.1016/j.gca.2008.06.023}

\bibitem[{A.~H. Davis {et~al.}(2023)Davis, Solomatova, Caracas, \& Campbell}]{davis_carbon_2023}
Davis, A.~H., Solomatova, N.~V., Caracas, R., \& Campbell, A.~J. 2023, \bibinfo{title}{Carbon {Storage} in {Earth}'s {Deep} {Interior} {Implied} by {Carbonate}-{Silicate}-{Iron} {Melt} {Miscibility},} Geochemistry, Geophysics, Geosystems, 24, e2023GC010896, \dodoi{10.1029/2023GC010896}

\bibitem[{K.~M. Deck \& E. Agol(2016)Deck \& Agol}]{deck_transit_2016}
Deck, K.~M., \& Agol, E. 2016, \bibinfo{title}{Transit {Timing} {Variations} for {Planets} near {Eccentricity}-type {Mean} {Motion} {Resonances},} The Astrophysical Journal, 821, 96, \dodoi{10.3847/0004-637X/821/2/96}

\bibitem[{J.-F. {Donati} {et~al.}(2018){Donati}, {Kouach}, {Lacombe}, {Baratchart}, {Doyon}, {Delfosse}, {Artigau}, {Moutou}, {H{\'e}brard}, {Bouchy}, {Bouvier}, {Alencar}, {Saddlemyer}, {Par{\`e}s}, {Rabou}, {Micheau}, {Dolon}, {Barrick}, {Hernandez}, {Wang}, {Reshetov}, {Striebig}, {Challita}, {Carmona}, {Tibault}, {Martioli}, {Figueira}, {Boisse}, \& {Pepe}}]{Donati2018}
{Donati}, J.-F., {Kouach}, D., {Lacombe}, M., {et~al.} 2018, \bibinfo{title}{{SPIRou: A NIR Spectropolarimeter/High-Precision Velocimeter for the CFHT},} in Handbook of Exoplanets, ed. H.~J. {Deeg} \& J.~A. {Belmonte}, 107, \dodoi{10.1007/978-3-319-55333-7_107}

\bibitem[{C. {Dong} {et~al.}(2018){Dong}, {Jin}, {Lingam}, {Airapetian}, {Ma}, \& {van der Holst}}]{Dong2018}
{Dong}, C., {Jin}, M., {Lingam}, M., {et~al.} 2018, \bibinfo{title}{{Atmospheric escape from the TRAPPIST-1 planets and implications for habitability},} Proceedings of the National Academy of Science, 115, 260, \dodoi{10.1073/pnas.1708010115}

\bibitem[{C. Dorn \& T. Lichtenberg(2021{\natexlab{a}})Dorn \& Lichtenberg}]{dorn_hidden_2021}
Dorn, C., \& Lichtenberg, T. 2021{\natexlab{a}}, \bibinfo{title}{Hidden water in magma ocean exoplanets,} arXiv:2110.15069 [astro-ph].
\newblock \url{http://arxiv.org/abs/2110.15069}

\bibitem[{C. Dorn \& T. Lichtenberg(2021{\natexlab{b}})Dorn \& Lichtenberg}]{dorn2021hidden}
Dorn, C., \& Lichtenberg, T. 2021{\natexlab{b}}, \bibinfo{title}{Hidden water in magma ocean exoplanets,} The Astrophysical Journal Letters, 922, L4

\bibitem[{C. Dorn {et~al.}(2026)Dorn, Werlen, \& Jordan}]{dorn_atmospheric_2026}
Dorn, C., Werlen, A., \& Jordan, S. 2026, Atmospheric diversity of sub-{Neptunes} from formation with rock, water, and soot, arXiv, \dodoi{10.48550/arXiv.2606.20464}

\bibitem[{D.~C. {Fabrycky} {et~al.}(2014){Fabrycky}, {Lissauer}, {Ragozzine}, {Rowe}, {Steffen}, {Agol}, {Barclay}, {Batalha}, {Borucki}, {Ciardi}, {Ford}, {Gautier}, {Geary}, {Holman}, {Jenkins}, {Li}, {Morehead}, {Morris}, {Shporer}, {Smith}, {Still}, \& {Van Cleve}}]{Fabrycky2014}
{Fabrycky}, D.~C., {Lissauer}, J.~J., {Ragozzine}, D., {et~al.} 2014, \bibinfo{title}{{Architecture of Kepler's Multi-transiting Systems. II. New Investigations with Twice as Many Candidates},} \apj, 790, 146, \dodoi{10.1088/0004-637X/790/2/146}

\bibitem[{T. {Fairnington} {et~al.}(2026){Fairnington}, {Radica}, {Bean}, {Kempton}, {Nixon}, \& {Zhang}}]{Fairnington2026}
{Fairnington}, T., {Radica}, M., {Bean}, J., {et~al.} 2026, \bibinfo{title}{{Resolving the Metal-Poor Tension of HAT-P-18b: A Volatile-Rich Atmosphere from 0.6-12 {\ensuremath{\mu}}m},} AASTCS 11: Exoplanet Atmospheres 2026, 58, 600.84

\bibitem[{A.~D. Feinstein {et~al.}(2023)Feinstein, Radica, Welbanks, Murray, Ohno, Coulombe, Espinoza, Bean, Teske, Benneke, Line, Rustamkulov, Saba, Tsiaras, Barstow, Fortney, Gao, Knutson, MacDonald, Mikal-Evans, Rackham, Taylor, Parmentier, Batalha, Berta-Thompson, Carter, Changeat, dos Santos, Gibson, Goyal, Kreidberg, López-Morales, Lothringer, Miguel, Molaverdikhani, Moran, Morello, Mukherjee, Sing, Stevenson, Wakeford, Ahrer, Alam, Alderson, Allen, Batalha, Bell, Blecic, Brande, Caceres, Casewell, Chubb, Crossfield, Crouzet, Cubillos, Decin, Désert, Harrington, Heng, Henning, Iro, Kempton, Kendrew, Kirk, Krick, Lagage, Lendl, Mancini, Mansfield, May, Mayne, Nikolov, Palle, Petit dit de~la Roche, Piaulet, Powell, Redfield, Rogers, Roman, Roy, Nixon, Schlawin, Tan, Tremblin, Turner, Venot, Waalkes, Wheatley, \& Zhang}]{feinstein_early_2023}
Feinstein, A.~D., Radica, M., Welbanks, L., {et~al.} 2023, \bibinfo{title}{Early {Release} {Science} of the exoplanet {WASP}-39b with {JWST} {NIRISS},} Nature, 614, 670, \dodoi{10.1038/s41586-022-05674-1}

\bibitem[{F. Feroz {et~al.}(2009)Feroz, Hobson, \& Bridges}]{feroz_multinest_2009}
Feroz, F., Hobson, M.~P., \& Bridges, M. 2009, \bibinfo{title}{{MULTINEST}: an efficient and robust {Bayesian} inference tool for cosmology and particle physics,} Monthly Notices of the Royal Astronomical Society, 398, 1601, \dodoi{10.1111/j.1365-2966.2009.14548.x}

\bibitem[{R.~A. Fischer {et~al.}(2020)Fischer, Cottrell, Hauri, Lee, \& Le~Voyer}]{fischer_carbon_2020}
Fischer, R.~A., Cottrell, E., Hauri, E., Lee, K. K.~M., \& Le~Voyer, M. 2020, \bibinfo{title}{The carbon content of {Earth} and its core,} Proceedings of the National Academy of Sciences, 117, 8743, \dodoi{10.1073/pnas.1919930117}

\bibitem[{D. {Foreman-Mackey} {et~al.}(2017){Foreman-Mackey}, {Agol}, {Angus}, \& {Ambikasaran}}]{celerite}
{Foreman-Mackey}, D., {Agol}, E., {Angus}, R., \& {Ambikasaran}, S. 2017, \bibinfo{title}{Fast and scalable Gaussian process modeling with applications to astronomical time series,} ArXiv.
\newblock \url{https://arxiv.org/abs/1703.09710}

\bibitem[{D. Foreman-Mackey {et~al.}(2013)Foreman-Mackey, Hogg, Lang, \& Goodman}]{foreman-mackey_emcee_2013}
Foreman-Mackey, D., Hogg, D.~W., Lang, D., \& Goodman, J. 2013, \bibinfo{title}{emcee: {The} {MCMC} {Hammer},} Publications of the Astronomical Society of the Pacific, 125, 306, \dodoi{10.1086/670067}

\bibitem[{H.-P. {Gail} \& M. {Trieloff}(2017){Gail} \& {Trieloff}}]{Gail2017}
{Gail}, H.-P., \& {Trieloff}, M. 2017, \bibinfo{title}{{Spatial distribution of carbon dust in the early solar nebula and the carbon content of planetesimals},} \aap, 606, A16, \dodoi{10.1051/0004-6361/201730480}

\bibitem[{P. Gao {et~al.}(2023)Gao, Piette, Steinrueck, Nixon, Zhang, Kempton, Bean, Rauscher, Parmentier, Batalha, Savel, Arnold, Roman, Malsky, \& Taylor}]{gao_hazy_2023}
Gao, P., Piette, A. A.~A., Steinrueck, M.~E., {et~al.} 2023, \bibinfo{title}{The {Hazy} and {Metal}-rich {Atmosphere} of {GJ} 1214 b {Constrained} by {Near}- and {Mid}-infrared {Transmission} {Spectroscopy},} The Astrophysical Journal, 951, 96, \dodoi{10.3847/1538-4357/acd16f}

\bibitem[{J. Goodman \& J. Weare(2010)Goodman \& Weare}]{goodman_ensemble_2010}
Goodman, J., \& Weare, J. 2010, \bibinfo{title}{Ensemble samplers with affine invariance,} Communications in Applied Mathematics and Computational Science, 5, 65, \dodoi{10.2140/camcos.2010.5.65}

\bibitem[{I.~E. Gordon {et~al.}(2022)Gordon, Rothman, Hargreaves, Hashemi, Karlovets, Skinner, Conway, Hill, Kochanov, Tan, Wcisło, Finenko, Nelson, Bernath, Birk, Boudon, Campargue, Chance, Coustenis, Drouin, Flaud, Gamache, Hodges, Jacquemart, Mlawer, Nikitin, Perevalov, Rotger, Tennyson, Toon, Tran, Tyuterev, Adkins, Baker, Barbe, Canè, Cs'asz'ar, Dudaryonok, Egorov, Fleisher, Fleurbaey, Foltynowicz, Furtenbacher, Harrison, Hartmann, Horneman, Huang, Karman, Karns, Kassi, Kleiner, Kofman, Kwabia-Tchana, Lavrentieva, Lee, Long, Lukashevskaya, Lyulin, Makhnev, Matt, Massie, Melosso, Mikhailenko, Mondelain, M"uller, Naumenko, Perrin, Polyansky, Raddaoui, Raston, Reed, Rey, Richard, T'obi'as, Sadiek, Schwenke, Starikova, Sung, Tamassia, Tashkun, Auwera, Vasilenko, Vigasin, Villanueva, Vispoel, Wagner, Yachmenev, \& Yurchenko}]{gordon_hitran2020_2022}
Gordon, I.~E., Rothman, L.~S., Hargreaves, R.~J., {et~al.} 2022, \bibinfo{title}{The {HITRAN2020} {Molecular} {Spectroscopic} {Database},} Journal of Quantitative Spectroscopy and Radiative Transfer, 277, 107949, \dodoi{10.1016/j.jqsrt.2021.107949}

\bibitem[{D. Grant \& H.~R. Wakeford(2022)Grant \& Wakeford}]{david_grant_2022_7437681}
Grant, D., \& Wakeford, H.~R. 2022, Exo-TiC/ExoTiC-LD: ExoTiC-LD v3.0.0, v3.0.0 Zenodo, \dodoi{10.5281/zenodo.7437681}

\bibitem[{D.~S. Grewal {et~al.}(2019)Grewal, Dasgupta, Sun, Tsuno, \& Costin}]{grewal_delivery_2019}
Grewal, D.~S., Dasgupta, R., Sun, C., Tsuno, K., \& Costin, G. 2019, \bibinfo{title}{Delivery of carbon, nitrogen, and sulfur to the silicate {Earth} by a giant impact,} Science Advances, 5, eaau3669, \dodoi{10.1126/sciadv.aau3669}

\bibitem[{S. Grimm {et~al.}(2026)Grimm, Steinmeyer, Werlen, Dorn, Schlichting, \& Young}]{grimm_new_2026}
Grimm, S., Steinmeyer, M.-L., Werlen, A., {et~al.} 2026, A {New} {Global} {Chemical} {Equilibrium} {Code}: {Refractory} {Element} {Signatures} in {Super}-{Earths} and {Sub}-{Neptunes}, arXiv, \dodoi{10.48550/arXiv.2605.07833}

\bibitem[{N. {Gromek} {et~al.}(2026){Gromek}, {Weisserman}, {Cloutier}, {L'Heureux}, \& {Cadieux}}]{Gromek2026}
{Gromek}, N., {Weisserman}, D., {Cloutier}, R., {L'Heureux}, A., \& {Cadieux}, C. 2026, \bibinfo{title}{{Elemental Abundances of Cool Stars: A Spectroscopic Framework Applied to Five Planet-Host Stars},} arXiv e-prints, arXiv:2609.24970.
\newblock \doarXiv{2609.24970}

\bibitem[{A. Gupta \& H.~E. Schlichting(2019)Gupta \& Schlichting}]{gupta_sculpting_2019}
Gupta, A., \& Schlichting, H.~E. 2019, \bibinfo{title}{Sculpting the valley in the radius distribution of small exoplanets as a by-product of planet formation: the core-powered mass-loss mechanism,} Monthly Notices of the Royal Astronomical Society, 487, 24, \dodoi{10.1093/mnras/stz1230}

\bibitem[{A. {Gupta} \& H.~E. {Schlichting}(2020){Gupta} \& {Schlichting}}]{Gupta_2020}
{Gupta}, A., \& {Schlichting}, H.~E. 2020, \bibinfo{title}{{Signatures of the core-powered mass-loss mechanism in the exoplanet population: dependence on stellar properties and observational predictions},} \mnras, 493, 792, \dodoi{10.1093/mnras/staa315}

\bibitem[{A. Gupta \& H.~E. Schlichting(2021)Gupta \& Schlichting}]{gupta_caught_2021}
Gupta, A., \& Schlichting, H.~E. 2021, \bibinfo{title}{Caught in the act: core-powered mass-loss predictions for observing atmospheric escape,} Monthly Notices of the Royal Astronomical Society, 504, 4634, \dodoi{10.1093/mnras/stab1128}

\bibitem[{B. Gustafsson {et~al.}(2008)Gustafsson, Edvardsson, Eriksson, Jørgensen, Nordlund, \& Plez}]{gustafsson_grid_2008}
Gustafsson, B., Edvardsson, B., Eriksson, K., {et~al.} 2008, \bibinfo{title}{A grid of {MARCS} model atmospheres for late-type stars. {I}. {Methods} and general properties,} Astronomy and Astrophysics, 486, 951, \dodoi{10.1051/0004-6361:200809724}

\bibitem[{G.~G. H.~Keppler {et~al.}(2019)H.~Keppler, Keppler, \& Golabek}]{h_keppler_graphite_2019}
H.~Keppler, G.~G., Keppler, H., \& Golabek, G. 2019, \bibinfo{title}{Graphite floatation on a magma ocean and the fate of carbon during core formation,} Geochemical Perspectives Letters, 11, 12.
\newblock \url{https://www.geochemicalperspectivesletters.org/article1918/}

\bibitem[{S. {Hadden} \& Y. {Lithwick}(2016){Hadden} \& {Lithwick}}]{Hadden2016}
{Hadden}, S., \& {Lithwick}, Y. 2016, \bibinfo{title}{{Numerical and Analytical Modeling of Transit Timing Variations},} \apj, 828, 44, \dodoi{10.3847/0004-637X/828/1/44}

\bibitem[{G.~J. {Harris} {et~al.}(2006){Harris}, {Tennyson}, {Kaminsky}, {Pavlenko}, \& {Jones}}]{Harris_HCN_2006}
{Harris}, G.~J., {Tennyson}, J., {Kaminsky}, B.~M., {Pavlenko}, Y.~V., \& {Jones}, H.~R.~A. 2006, \bibinfo{title}{{Improved HCN/HNC linelist, model atmospheres and synthetic spectra for WZ Cas},} \mnras, 367, 400, \dodoi{10.1111/j.1365-2966.2005.09960.x}

\bibitem[{C. {Helling} {et~al.}(2017){Helling}, {Tootill}, {Woitke}, \& {Lee}}]{Helling2017}
{Helling}, C., {Tootill}, D., {Woitke}, P., \& {Lee}, E. 2017, \bibinfo{title}{{Dust in brown dwarfs and extrasolar planets. V. Cloud formation in carbon- and oxygen-rich environments},} \aap, 603, A123, \dodoi{10.1051/0004-6361/201629696}

\bibitem[{N.~R. Hinkel {et~al.}(2022)Hinkel, Young, \& Wheeler}]{hinkel_concise_2022}
Hinkel, N.~R., Young, P.~A., \& Wheeler, III, C.~H. 2022, \bibinfo{title}{A {Concise} {Treatise} on {Converting} {Stellar} {Mass} {Fractions} to {Abundances} to {Molar} {Ratios},} The Astronomical Journal, 164, 256, \dodoi{10.3847/1538-3881/ac9bfa}

\bibitem[{M. Holmberg \& N. Madhusudhan(2024)Holmberg \& Madhusudhan}]{holmberg_possible_2024}
Holmberg, M., \& Madhusudhan, N. 2024, Possible {Hycean} conditions in the sub-{Neptune} {TOI}-270 d, \dodoi{10.1051/0004-6361/202348238}

\bibitem[{K. {Horne}(1986){Horne}}]{Horne1986optimalextraction}
{Horne}, K. 1986, \bibinfo{title}{{An optimal extraction algorithm for CCD spectroscopy.},} \pasp, 98, 609, \dodoi{10.1086/131801}

\bibitem[{A.~W. {Howard} {et~al.}(2025){Howard}, {Sinukoff}, {Blunt}, {Petigura}, {Crossfield}, {Isaacson}, {Kosiarek}, {Rubenzahl}, {Brewer}, {Fulton}, {Dressing}, {Hirsch}, {Knutson}, {Livingston}, {Mills}, {Roy}, {Weiss}, {Benneke}, {Ciardi}, {Christiansen}, {Cochran}, {Crepp}, {Gonzales}, {Hansen}, {Hardegree-Ullman}, {Howell}, {L{\'e}pine}, {Martinez}, {Rogers}, {Schlieder}, {Werner}, {Polanski}, {Angelo}, {Beard}, {Behmard}, {Bouma}, {Brinkman}, {Chontos}, {Dai}, {Dalba}, {Giacalone}, {Grunblatt}, {Hill}, {Kane}, {Lubin}, {Mayo}, {Mocnik}, {Murphy}, {Rice}, {Rosenthal}, {Tyler}, {Van Zandt}, \& {Yee}}]{Howard2025}
{Howard}, A.~W., {Sinukoff}, E., {Blunt}, S., {et~al.} 2025, \bibinfo{title}{{Planet Masses, Radii, and Orbits from NASA's K2 Mission},} \apjs, 278, 52, \dodoi{10.3847/1538-4365/adc5e4}

\bibitem[{T.-O. Husser {et~al.}(2013)Husser, Wende-von Berg, Dreizler, Homeier, Reiners, Barman, \& Hauschildt}]{husser_new_2013}
Husser, T.-O., Wende-von Berg, S., Dreizler, S., {et~al.} 2013, \bibinfo{title}{A new extensive library of {PHOENIX} stellar atmospheres and synthetic spectra,} Astronomy \&amp; Astrophysics, Volume 553, id.A6, {\textless}NUMPAGES{\textgreater}9{\textless}/NUMPAGES{\textgreater} pp., 553, A6, \dodoi{10.1051/0004-6361/201219058}

\bibitem[{J. Ih {et~al.}(2026)Ih, Kanodia, Wogan, Fortune, Allen, Diamond-Lowe, Greklek-McKeon, Holmberg, Brande, Ashtari, Bennett, Lustig-Yaeger, Mansfield, Lin, Cañas, Han, Libby-Roberts, \& Mahadevan}]{ih_jwst_2026}
Ih, J., Kanodia, S., Wogan, N., {et~al.} 2026, {JWST} {Rocky} {Worlds} {DDT} {Observations} of {LTT} 1445 {A} b and c in {Secondary} {Eclipse} {Using} {MIRI}/{F1500W}, arXiv, \dodoi{10.48550/arXiv.2611.17300}

\bibitem[{H. Innes {et~al.}(2023)Innes, Tsai, \& Pierrehumbert}]{innes_runaway_2023}
Innes, H., Tsai, S.-M., \& Pierrehumbert, R.~T. 2023, \bibinfo{title}{The {Runaway} {Greenhouse} {Effect} on {Hycean} {Worlds},} The Astrophysical Journal, 953, 168, \dodoi{10.3847/1538-4357/ace346}

\bibitem[{F. Jahandar {et~al.}(2024)Jahandar, Doyon, Artigau, Cook, Cadieux, Donati, Cowan, Cloutier, Pelletier, Alves-Brito, Martins, Shang, \& Carmona}]{jahandar_chemical_2024}
Jahandar, F., Doyon, R., Artigau, {\'E}., {et~al.} 2024, Chemical {Fingerprints} of {M} {Dwarfs}: {High}-{Resolution} {Spectroscopy} on 31 {M} {Dwarfs} with {SPIRou}, arXiv, \dodoi{10.48550/arXiv.2411.07377}

\bibitem[{X. Ji {et~al.}(2025)Ji, Chatterjee, Coy, \& Kite}]{ji_cosmic_2025}
Ji, X., Chatterjee, R.~D., Coy, B.~P., \& Kite, E. 2025, \bibinfo{title}{The {Cosmic} {Shoreline} {Revisited}: {A} {Metric} for {Atmospheric} {Retention} {Informed} by {Hydrodynamic} {Escape},} The Astrophysical Journal, 992, 198, \dodoi{10.3847/1538-4357/adfe69}

\bibitem[{T. {Karman} {et~al.}(2019){Karman}, {Gordon}, {van der Avoird}, {Baranov}, {Boulet}, {Drouin}, {Groenenboom}, {Gustafsson}, {Hartmann}, {Kurucz}, {Rothman}, {Sun}, {Sung}, {Thalman}, {Tran}, {Wishnow}, {Wordsworth}, {Vigasin}, {Volkamer}, \& {van der Zande}}]{Karman2019}
{Karman}, T., {Gordon}, I.~E., {van der Avoird}, A., {et~al.} 2019, \bibinfo{title}{{Update of the HITRAN collision-induced absorption section},} \icarus, 328, 160, \dodoi{10.1016/j.icarus.2019.02.034}

\bibitem[{D. Kasper {et~al.}(2020)Kasper, Bean, Oklopčić, Malsky, Kempton, Désert, Rogers, \& Mansfield}]{kasper_non-detection_2020}
Kasper, D., Bean, J.~L., Oklopčić, A., {et~al.} 2020, \bibinfo{title}{Non-detection of {Helium} in the {Upper} {Atmospheres} of {Three} {Sub}-{Neptune} {Exoplanets},} The Astronomical Journal, 160, 258, \dodoi{10.3847/1538-3881/abbee6}

\bibitem[{E.~S. Kite \& M. Barnett(2020)Kite \& Barnett}]{kite_exoplanet_2020}
Kite, E.~S., \& Barnett, M. 2020, \bibinfo{title}{Exoplanet secondary atmosphere loss and revival,} arXiv e-prints, 2006, arXiv:2006.02589.
\newblock \url{http://adsabs.harvard.edu/abs/2020arXiv200602589K}

\bibitem[{E.~S. Kite \& L. Schaefer(2021)Kite \& Schaefer}]{kite_water_2021}
Kite, E.~S., \& Schaefer, L. 2021, \bibinfo{title}{Water on {Hot} {Rocky} {Exoplanets},} The Astrophysical Journal, 909, L22, \dodoi{10.3847/2041-8213/abe7dc}

\bibitem[{D. Kitzmann {et~al.}(2023)Kitzmann, Stock, \& Patzer}]{kitzmann_fastchem_2023}
Kitzmann, D., Stock, J.~W., \& Patzer, A. B.~C. 2023, {FastChem} {Cond}: {Equilibrium} chemistry with condensation and rainout for cool planetary and stellar environments, arXiv, \dodoi{10.48550/arXiv.2309.02337}

\bibitem[{D. Kitzmann {et~al.}(2026)Kitzmann, Stock, \& Patzer}]{kitzmann_fastchem_2026}
Kitzmann, D., Stock, J.~W., \& Patzer, A. B.~C. 2026, {FastChem} 4: {New} chemical elements and improved convergence behaviour, \url{https://arxiv.org/abs/2605.18264v1}

\bibitem[{M.~R. Kosiarek {et~al.}(2020)Kosiarek, Berardo, Crossfield, Laguna, Murphy, Howell, Henry, Isaacson, Weiss, Petigura, Fulton, Behmard, Hirsch, Teske, Burt, Mills, Chontos, Mocnik, Howard, Werner, Livingston, Krick, Beichman, Gorjian, Kreidberg, Christiansen, Morales, Scott, Crane, Rosenthal, Grunblatt, Rubenzahl, Dalba, Giacalone, Villanueva, Liu, Dai, Hill, Rice, Kane, \& Mayo}]{kosiarek_physical_2020}
Kosiarek, M.~R., Berardo, D.~A., Crossfield, I. J.~M., {et~al.} 2020, \bibinfo{title}{Physical {Parameters} of the {Multi}-{Planet} {Systems} {HD} 106315 and {GJ} 9827,} arXiv:2009.03398 [astro-ph].
\newblock \url{http://arxiv.org/abs/2009.03398}

\bibitem[{M.~R. Kosiarek {et~al.}(2021)Kosiarek, Berardo, Crossfield, Laguna, Piaulet, Akana~Murphy, Howell, Henry, Isaacson, Fulton, Weiss, Petigura, Behmard, Hirsch, Teske, Burt, Mills, Chontos, Močnik, Howard, Werner, Livingston, Krick, Beichman, Gorjian, Kreidberg, Morley, Christiansen, Morales, Scott, Crane, Wang, Shectman, Rosenthal, Grunblatt, Rubenzahl, Dalba, Giacalone, Villanueva, Liu, Dai, Hill, Rice, Kane, \& Mayo}]{kosiarek_physical_2021}
Kosiarek, M.~R., Berardo, D.~A., Crossfield, I. J.~M., {et~al.} 2021, \bibinfo{title}{Physical {Parameters} of the {Multiplanet} {Systems} {HD} 106315 and {GJ} 9827,} The Astronomical Journal, 161, 47, \dodoi{10.3847/1538-3881/abca39}

\bibitem[{L. Kreidberg(2015)Kreidberg}]{kreidberg_batman_2015}
Kreidberg, L. 2015, \bibinfo{title}{batman: {BAsic} {Transit} {Model} {cAlculatioN} in {Python},} Publications of the Astronomical Society of the Pacific, 127, 1161, \dodoi{10.1086/683602}

\bibitem[{L. Kreidberg {et~al.}(2014)Kreidberg, Bean, Désert, Benneke, Deming, Stevenson, Seager, Berta-Thompson, Seifahrt, \& Homeier}]{kreidberg_clouds_2014}
Kreidberg, L., Bean, J.~L., Désert, J.-M., {et~al.} 2014, \bibinfo{title}{Clouds in the atmosphere of the super-{Earth} exoplanet {GJ1214b},} Nature, 505, 69, \dodoi{10.1038/nature12888}

\bibitem[{M.~E. Kress {et~al.}(2010)Kress, Tielens, \& Frenklach}]{kress_soot_2010}
Kress, M.~E., Tielens, A. G. G.~M., \& Frenklach, M. 2010, \bibinfo{title}{The ‘soot line’: {Destruction} of presolar polycyclic aromatic hydrocarbons in the terrestrial planet-forming region of disks,} Advances in Space Research, 46, 44, \dodoi{10.1016/j.asr.2010.02.004}

\bibitem[{V. Krishnamurthy {et~al.}(2023)Krishnamurthy, Hirano, Gaidos, Sato, Kopparapu, Barclay, Garcia-Sage, Harakawa, Hodapp, Jacobson, Konishi, Kotani, Kudo, Kurokawa, Kuzuhara, Lopez, Nishikawa, Omiya, Schlieder, Serizawa, Tamura, Ueda, \& Vievard}]{krishnamurthy_absence_2023}
Krishnamurthy, V., Hirano, T., Gaidos, E., {et~al.} 2023, \bibinfo{title}{Absence of extended atmospheres in low-mass star radius-gap planets,} Monthly Notices of the Royal Astronomical Society, 521, 1210, \dodoi{10.1093/mnras/stad404}

\bibitem[{J. Krissansen-Totton {et~al.}(2021)Krissansen-Totton, Fortney, Nimmo, \& Wogan}]{krissansen-totton_oxygen_2021}
Krissansen-Totton, J., Fortney, J.~J., Nimmo, F., \& Wogan, N. 2021, \bibinfo{title}{Oxygen {False} {Positives} on {Habitable} {Zone} {Planets} {Around} {Sun}-{Like} {Stars},} AGU Advances, 2, e00294, \dodoi{10.1029/2020AV000294}

\bibitem[{J. Krissansen-Totton {et~al.}(2024)Krissansen-Totton, Wogan, Thompson, \& Fortney}]{krissansen-totton_erosion_2024}
Krissansen-Totton, J., Wogan, N., Thompson, M., \& Fortney, J.~J. 2024, \bibinfo{title}{The erosion of large primary atmospheres typically leaves behind substantial secondary atmospheres on temperate rocky planets,} Nature Communications, 15, 8374, \dodoi{10.1038/s41467-024-52642-6}

\bibitem[{D. Kubyshkina {et~al.}(2026)Kubyshkina, Egger, \& Piaulet-Ghorayeb}]{kubyshkina_escape_2026}
Kubyshkina, D., Egger, J.~A., \& Piaulet-Ghorayeb, C. 2026, Escape of {Water}- and {Metal}-enriched {Atmospheres} from compact {Hot} mini-{Neptunes} with {CHAIN}, arXiv.
\newblock \url{https://ui.adsabs.harvard.edu/abs/2026arXiv260602083K}

\bibitem[{D. Kubyshkina {et~al.}(2018)Kubyshkina, Fossati, Erkaev, Johnstone, Cubillos, Kislyakova, Lammer, Lendl, \& Odert}]{kubyshkina_grid_2018}
Kubyshkina, D., Fossati, L., Erkaev, N.~V., {et~al.} 2018, \bibinfo{title}{Grid of upper atmosphere models for 1-40 {M}⊕ planets: application to {CoRoT}-7 b and {HD} 219134 b,c,} Astronomy and Astrophysics, 619, A151, \dodoi{10.1051/0004-6361/201833737}

\bibitem[{A. Lecavelier Des~Etangs {et~al.}(2008)Lecavelier Des~Etangs, Pont, Vidal-Madjar, \& Sing}]{lecavelier_des_etangs_rayleigh_2008}
Lecavelier Des~Etangs, A., Pont, F., Vidal-Madjar, A., \& Sing, D. 2008, \bibinfo{title}{Rayleigh scattering in the transit spectrum of {HD} 189733b,} Astronomy and Astrophysics, 481, L83, \dodoi{10.1051/0004-6361:200809388}

\bibitem[{G. {Li} {et~al.}(2015){Li}, {Gordon}, {Rothman}, {Tan}, {Hu}, {Kassi}, {Campargue}, \& {Medvedev}}]{Li2015}
{Li}, G., {Gordon}, I.~E., {Rothman}, L.~S., {et~al.} 2015, \bibinfo{title}{{Rovibrational Line Lists for Nine Isotopologues of the CO Molecule in the X $^{1}${\ensuremath{\Sigma}}$^{+}$ Ground Electronic State},} \apjs, 216, 15, \dodoi{10.1088/0067-0049/216/1/15}

\bibitem[{H. Li {et~al.}(2025)Li, He, Wang, Yang, Liu, Wang, Luo, Moran, Pesciotta, Hörst, Moses, \& Vuitton}]{li_impact_2025}
Li, H., He, C., Wang, S., {et~al.} 2025, \bibinfo{title}{The impact of organic hazes and graphite on the observation of {CO2}-rich sub-{Neptune} atmospheres,} \dodoi{10.48550/arXiv.2508.07161}

\bibitem[{J. Li {et~al.}(2021)Li, Bergin, Blake, Ciesla, \& Hirschmann}]{li_earths_2021}
Li, J., Bergin, E.~A., Blake, G.~A., Ciesla, F.~J., \& Hirschmann, M.~M. 2021, \bibinfo{title}{Earth’s carbon deficit caused by early loss through irreversible sublimation,} Science Advances, 7, eabd3632, \dodoi{10.1126/sciadv.abd3632}

\bibitem[{J. Li {et~al.}(2026)Li, Bergin, Hirschmann, Blake, Ciesla, \& Kempton}]{li_soot_2026}
Li, J., Bergin, E.~A., Hirschmann, M.~M., {et~al.} 2026, \bibinfo{title}{Soot {Planets} {Instead} of {Water} {Worlds},} The Astrophysical Journal, 997, L29, \dodoi{10.3847/2041-8213/ae29a6}

\bibitem[{T. Lichtenberg(2021)Lichtenberg}]{lichtenberg_redox_2021}
Lichtenberg, T. 2021, \bibinfo{title}{Redox {Hysteresis} of {Super}-{Earth} {Exoplanets} from {Magma} {Ocean} {Circulation},} The Astrophysical Journal, 914, L4, \dodoi{10.3847/2041-8213/ac0146}

\bibitem[{T. Lichtenberg {et~al.}(2021)Lichtenberg, Dr{\k{a}}{\.{z}}kowska, Schönbächler, Golabek, \& Hands}]{lichtenberg_bifurcation_2021}
Lichtenberg, T., Dr{\k{a}}{\.{z}}kowska, J., Schönbächler, M., Golabek, G.~J., \& Hands, T.~O. 2021, \bibinfo{title}{Bifurcation of planetary building blocks during {Solar} {System} formation,} Science, 371, 365, \dodoi{10.1126/science.abb3091}

\bibitem[{T. Lichtenberg {et~al.}(2025)Lichtenberg, Shorttle, Teske, \& Kempton}]{lichtenberg_constraining_2025}
Lichtenberg, T., Shorttle, O., Teske, J., \& Kempton, E. M.~R. 2025, \bibinfo{title}{Constraining exoplanet interiors using observations of their atmospheres,} \dodoi{10.48550/arXiv.2510.08844}

\bibitem[{Z. Lin \& S. Seager(2025)Lin \& Seager}]{lin_carbon-rich_2025}
Lin, Z., \& Seager, S. 2025, \bibinfo{title}{Carbon-rich {Sub}-{Neptune} {Interiors} {Are} {Compatible} with {JWST} {Observations},} The Astrophysical Journal, 990, L35, \dodoi{10.3847/2041-8213/adfcc8}

\bibitem[{M.~R. Line \& V. Parmentier(2016)Line \& Parmentier}]{line_influence_2016}
Line, M.~R., \& Parmentier, V. 2016, \bibinfo{title}{The {Influence} of {Nonuniform} {Cloud} {Cover} on {Transit} {Transmission} {Spectra},} The Astrophysical Journal, 820, 78, \dodoi{10.3847/0004-637X/820/1/78}

\bibitem[{Y. Lithwick {et~al.}(2012)Lithwick, Xie, \& Wu}]{lithwick_extracting_2012}
Lithwick, Y., Xie, J., \& Wu, Y. 2012, \bibinfo{title}{Extracting {Planet} {Mass} and {Eccentricity} from {TTV} {Data},} The Astrophysical Journal, 761, 122, \dodoi{10.1088/0004-637X/761/2/122}

\bibitem[{E.~D. Lopez \& J.~J. Fortney(2014)Lopez \& Fortney}]{lopez_understanding_2014}
Lopez, E.~D., \& Fortney, J.~J. 2014, \bibinfo{title}{Understanding the {Mass}-{Radius} {Relation} for {Sub}-{Neptunes}: {Radius} as a {Proxy} for {Composition},} The Astrophysical Journal, 792, 1, \dodoi{10.1088/0004-637X/792/1/1}

\bibitem[{A. Louca \& Y. Miguel(2025)Louca \& Miguel}]{louca_metal-enriched_2025}
Louca, A., \& Miguel, Y. 2025, \bibinfo{title}{Metal-{Enriched} {Atmospheres} in {Warm} ({Super}- and {Sub}-){Neptunes} {Induced} by {Extreme} {Atmospheric} {Escape},} \dodoi{10.48550/arXiv.2505.06013}

\bibitem[{A.~J. Louca {et~al.}(2023)Louca, Miguel, \& Kubyshkina}]{louca_metallicity_2023}
Louca, A.~J., Miguel, Y., \& Kubyshkina, D. 2023, \bibinfo{title}{Metallicity and {Spectral} {Evolution} of {WASP}-39 b: {The} {Limited} {Role} of {Hydrodynamic} {Escape},} \url{http://arxiv.org/abs/2309.09605}

\bibitem[{R. Luque \& E. Pallé(2022)Luque \& Pallé}]{luque_density_2022}
Luque, R., \& Pallé, E. 2022, \bibinfo{title}{Density, not radius, separates rocky and water-rich small planets orbiting {M} dwarf stars,} Science, 377, 1211, \dodoi{10.1126/science.abl7164}

\bibitem[{A. Léger {et~al.}(2004)Léger, Selsis, Sotin, Guillot, Despois, Mawet, Ollivier, Labèque, Valette, Brachet, Chazelas, \& Lammer}]{leger_new_2004}
Léger, A., Selsis, F., Sotin, C., {et~al.} 2004, \bibinfo{title}{A new family of planets? ``{Ocean}-{Planets}'',} Icarus, 169, 499, \dodoi{10.1016/j.icarus.2004.01.001}

\bibitem[{R.~J. {MacDonald}(2023){MacDonald}}]{macdonald_poseidon_2023}
{MacDonald}, R.~J. 2023, \bibinfo{title}{{POSEIDON: A Multidimensional Atmospheric Retrieval Code for Exoplanet Spectra},} The Journal of Open Source Software, 8, 4873, \dodoi{10.21105/joss.04873}

\bibitem[{R.~J. {MacDonald} \& N.~K. {Lewis}(2022){MacDonald} \& {Lewis}}]{macdonald_trident_2022}
{MacDonald}, R.~J., \& {Lewis}, N.~K. 2022, \bibinfo{title}{{TRIDENT: A Rapid 3D Radiative-transfer Model for Exoplanet Transmission Spectra},} \apj, 929, 20, \dodoi{10.3847/1538-4357/ac47fe}

\bibitem[{R.~J. {MacDonald} \& N. {Madhusudhan}(2017){MacDonald} \& {Madhusudhan}}]{macdonald_hd_209458b_2017}
{MacDonald}, R.~J., \& {Madhusudhan}, N. 2017, \bibinfo{title}{{HD 209458b in new light: evidence of nitrogen chemistry, patchy clouds and sub-solar water},} \mnras, 469, 1979, \dodoi{10.1093/mnras/stx804}

\bibitem[{N. Madhusudhan {et~al.}(2023)Madhusudhan, Sarkar, Constantinou, Holmberg, Piette, \& Moses}]{madhusudhan_carbon-bearing_2023}
Madhusudhan, N., Sarkar, S., Constantinou, S., {et~al.} 2023, \bibinfo{title}{Carbon-bearing {Molecules} in a {Possible} {Hycean} {Atmosphere},} The Astrophysical Journal, 956, L13, \dodoi{10.3847/2041-8213/acf577}

\bibitem[{S.~P. {Marcum} {et~al.}(2026){Marcum}, {Stixrude}, \& {Young}}]{Marcum2026}
{Marcum}, S.~P., {Stixrude}, L., \& {Young}, E.~D. 2026, \bibinfo{title}{{An Equation of State for Supercritical Silicate-Hydrogen Mixtures at Sub-Neptune Interior Conditions},} arXiv e-prints, arXiv:2608.27401, \dodoi{10.48550/arXiv.2608.27401}

\bibitem[{R. Mbarek \& E.~M.-R. Kempton(2016)Mbarek \& Kempton}]{mbarek_clouds_2016}
Mbarek, R., \& Kempton, E. M.-R. 2016, \bibinfo{title}{Clouds in {Super}-{Earth} {Atmospheres}: {Chemical} {Equilibrium} {Calculations},} The Astrophysical Journal, 827, 121, \dodoi{10.3847/0004-637X/827/2/121}

\bibitem[{W.~F. McDonough(2003)McDonough}]{mcdonough_215_2003}
McDonough, W.~F. 2003, \bibinfo{title}{2.15 - {Compositional} {Model} for the {Earth}'s {Core},} in Treatise on {Geochemistry}, ed. H.~D. Holland \& K.~K. Turekian (Oxford: Pergamon), 547--568, \dodoi{10.1016/B0-08-043751-6/02015-6}

\bibitem[{S.~C. Millholland \& J.~N. Winn(2021)Millholland \& Winn}]{millholland_split_2021}
Millholland, S.~C., \& Winn, J.~N. 2021, \bibinfo{title}{Split {Peas} in a {Pod}: {Intra}-system {Uniformity} of {Super}-{Earths} and {Sub}-{Neptunes},} The Astrophysical Journal, 920, L34, \dodoi{10.3847/2041-8213/ac2c77}

\bibitem[{C. Mordasini \& R. Burn(2024)Mordasini \& Burn}]{mordasini_planet_2024}
Mordasini, C., \& Burn, R. 2024, Planet formation -- observational constraints, physical processes, and compositional patterns, arXiv.
\newblock \url{http://arxiv.org/abs/2404.15555}

\bibitem[{J.~I. Moses {et~al.}(2013)Moses, Line, Visscher, Richardson, Nettelmann, Fortney, Barman, Stevenson, \& Madhusudhan}]{moses_compositional_2013}
Moses, J.~I., Line, M.~R., Visscher, C., {et~al.} 2013, \bibinfo{title}{Compositional {Diversity} in the {Atmospheres} of {Hot} {Neptunes}, with {Application} to {GJ} 436b,} The Astrophysical Journal, 777, 34, \dodoi{10.1088/0004-637X/777/1/34}

\bibitem[{S. Mukherjee {et~al.}(2024)Mukherjee, Fortney, Wogan, Sing, \& Ohno}]{mukherjee_effects_2024}
Mukherjee, S., Fortney, J.~J., Wogan, N.~F., Sing, D.~K., \& Ohno, K. 2024, Effects of {Planetary} {Parameters} on {Disequilibrium} {Chemistry} in {Irradiated} {Planetary} {Atmospheres}: {From} {Gas} {Giants} to {Sub}-{Neptunes}, arXiv, \dodoi{10.48550/arXiv.2410.17169}

\bibitem[{R.~A. Murray-Clay {et~al.}(2009)Murray-Clay, Chiang, \& Murray}]{murray-clay_atmospheric_2009}
Murray-Clay, R.~A., Chiang, E.~I., \& Murray, N. 2009, \bibinfo{title}{Atmospheric {Escape} {From} {Hot} {Jupiters},} The Astrophysical Journal, 693, 23, \dodoi{10.1088/0004-637X/693/1/23}

\bibitem[{ {National Academies of Sciences, Engineering, and Medicine}(2021){National Academies of Sciences, Engineering, and Medicine}}]{AstroDecadal2020}
{National Academies of Sciences, Engineering, and Medicine}. 2021, {Pathways to Discovery in Astronomy and Astrophysics for the 2020s}, \dodoi{10.17226/26141}

\bibitem[{B.~K. {Nguyen} {et~al.}(2026){Nguyen}, {Schaefer}, {Ji}, {Theissen}, {Dai}, {Peng}, {Tang}, {Zorzi}, {Hill}, \& {Weiner Mansfield}}]{nguyen_shoreline_2026}
{Nguyen}, B.~K., {Schaefer}, L.~K., {Ji}, X., {et~al.} 2026, \bibinfo{title}{{An Evolving Cosmic Shoreline and Sandbar Bounding the Rocky Airless Valley},} arXiv e-prints, arXiv:2608.23912.
\newblock \doarXiv{2608.23912}

\bibitem[{H. Nicholls {et~al.}(2025)Nicholls, Lichtenberg, Chatterjee, Guimond, Postolec, \& Pierrehumbert}]{nicholls_volatile-rich_2025}
Nicholls, H., Lichtenberg, T., Chatterjee, R.~D., {et~al.} 2025, \bibinfo{title}{Volatile-rich evolution of molten super-{Earth} {L} 98-59 d,} \dodoi{10.48550/arXiv.2507.02656}

\bibitem[{H. {Nicholls} {et~al.}(2026){Nicholls}, {Lichtenberg}, {Chatterjee}, {Guimond}, {Postolec}, \& {Pierrehumbert}}]{Nicholls2026}
{Nicholls}, H., {Lichtenberg}, T., {Chatterjee}, R.~D., {et~al.} 2026, \bibinfo{title}{{Volatile-rich evolution of molten super-Earth L 98-59 d},} Nature Astronomy, 10, 809, \dodoi{10.1038/s41550-026-02815-8}

\bibitem[{H. Nicholls {et~al.}(2026)Nicholls, Lichtenberg, Riegler, Calder, \& Fortuin}]{nicholls_constraining_2026}
Nicholls, H., Lichtenberg, T., Riegler, B., Calder, R., \& Fortuin, V. 2026, Constraining the lives and times of exoplanets through evolutionary {Bayesian} retrievals, arXiv, \dodoi{10.48550/arXiv.2607.25845}

\bibitem[{P. Niraula {et~al.}(2017)Niraula, Redfield, Dai, Barragán, Gandolfi, Cauley, Hirano, Korth, Smith, Prieto-Arranz, Grziwa, Fridlund, Persson, Justesen, Winn, Albrecht, Cochran, Csizmadia, Duvvuri, Endl, Hatzes, Livingston, Narita, Nespral, Nowak, Pätzold, Palle, \& Van~Eylen}]{niraula_three_2017}
Niraula, P., Redfield, S., Dai, F., {et~al.} 2017, \bibinfo{title}{Three {Super}-{Earths} {Transiting} the {Nearby} {Star} {GJ} 9827,} The Astronomical Journal, 154, 266, \dodoi{10.3847/1538-3881/aa957c}

\bibitem[{M.~C. {Nixon} \& N. {Madhusudhan}(2020){Nixon} \& {Madhusudhan}}]{Nixon2020}
{Nixon}, M.~C., \& {Madhusudhan}, N. 2020, \bibinfo{title}{{Assessment of supervised machine learning for atmospheric retrieval of exoplanets},} \mnras, 496, 269, \dodoi{10.1093/mnras/staa1150}

\bibitem[{M.~C. {Nixon} \& N. {Madhusudhan}(2022){Nixon} \& {Madhusudhan}}]{Nixon2022}
{Nixon}, M.~C., \& {Madhusudhan}, N. 2022, \bibinfo{title}{{Aura-3D: A Three-dimensional Atmospheric Retrieval Framework for Exoplanet Transmission Spectra},} \apj, 935, 73, \dodoi{10.3847/1538-4357/ac7c09}

\bibitem[{M.~C. {Nixon} {et~al.}(2024){Nixon}, {Welbanks}, {McGill}, \& {Kempton}}]{Nixon2024}
{Nixon}, M.~C., {Welbanks}, L., {McGill}, P., \& {Kempton}, E. M.-R. 2024, \bibinfo{title}{{Methods for Incorporating Model Uncertainty into Exoplanet Atmospheric Analysis},} \apj, 966, 156, \dodoi{10.3847/1538-4357/ad354e}

\bibitem[{M.~C. Nixon {et~al.}(2025)Nixon, Somers, Savel, Ih, Kempton, Young, Schlichting, Lichtenberg, Welbanks, Misener, Piette, \& Wogan}]{nixon_magma_2025}
Nixon, M.~C., Somers, R.~S., Savel, A.~B., {et~al.} 2025, \bibinfo{title}{Magma ocean interactions can explain {JWST} observations of the sub-{Neptune} {TOI}-270 d,} \dodoi{10.48550/arXiv.2510.07367}

\bibitem[{K. Ohno {et~al.}(2024)Ohno, Schlawin, Bell, Murphy, Beatty, Welbanks, Greene, Fortney, Parmentier, Edelman, Mehta, \& Rieke}]{ohno_possible_2024}
Ohno, K., Schlawin, E., Bell, T.~J., {et~al.} 2024, \bibinfo{title}{A {Possible} {Metal}-{Dominated} {Atmosphere} {Below} the {Thick} {Aerosols} of {GJ} 1214 b {Suggested} by its {JWST} {Panchromatic} {Transmission} {Spectrum},} \dodoi{10.48550/arXiv.2410.10186}

\bibitem[{C.~W. {Ormel} {et~al.}(2021){Ormel}, {Vazan}, \& {Brouwers}}]{Ormel2021}
{Ormel}, C.~W., {Vazan}, A., \& {Brouwers}, M.~G. 2021, \bibinfo{title}{{How planets grow by pebble accretion. III. Emergence of an interior composition gradient},} \aap, 647, A175, \dodoi{10.1051/0004-6361/202039706}

\bibitem[{J.~E. Owen \& A.~P. Jackson(2012)Owen \& Jackson}]{owen_planetary_2012}
Owen, J.~E., \& Jackson, A.~P. 2012, \bibinfo{title}{Planetary evaporation by {UV} \&amp; {X}-ray radiation: basic hydrodynamics,} Monthly Notices of the Royal Astronomical Society, 425, 2931, \dodoi{10.1111/j.1365-2966.2012.21481.x}

\bibitem[{J.~E. Owen \& Y. Wu(2016)Owen \& Wu}]{owen_atmospheres_2016}
Owen, J.~E., \& Wu, Y. 2016, \bibinfo{title}{Atmospheres of {Low}-mass {Planets}: {The} "{Boil}-off",} The Astrophysical Journal, 817, 107, \dodoi{10.3847/0004-637X/817/2/107}

\bibitem[{J.~E. Owen \& Y. Wu(2017)Owen \& Wu}]{owen_evaporation_2017}
Owen, J.~E., \& Wu, Y. 2017, \bibinfo{title}{The {Evaporation} {Valley} in the {Kepler} {Planets},} The Astrophysical Journal, 847, 29, \dodoi{10.3847/1538-4357/aa890a}

\bibitem[{V.~M. Passegger {et~al.}(2024)Passegger, Mascareño, Allart, Hernández, Lovis, Lavie, Silva, Müller, Tabernero, Cristiani, Pepe, Rebolo, Santos, Adibekyan, Alibert, Prieto, Barros, Bouchy, Castro-González, D'Odorico, Dumusque, Di~Marcantonio, Ehrenreich, Figueira, Santos, Curto, Martins, Mehner, Micela, Molaro, Nari, Nunes, Pallé, Poretti, Rodrigues, Sousa, Sozzetti, Udry, \& Osorio}]{passegger_compact_2024}
Passegger, V.~M., Mascareño, A.~S., Allart, R., {et~al.} 2024, The compact multi-planet system {GJ} 9827 revisited with {ESPRESSO}, arXiv.
\newblock \url{http://arxiv.org/abs/2401.06276}

\bibitem[{B. {Paxton} {et~al.}(2011){Paxton}, {Bildsten}, {Dotter}, {Herwig}, {Lesaffre}, \& {Timmes}}]{Paxton2011}
{Paxton}, B., {Bildsten}, L., {Dotter}, A., {et~al.} 2011, \bibinfo{title}{{Modules for Experiments in Stellar Astrophysics (MESA)},} \apjs, 192, 3, \dodoi{10.1088/0067-0049/192/1/3}

\bibitem[{S. Pelletier {et~al.}(2021)Pelletier, Benneke, Darveau-Bernier, Boucher, Cook, Piaulet, Coulombe, Artigau, Lafrenière, Delisle, Allart, Doyon, Donati, Fouqué, Moutou, Cadieux, Delfosse, Hébrard, Martins, Martioli, \& Vandal}]{pelletier_where_2021}
Pelletier, S., Benneke, B., Darveau-Bernier, A., {et~al.} 2021, \bibinfo{title}{Where {Is} the {Water}? {Jupiter}-like {C}/{H} {Ratio} but {Strong} {H2O} {Depletion} {Found} on τ {Boötis} b {Using} {SPIRou},} The Astronomical Journal, 162, 73, \dodoi{10.3847/1538-3881/ac0428}

\bibitem[{B. Peng \& D. Valencia(2024)Peng \& Valencia}]{peng_puffy_2024}
Peng, B., \& Valencia, D. 2024, \bibinfo{title}{Puffy {Venuses}: the {Mass}-{Radius} {Impact} of {Carbon}-{Rich} {Atmospheres} on {Lava} {Worlds},} \url{http://arxiv.org/abs/2405.08998}

\bibitem[{C. Piaulet {et~al.}(2021)Piaulet, Benneke, Rubenzahl, Howard, Lee, Thorngren, Angus, Peterson, Schlieder, Werner, Kreidberg, Jaouni, Crossfield, Ciardi, Petigura, Livingston, Dressing, Fulton, Beichman, Christiansen, Gorjian, Hardegree-Ullman, Krick, \& Sinukoff}]{piaulet_wasp-107bs_2021}
Piaulet, C., Benneke, B., Rubenzahl, R.~A., {et~al.} 2021, \bibinfo{title}{{WASP}-107b’s {Density} {Is} {Even} {Lower}: {A} {Case} {Study} for the {Physics} of {Planetary} {Gas} {Envelope} {Accretion} and {Orbital} {Migration},} The Astronomical Journal, 161, 70, \dodoi{10.3847/1538-3881/abcd3c}

\bibitem[{C. Piaulet {et~al.}(2023)Piaulet, Benneke, Almenara, Dragomir, Knutson, Thorngren, Peterson, Crossfield, Kempton, Kubyshkina, Howard, Angus, Isaacson, Weiss, Beichman, Fortney, Fossati, Lammer, McCullough, Morley, \& Wong}]{piaulet_evidence_2023}
Piaulet, C., Benneke, B., Almenara, J.~M., {et~al.} 2023, \bibinfo{title}{Evidence for the volatile-rich composition of a 1.5-{Earth}-radius planet,} Nature Astronomy, 7, 206, \dodoi{10.1038/s41550-022-01835-4}

\bibitem[{C. Piaulet-Ghorayeb(2025)Piaulet-Ghorayeb}]{piaulet-ghorayeb_stctm_2025}
Piaulet-Ghorayeb, C. 2025, {STCTM}: a forward modeling and retrieval framework for stellar contamination and stellar spectra, arXiv, \dodoi{10.48550/arXiv.2508.19297}

\bibitem[{C. Piaulet-Ghorayeb {et~al.}(2025{\natexlab{a}})Piaulet-Ghorayeb, Thorngren, Kempton, Lipper, Rogers, Correa~Horta, \& Sun}]{piaulet-ghorayeb_window_2025}
Piaulet-Ghorayeb, C., Thorngren, D.~P., Kempton, E. M.-R., {et~al.} 2025{\natexlab{a}}, A window for water-hydrogen demixing on warm metal-rich sub-{Neptunes}, arXiv, \dodoi{10.48550/arXiv.2512.01805}

\bibitem[{C. Piaulet-Ghorayeb {et~al.}(2024)Piaulet-Ghorayeb, Benneke, Radica, Raul, Coulombe, Ahrer, Kubyshkina, Howard, Krissansen-Totton, MacDonald, Roy, Louca, Christie, Fournier-Tondreau, Allart, Miguel, Schlichting, Welbanks, Cadieux, Dorn, Evans-Soma, Fortney, Pierrehumbert, Lafrenière, Acuña, Komacek, Innes, Beatty, Cloutier, Doyon, Gagnebin, Gapp, \& Knutson}]{piaulet-ghorayeb_jwstniriss_2024}
Piaulet-Ghorayeb, C., Benneke, B., Radica, M., {et~al.} 2024, \bibinfo{title}{{JWST}/{NIRISS} {Reveals} the {Water}-rich "{Steam} {World}" {Atmosphere} of {GJ} 9827 d,} The Astrophysical Journal, 974, L10, \dodoi{10.3847/2041-8213/ad6f00}

\bibitem[{C. Piaulet-Ghorayeb {et~al.}(2025{\natexlab{b}})Piaulet-Ghorayeb, Benneke, Turbet, Moore, Roy, Lim, Doyon, Fauchez, Albert, Radica, Coulombe, Lafrenière, Cowan, Belzile, Musfirat, Kaur, L'Heureux, Johnstone, MacDonald, Allart, Dang, Kaltenegger, Pelletier, Rowe, Taylor, \& Turner}]{piaulet-ghorayeb_strict_2025}
Piaulet-Ghorayeb, C., Benneke, B., Turbet, M., {et~al.} 2025{\natexlab{b}}, \bibinfo{title}{Strict {Limits} on {Potential} {Secondary} {Atmospheres} on the {Temperate} {Rocky} {Exo}-{Earth} {TRAPPIST}-1 d,} The Astrophysical Journal, 989, 181, \dodoi{10.3847/1538-4357/adf207}

\bibitem[{R.~T. Pierrehumbert(2023)Pierrehumbert}]{pierrehumbert_runaway_2023}
Pierrehumbert, R.~T. 2023, \bibinfo{title}{The {Runaway} {Greenhouse} on {Sub}-{Neptune} {Waterworlds},} The Astrophysical Journal, 944, 20, \dodoi{10.3847/1538-4357/acafdf}

\bibitem[{A. Pinhas {et~al.}(2018)Pinhas, Rackham, Madhusudhan, \& Apai}]{pinhas_retrieval_2018}
Pinhas, A., Rackham, B.~V., Madhusudhan, N., \& Apai, D. 2018, \bibinfo{title}{Retrieval of planetary and stellar properties in transmission spectroscopy with {AURA},} Monthly Notices of the Royal Astronomical Society, 480, 5314, \dodoi{10.1093/mnras/sty2209}

\bibitem[{B. Plez(2012)Plez}]{plez_turbospectrum_2012}
Plez, B. 2012, \bibinfo{title}{Turbospectrum: {Code} for spectral synthesis,} Astrophysics Source Code Library, ascl:1205.004.
\newblock \url{https://ui.adsabs.harvard.edu/abs/2012ascl.soft05004P}

\bibitem[{M. {Plotnykov} \& D. {Valencia}(2020){Plotnykov} \& {Valencia}}]{Plotnykov2020}
{Plotnykov}, M., \& {Valencia}, D. 2020, \bibinfo{title}{{Chemical fingerprints of formation in rocky super-Earths' data},} \mnras, 499, 932, \dodoi{10.1093/mnras/staa2615}

\bibitem[{M. Plotnykov {et~al.}(2025)Plotnykov, Valencia, Ross, Reggiani, \& Schlaufman}]{plotnykov_evidence_2025}
Plotnykov, M., Valencia, D., Ross, A., Reggiani, H., \& Schlaufman, K.~C. 2025, Evidence of 1:1 slope between rocky {Super}-{Earths} and their host stars, \dodoi{10.3847/1538-4357/ae200b}

\bibitem[{O.~L. Polyansky {et~al.}(2018)Polyansky, Kyuberis, Zobov, Tennyson, Yurchenko, \& Lodi}]{ExoMol_H2O}
Polyansky, O.~L., Kyuberis, A.~A., Zobov, N.~F., {et~al.} 2018, \bibinfo{title}{{ExoMol molecular line lists XXX: a complete high-accuracy line list for water},} Mon. Not. R. Astron. Soc., 480, 2597, \dodoi{10.1093/mnras/sty1877}

\bibitem[{E. Postolec {et~al.}(2026)Postolec, Lichtenberg, Nicholls, Soucasse, \& van~der Tak}]{postolec_atmospheric_2026}
Postolec, E., Lichtenberg, T., Nicholls, H., Soucasse, L., \& van~der Tak, F. 2026, Atmospheric evolution through outgassing and escape on young molten rocky exoplanets, arXiv, \dodoi{10.48550/arXiv.2607.15011}

\bibitem[{J. Prieto-Arranz {et~al.}(2018)Prieto-Arranz, Palle, Gandolfi, Barragán, Guenther, Dai, Fridlund, Hirano, Livingston, Luque, Niraula, Persson, Redfield, Albrecht, Alonso, Antoniciello, Cabrera, Cochran, Csizmadia, Deeg, Eigmüller, Endl, Erikson, Everett, Fukui, Grziwa, Hatzes, Hidalgo, Hjorth, Korth, Lorenzo-Oliveira, Murgas, Narita, Nespral, Nowak, Pätzold, Montañez~Rodríguez, Rauer, Ribas, Smith, Trifonov, Van~Eylen, \& Winn}]{prieto-arranz_mass_2018}
Prieto-Arranz, J., Palle, E., Gandolfi, D., {et~al.} 2018, \bibinfo{title}{Mass determination of the 1:3:5 near-resonant planets transiting {GJ} 9827 ({K2}-135),} Astronomy and Astrophysics, 618, A116, \dodoi{10.1051/0004-6361/201832872}

\bibitem[{M. Radica(2024)Radica}]{radica_exotedrf_2024}
Radica, M. 2024, {exoTEDRF}: {An} {EXOplanet} {Transit} and {Eclipse} {Data} {Reduction} {Framework}, arXiv.
\newblock \url{http://arxiv.org/abs/2407.17541}

\bibitem[{M. Radica {et~al.}(2023)Radica, Welbanks, Espinoza, Taylor, Coulombe, Feinstein, Goyal, Scarsdale, Albert, Baghel, Bean, Blecic, Lafrenière, MacDonald, Zamyatina, Allart1, Artigau, Batalha, Cook, Cowan, Dang, Doyon, Fournier-Tondreau, Johnstone, Line, Moran, Mukherjee, Pelletier, Roy, Talens, Filippazzo, Pontoppidan, \& Volk}]{radica_awesome_2023}
Radica, M., Welbanks, L., Espinoza, N., {et~al.} 2023, \bibinfo{title}{Awesome {SOSS}: transmission spectroscopy of {WASP}-96b with {NIRISS}/{SOSS},} Monthly Notices of the Royal Astronomical Society, 524, 835, \dodoi{10.1093/mnras/stad1762}

\bibitem[{M. Radica {et~al.}(2024)Radica, Coulombe, Taylor, Albert, Allart, Benneke, Cowan, Dang, Lafrenière, Thorngren, Artigau, Doyon, Flagg, Johnstone, Pelletier, \& Roy}]{radica_muted_2024}
Radica, M., Coulombe, L.-P., Taylor, J., {et~al.} 2024, \bibinfo{title}{Muted {Features} in the {JWST} {NIRISS} {Transmission} {Spectrum} of {Hot} {Neptune} {LTT} 9779b,} The Astrophysical Journal, 962, L20, \dodoi{10.3847/2041-8213/ad20e4}

\bibitem[{M. Radica {et~al.}(2026)Radica, Taylor, Rotman, Blecic, Welbanks, Ahrer, Christie, Coulombe, Lowry, Murphy, Feinstein, Lafreniere, MacDonald, Mayne, Tsai, \& Zamyatina}]{radica_super-solar_2026}
Radica, M., Taylor, J., Rotman, Y., {et~al.} 2026, Super-{Solar} {Metallicity} and {Tentative} {Evidence} for {Photochemistry} on {WASP}-96b from {JWST} and {Ground}-{Based} {VLT} {Transmission} {Spectroscopy}, arXiv, \dodoi{10.48550/arXiv.2604.05049}

\bibitem[{S. Redfield {et~al.}(2024)Redfield, Batalha, Benneke, Biller, Espinoza, France, Konopacky, Kreidberg, Rauscher, \& Sing}]{redfield_report_2024}
Redfield, S., Batalha, N., Benneke, B., {et~al.} 2024, Report of the {Working} {Group} on {Strategic} {Exoplanet} {Initiatives} with {HST} and {JWST}, \dodoi{10.48550/arXiv.2404.02932}

\bibitem[{H. Rein \& S.-F. Liu(2012)Rein \& Liu}]{rein_rebound_2012}
Rein, H., \& Liu, S.-F. 2012, \bibinfo{title}{{REBOUND}: an open-source multi-purpose {N}-body code for collisional dynamics,} Astronomy and Astrophysics, 537, A128, \dodoi{10.1051/0004-6361/201118085}

\bibitem[{H. {Rein} \& D.~S. {Spiegel}(2015){Rein} \& {Spiegel}}]{rein_IAS15_2015}
{Rein}, H., \& {Spiegel}, D.~S. 2015, \bibinfo{title}{{IAS15: a fast, adaptive, high-order integrator for gravitational dynamics, accurate to machine precision over a billion orbits},} \mnras, 446, 1424, \dodoi{10.1093/mnras/stu2164}

\bibitem[{H. Rein \& D. Tamayo(2015)Rein \& Tamayo}]{rein_whfast_2015}
Rein, H., \& Tamayo, D. 2015, \bibinfo{title}{{WHFAST}: a fast and unbiased implementation of a symplectic {Wisdom}-{Holman} integrator for long-term gravitational simulations,} Monthly Notices of the Royal Astronomical Society, 452, 376, \dodoi{10.1093/mnras/stv1257}

\bibitem[{C. {Richard} {et~al.}(2012){Richard}, {Gordon}, {Rothman}, {Abel}, {Frommhold}, {Gustafsson}, {Hartmann}, {Hermans}, {Lafferty}, {Orton}, \& et~al.}]{Richard2012}
{Richard}, C., {Gordon}, I.~E., {Rothman}, L.~S., {et~al.} 2012, \bibinfo{title}{{New section of the HITRAN database: Collision-induced absorption (CIA)},} \jqsrt, 113, 1276, \dodoi{10.1016/j.jqsrt.2011.11.004}

\bibitem[{F.~E. Rigby {et~al.}(2025)Rigby, Madhusudhan, Sarkar, Pica-Ciamarra, Holmberg, \& Moses}]{rigby_jwst_2025}
Rigby, F.~E., Madhusudhan, N., Sarkar, S., {et~al.} 2025, \bibinfo{title}{A {JWST} {Transmission} {Spectrum} of the {Temperate} {Sub}-{Neptune} {TOI}-732 c,} The Astrophysical Journal, 995, L70, \dodoi{10.3847/2041-8213/ae247d}

\bibitem[{J.~E. Rodriguez {et~al.}(2018)Rodriguez, Vanderburg, Eastman, Mann, Crossfield, Ciardi, Latham, \& Quinn}]{rodriguez_system_2018}
Rodriguez, J.~E., Vanderburg, A., Eastman, J.~D., {et~al.} 2018, \bibinfo{title}{A {System} of {Three} {Super} {Earths} {Transiting} the {Late} {K}-{Dwarf} {GJ} 9827 at 30 pc,} The Astronomical Journal, 155, 72, \dodoi{10.3847/1538-3881/aaa292}

\bibitem[{J.~G. Rogers {et~al.}(2024)Rogers, Dorn, Aditya~Raj, Schlichting, \& Young}]{rogers_most_2024}
Rogers, J.~G., Dorn, C., Aditya~Raj, V., Schlichting, H.~E., \& Young, E.~D. 2024, Most {Super}-{Earths} {Have} {Less} {Than} 3\% {Water}, \dodoi{10.48550/arXiv.2409.17394}

\bibitem[{J.~G. Rogers {et~al.}(2021)Rogers, Gupta, Owen, \& Schlichting}]{rogers_photoevaporation_2021}
Rogers, J.~G., Gupta, A., Owen, J.~E., \& Schlichting, H.~E. 2021, \bibinfo{title}{Photoevaporation versus core-powered mass-loss: model comparison with the {3D} radius gap,} Monthly Notices of the Royal Astronomical Society, 508, 5886, \dodoi{10.1093/mnras/stab2897}

\bibitem[{J.~G. Rogers {et~al.}(2023)Rogers, Schlichting, \& Owen}]{rogers_conclusive_2023}
Rogers, J.~G., Schlichting, H.~E., \& Owen, J.~E. 2023, Conclusive evidence for a population of water-worlds around {M}-dwarfs remains elusive, arXiv.
\newblock \url{http://arxiv.org/abs/2301.04321}

\bibitem[{L.~A. Rogers \& S. Seager(2010)Rogers \& Seager}]{rogers_framework_2010}
Rogers, L.~A., \& Seager, S. 2010, \bibinfo{title}{A {Framework} for {Quantifying} the {Degeneracies} of {Exoplanet} {Interior} {Compositions},} The Astrophysical Journal, 712, 974, \dodoi{10.1088/0004-637X/712/2/974}

\bibitem[{L.~S. {Rothman} {et~al.}(2010){Rothman}, {Gordon}, {Barber}, {Dothe}, {Gamache}, {Goldman}, {Perevalov}, {Tashkun}, \& {Tennyson}}]{Rothman2010}
{Rothman}, L.~S., {Gordon}, I.~E., {Barber}, R.~J., {et~al.} 2010, \bibinfo{title}{{HITEMP, the high-temperature molecular spectroscopic database},} \jqsrt, 111, 2139, \dodoi{10.1016/j.jqsrt.2010.05.001}

\bibitem[{P.-A. Roy {et~al.}(2023)Roy, Benneke, Piaulet, Gully-Santiago, Crossfield, Morley, Kreidberg, Mikal-Evans, Brande, Delisle, Greene, Hardegree-Ullman, Barman, Christiansen, Dragomir, Fortney, Howard, Kosiarek, \& Lothringer}]{roy_water_2023}
Roy, P.-A., Benneke, B., Piaulet, C., {et~al.} 2023, \bibinfo{title}{Water {Absorption} in the {Transmission} {Spectrum} of the {Water} {World} {Candidate} {GJ} 9827 d,} The Astrophysical Journal Letters, 954, L52, \dodoi{10.3847/2041-8213/acebf0}

\bibitem[{P.-A. Roy {et~al.}(2025)Roy, Benneke, Fournier-Tondreau, Coulombe, Piaulet-Ghorayeb, Lafrenière, Allart, Cowan, Dang, Johnstone, Langeveld, Pelletier, Radica, Taylor, Albert, Doyon, Flagg, Jayawardhana, MacDonald, \& Turner}]{roy_diversity_2025}
Roy, P.-A., Benneke, B., Fournier-Tondreau, M., {et~al.} 2025, \bibinfo{title}{Diversity in the haziness and chemistry of temperate sub-{Neptunes},} Nature Astronomy, \dodoi{10.1038/s41550-025-02723-3}

\bibitem[{T. {Ryabchikova} {et~al.}(2015){Ryabchikova}, {Piskunov}, {Kurucz}, {Stempels}, {Heiter}, {Pakhomov}, \& {Barklem}}]{Ryabchikova_2015}
{Ryabchikova}, T., {Piskunov}, N., {Kurucz}, R.~L., {et~al.} 2015, \bibinfo{title}{{A major upgrade of the VALD database},} \physscr, 90, 054005, \dodoi{10.1088/0031-8949/90/5/054005}

\bibitem[{E. Schlawin {et~al.}(2024)Schlawin, Ohno, Bell, Murphy, Welbanks, Beatty, Greene, Fortney, Parmentier, Edelman, Gill, Anderson, Wheatley, Henry, Mehta, Kreidberg, \& Rieke}]{schlawin_possible_2024}
Schlawin, E., Ohno, K., Bell, T.~J., {et~al.} 2024, \bibinfo{title}{Possible {Carbon} {Dioxide} {Above} the {Thick} {Aerosols} of {GJ} 1214 b,} \dodoi{10.48550/arXiv.2410.10183}

\bibitem[{H.~E. {Schlichting} \& S. {Mukhopadhyay}(2018){Schlichting} \& {Mukhopadhyay}}]{Schlichting2018}
{Schlichting}, H.~E., \& {Mukhopadhyay}, S. 2018, \bibinfo{title}{{Atmosphere Impact Losses},} \ssr, 214, 34, \dodoi{10.1007/s11214-018-0471-z}

\bibitem[{H.~E. Schlichting \& E.~D. Young(2022)Schlichting \& Young}]{schlichting_chemical_2022}
Schlichting, H.~E., \& Young, E.~D. 2022, \bibinfo{title}{Chemical {Equilibrium} between {Cores}, {Mantles}, and {Atmospheres} of {Super}-{Earths} and {Sub}-{Neptunes} and {Implications} for {Their} {Compositions}, {Interiors}, and {Evolution},} The Planetary Science Journal, 3, 127, \dodoi{10.3847/PSJ/ac68e6}

\bibitem[{A. Sharma {et~al.}(2024)Sharma, Stonkutė, Drazdauska, Minkevičiūtė, Mikolaitis, Tautvaišienė, \& Narbuntas}]{sharma_chemical_2024}
Sharma, A., Stonkutė, E., Drazdauska, A., {et~al.} 2024, \bibinfo{title}{Chemical composition of planetary hosts: {C}, {N}, and α-element abundances,} Astronomy and Astrophysics, 691, A160, \dodoi{10.1051/0004-6361/202451889}

\bibitem[{S.~W. Sharpe {et~al.}(2004)Sharpe, Johnson, Sams, Chu, Rhoderick, \& Johnson}]{sharpe_gas-phase_2004}
Sharpe, S.~W., Johnson, T.~J., Sams, R.~L., {et~al.} 2004, \bibinfo{title}{Gas-{Phase} {Databases} for {Quantitative} {Infrared} {Spectroscopy},} Applied Spectroscopy, 58, 1452, \dodoi{10.1366/0003702042641281}

\bibitem[{O. Shorttle {et~al.}(2024)Shorttle, Jordan, Nicholls, Lichtenberg, \& Bower}]{shorttle_distinguishing_2024}
Shorttle, O., Jordan, S., Nicholls, H., Lichtenberg, T., \& Bower, D.~J. 2024, \bibinfo{title}{Distinguishing {Oceans} of {Water} from {Magma} on {Mini}-{Neptune} {K2}-18b,} The Astrophysical Journal, 962, L8, \dodoi{10.3847/2041-8213/ad206e}

\bibitem[{J. Skilling(2004)Skilling}]{skilling_nested_2004}
Skilling, J. 2004, \bibinfo{title}{Nested {Sampling},} 735, 395, \dodoi{10.1063/1.1835238}

\bibitem[{J. Skilling(2006)Skilling}]{skilling_nested_2006}
Skilling, J. 2006, \bibinfo{title}{Nested sampling for general {Bayesian} computation,} Bayesian Analysis, 1, 833, \dodoi{10.1214/06-BA127}

\bibitem[{N.~V. Solomatova \& R. Caracas(2021)Solomatova \& Caracas}]{solomatova_genesis_2021}
Solomatova, N.~V., \& Caracas, R. 2021, \bibinfo{title}{Genesis of a {CO2}-rich and {H2O}-depleted atmosphere from {Earth}'s early global magma ocean,} Science Advances, 7, eabj0406, \dodoi{10.1126/sciadv.abj0406}

\bibitem[{V. Sparrman \& A. Johansen(2026)Sparrman \& Johansen}]{sparrman_when_2026}
Sparrman, V., \& Johansen, A. 2026, When the core ate the atmosphere, arXiv, \dodoi{10.48550/arXiv.2610.27788}

\bibitem[{M.-L. Steinmeyer {et~al.}(2026)Steinmeyer, Dorn, Werlen, \& Grimm}]{steinmeyer_coupled_2026}
Steinmeyer, M.-L., Dorn, C., Werlen, A., \& Grimm, S.~L. 2026, \bibinfo{title}{Coupled thermal-chemical evolution models of sub-{Neptunes} reveal atmospheric signatures of their formation location,} \dodoi{10.48550/arXiv.2601.21377}

\bibitem[{Q. Sun(2026)Sun}]{sun_jwst_2026}
Sun, Q. 2026, \bibinfo{title}{{JWST} {Exoplanetary} {Worlds} and {Elemental} {Survey} ({JEWELS}). {I}. {High}-precision {Chemical} {Abundances} of 20 {FGK} {Planet}-hosting {Stars} from {JWST} {Cycle} 2,} The Astrophysical Journal Supplement Series, 282, 37, \dodoi{10.3847/1538-4365/ae2a29}

\bibitem[{D. {Tamayo} {et~al.}(2020){Tamayo}, {Cranmer}, {Hadden}, {Rein}, {Battaglia}, {Obertas}, {Armitage}, {Ho}, {Spergel}, {Gilbertson}, {Hussain}, {Silburt}, {Jontof-Hutter}, \& {Menou}}]{Tamayo2020}
{Tamayo}, D., {Cranmer}, M., {Hadden}, S., {et~al.} 2020, \bibinfo{title}{{Predicting the long-term stability of compact multi-planet systems},} in AAS/Division of Dynamical Astronomy Meeting, Vol.~52, AAS/Division of Dynamical Astronomy Meeting \#51, 300.01

\bibitem[{Y. Tang {et~al.}(2024)Tang, Fortney, Nimmo, Thorngren, Ohno, \& Murray-Clay}]{tang_reassessing_2024}
Tang, Y., Fortney, J.~J., Nimmo, F., {et~al.} 2024, Reassessing {Sub}-{Neptune} {Structure}, {Radii}, and {Thermal} {Evolution}, arXiv.
\newblock \url{http://arxiv.org/abs/2410.21584}

\bibitem[{J. {Terragni} {et~al.}(2025){Terragni}, {Gordon}, {Adkins}, {Boulet}, {Campargue}, {Chistikov}, {Finenko}, {Finkenzeller}, {Fleurbaey}, {Hargreaves}, {Hanson}, {Hartmann}, {Klingberg}, {Kohler}, {Koroleva}, {Mondelain}, {Piccioni}, {Stefani}, {Strand}, {Tran}, {Turbet}, {Vigasin}, {Vitali}, {Volkamer}, \& {Wei}}]{Terragni2025}
{Terragni}, J., {Gordon}, I.~E., {Adkins}, E.~M., {et~al.} 2025, \bibinfo{title}{{Collision induced absorption in HITRAN2024: Enhanced and improved data for atmospheric and planetary studies},} \jqsrt, 347, 109631, \dodoi{10.1016/j.jqsrt.2025.109631}

\bibitem[{R. Townsend \& A. Lopez(2023)Townsend \& Lopez}]{townsend_msg_2023}
Townsend, R., \& Lopez, A. 2023, \bibinfo{title}{{MSG}: {A} software package for interpolating stellar spectra in pre-calculated grids,} Journal of Open Source Software, 8, 4602, \dodoi{10.21105/joss.04602}

\bibitem[{S.-M. Tsai {et~al.}(2023)Tsai, Lee, Powell, Gao, Zhang, Moses, Hébrard, Venot, Parmentier, Jordan, Hu, Alam, Alderson, Batalha, Bean, Benneke, Bierson, Brady, Carone, Carter, Chubb, Inglis, Leconte, Line, López-Morales, Miguel, Molaverdikhani, Rustamkulov, Sing, Stevenson, Wakeford, Yang, Aggarwal, Baeyens, Barat, de~Val-Borro, Daylan, Fortney, France, Goyal, Grant, Kirk, Kreidberg, Louca, Moran, Mukherjee, Nasedkin, Ohno, Rackham, Redfield, Taylor, Tremblin, Visscher, Wallack, Welbanks, Youngblood, Ahrer, Batalha, Behr, Berta-Thompson, Blecic, Casewell, Crossfield, Crouzet, Cubillos, Decin, Désert, Feinstein, Gibson, Harrington, Heng, Henning, Kempton, Krick, Lagage, Lendl, Lothringer, Mansfield, Mayne, Mikal-Evans, Palle, Schlawin, Shorttle, Wheatley, \& Yurchenko}]{tsai_photochemically_2023}
Tsai, S.-M., Lee, E. K.~H., Powell, D., {et~al.} 2023, \bibinfo{title}{Photochemically produced {SO2} in the atmosphere of {WASP}-39b,} Nature, 617, 483, \dodoi{10.1038/s41586-023-05902-2}

\bibitem[{J.~M. Tucker \& S. Mukhopadhyay(2014)Tucker \& Mukhopadhyay}]{tucker_evidence_2014}
Tucker, J.~M., \& Mukhopadhyay, S. 2014, \bibinfo{title}{Evidence for multiple magma ocean outgassing and atmospheric loss episodes from mantle noble gases,} Earth and Planetary Science Letters, 393, 254, \dodoi{10.1016/j.epsl.2014.02.050}

\bibitem[{D.~S. {Underwood} {et~al.}(2016{\natexlab{a}}){Underwood}, {Tennyson}, {Yurchenko}, {Huang}, {Schwenke}, {Lee}, {Clausen}, \& {Fateev}}]{Exomol_SO2}
{Underwood}, D.~S., {Tennyson}, J., {Yurchenko}, S.~N., {et~al.} 2016{\natexlab{a}}, \bibinfo{title}{{ExoMol molecular line lists - XIV. The rotation-vibration spectrum of hot SO$_{2}$},} \mnras, 459, 3890, \dodoi{10.1093/mnras/stw849}

\bibitem[{D.~S. {Underwood} {et~al.}(2016{\natexlab{b}}){Underwood}, {Tennyson}, {Yurchenko}, {Huang}, {Schwenke}, {Lee}, {Clausen}, \& {Fateev}}]{Underwood2016}
{Underwood}, D.~S., {Tennyson}, J., {Yurchenko}, S.~N., {et~al.} 2016{\natexlab{b}}, \bibinfo{title}{{ExoMol molecular line lists - XIV. The rotation-vibration spectrum of hot SO$_{2}$},} \mnras, 459, 3890, \dodoi{10.1093/mnras/stw849}

\bibitem[{M. Valatsou {et~al.}(2026)Valatsou, Dorn, Marty, \& Owen}]{valatsou_oxygen_2026}
Valatsou, M., Dorn, C., Marty, P., \& Owen, J.~E. 2026, Oxygen left behind: {Atmospheric} {Enrichment} due to {Fractionation} in {Sub}-{Neptunes} using {BOREAS}, arXiv, \dodoi{10.48550/arXiv.2602.12201}

\bibitem[{G. {Van Looveren} {et~al.}(2025){Van Looveren}, {Boro Saikia}, {Herbort}, {Schleich}, {G{\"u}del}, {Johnstone}, \& {Kislyakova}}]{van_looveren_hazard_2025}
{Van Looveren}, G., {Boro Saikia}, S., {Herbort}, O., {et~al.} 2025, \bibinfo{title}{{Habitable Zone and Atmosphere Retention Distance (HaZARD): Stellar-evolution-dependent loss models of secondary atmospheres},} \aap, 694, A310, \dodoi{10.1051/0004-6361/202452998}

\bibitem[{J. Venturini {et~al.}(2020)Venturini, Guilera, Haldemann, Ronco, \& Mordasini}]{venturini_nature_2020}
Venturini, J., Guilera, O.~M., Haldemann, J., Ronco, M.~P., \& Mordasini, C. 2020, \bibinfo{title}{The nature of the radius valley - {Hints} from formation and evolution models,} Astronomy \& Astrophysics, 643, L1, \dodoi{10.1051/0004-6361/202039141}

\bibitem[{J. Venturini {et~al.}(2024)Venturini, Ronco, Guilera, Haldemann, Mordasini, \& Bertolami}]{venturini_fading_2024}
Venturini, J., Ronco, M.~P., Guilera, O.~M., {et~al.} 2024, \bibinfo{title}{A fading radius valley towards {M}-dwarfs, a persistent density valley across stellar types,} \url{http://arxiv.org/abs/2404.01967}

\bibitem[{N.~L. Wallack {et~al.}(2024)Wallack, Batalha, Alderson, Scarsdale, Redai, Aguichine, Alam, Gao, Wolfgang, Batalha, Kirk, López-Morales, Moran, Teske, Wakeford, \& Wogan}]{wallack_jwst_2024}
Wallack, N.~L., Batalha, N.~E., Alderson, L., {et~al.} 2024, \bibinfo{title}{{JWST} {COMPASS}: {A} {NIRSpec}/{G395H} {Transmission} {Spectrum} of the {Sub}-{Neptune} {TOI}-836c,} \url{http://arxiv.org/abs/2404.01264}

\bibitem[{S. Wang {et~al.}(2025)Wang, Pan, Dong, Zhao, \& Ji}]{wang_formation_2025}
Wang, S., Pan, M., Dong, Y., Zhao, G., \& Ji, J. 2025, Formation of {Ultra}-short-period {Planet} in {Hot} {Jupiter} {Systems}: {Application} to {WASP}-47, arXiv, \dodoi{10.48550/arXiv.2503.00872}

\bibitem[{A.~J. Watson {et~al.}(1981)Watson, Donahue, \& Walker}]{watson_dynamics_1981}
Watson, A.~J., Donahue, T.~M., \& Walker, J. C.~G. 1981, \bibinfo{title}{The dynamics of a rapidly escaping atmosphere - {Applications} to the evolution of earth and {Venus},} Icarus, 48, 150, \dodoi{10.1016/0019-1035(81)90101-9}

\bibitem[{D. {Weisserman} {et~al.}(2026){Weisserman}, {Gromek}, {Cloutier}, {Bali}, {Cadieux}, {Plotnykov}, {L'Heureux}, {Srivastava}, {Carmona}, {Frensch}, {Artigau}, {Baron}, {Barros}, {Benneke}, {Bonfils}, {Bouchy}, {Bryan}, {Cook}, {Cowan}, {Cristo}, {Delfosse}, {Doyon}, {Dumusque}, {Ehrenreich}, {Gonz{\'a}lez Hern{\'a}ndez}, {Lafreni{\`e}re}, {de Castro Le{\~a}o}, {Lovis}, {Malo}, {Canto Martins}, {Su{\'a}rez Mascare{\~n}o}, {De Medeiros}, {Melo}, {Mignon}, {Mordasini}, {Pepe}, {Rebolo}, {Rowe}, {Santos}, {S{\'e}gransan}, {Udry}, {Valencia}, {Wade}, {Aguiar}, {Allart}, {Bazinet}, {Delisle}, {B{\'e}langer}, {Blackman}, {Bourrier}, {Branco}, {Bruniquel}, {Carteret}, {Cointepas}, {Darveau-Bernier}, {Dauplaise}, {Delgado-Mena}, {Dorn}, {Doshi}, {Faria}, {Fontinele}, {Forveille}, {Gagn{\'e}}, {Genest}, {Glover}, {de Lima Gomes}, {Grieves}, {Hobson}, {Hoeijmakers}, {Jahandar}, {Krishnamurthy}, {Lamontagne}, {Larue}, {Leath}, {Lim}, {Lipper}, {Messamah}, {Messias}, {Monteiro}, {Moranta}, {Al Moulla}, {Mounzer},
  {Mraz}, {Nari}, {Nielsen}, {Osborn}, {Otegi}, {Parc}, {Pelletier}, {Pereira}, {Piaulet-Ghorayeb}, {Rosener}, {Seidel}, {Gomes da Silva}, {Costa Silva}, {Stefanov}, {Teixeira}, {Vandal}, {Vaulato}, {Wardenier}, \& {Yariv}}]{Weisserman2026}
{Weisserman}, D., {Gromek}, N., {Cloutier}, R., {et~al.} 2026, \bibinfo{title}{{Super-Earth masses and stellar abundances from NIRPS reveal tentative evidence for water-rich formation around M dwarfs},} \aap, 709, A165, \dodoi{10.1051/0004-6361/202558011}

\bibitem[{L. {Welbanks} \& N. {Madhusudhan}(2019){Welbanks} \& {Madhusudhan}}]{Welbanks2019a}
{Welbanks}, L., \& {Madhusudhan}, N. 2019, \bibinfo{title}{{On Degeneracies in Retrievals of Exoplanetary Transmission Spectra},} \aj, 157, 206, \dodoi{10.3847/1538-3881/ab14de}

\bibitem[{L. {Welbanks} \& N. {Madhusudhan}(2021){Welbanks} \& {Madhusudhan}}]{Welbanks2021}
{Welbanks}, L., \& {Madhusudhan}, N. 2021, \bibinfo{title}{{Aurora: A Generalized Retrieval Framework for Exoplanetary Transmission Spectra},} \apj, 913, 114, \dodoi{10.3847/1538-4357/abee94}

\bibitem[{A. Werlen {et~al.}(2025)Werlen, Dorn, Schlichting, Grimm, \& Young}]{werlen_atmospheric_2025}
Werlen, A., Dorn, C., Schlichting, H.~E., Grimm, S.~L., \& Young, E.~D. 2025, \bibinfo{title}{Atmospheric {C}/{O} {Ratios} of {Sub}-{Neptunes} with {Magma} {Oceans}: {Homemade} rather than {Inherited},} \dodoi{10.48550/arXiv.2504.20450}

\bibitem[{A. Werlen {et~al.}(2026)Werlen, Young, Schlichting, Dorn, \& Shahar}]{werlen_effects_2026}
Werlen, A., Young, E.~D., Schlichting, H.~E., Dorn, C., \& Shahar, A. 2026, \bibinfo{title}{The {Effects} of {Non}-ideal {Mixing} in {Planetary} {Magma} {Oceans} and {Atmospheres},} The Astrophysical Journal, 999, 178, \dodoi{10.3847/1538-4357/ae434d}

\bibitem[{N.~F. Wogan {et~al.}(2024)Wogan, Batalha, Zahnle, Krissansen-Totton, Tsai, \& Hu}]{wogan_jwst_2024}
Wogan, N.~F., Batalha, N.~E., Zahnle, K., {et~al.} 2024, \bibinfo{title}{{JWST} observations of {K2}-18b can be explained by a gas-rich mini-{Neptune} with no habitable surface,} \dodoi{10.48550/arXiv.2401.11082}

\bibitem[{I. Wong {et~al.}(2020)Wong, Benneke, Shporer, Fetherolf, Kane, Ricker, Vanderspek, Seager, Winn, Collins, Mireles, Morris, Tenenbaum, Ting, Rinehart, \& Villaseñor}]{wong_tess_2020}
Wong, I., Benneke, B., Shporer, A., {et~al.} 2020, \bibinfo{title}{{TESS} {Phase} {Curve} of the {Hot} {Jupiter} {WASP}-19b,} The Astronomical Journal, 159, 104, \dodoi{10.3847/1538-3881/ab6d6e}

\bibitem[{B.~J. Wood {et~al.}(2013)Wood, Li, \& Shahar}]{wood_carbon_2013}
Wood, B.~J., Li, J., \& Shahar, A. 2013, \bibinfo{title}{Carbon in the {Core}: {Its} {Influence} on the {Properties} of {Core} and {Mantle},} Reviews in Mineralogy and Geochemistry, 75, 231, \dodoi{10.2138/rmg.2013.75.8}

\bibitem[{B.~J. {Wood} {et~al.}(2013){Wood}, {Li}, \& {Shahar}}]{Wood2013}
{Wood}, B.~J., {Li}, J., \& {Shahar}, A. 2013, \bibinfo{title}{{Carbon in the Core: Its Influence on the Properties of Core and Mantle},} Reviews in Mineralogy and Geochemistry, 75, 231, \dodoi{10.2138/rmg.2013.75.8}

\bibitem[{R. {Wordsworth} \& L. {Kreidberg}(2022){Wordsworth} \& {Kreidberg}}]{Wordsworth2022}
{Wordsworth}, R., \& {Kreidberg}, L. 2022, \bibinfo{title}{{Atmospheres of Rocky Exoplanets},} \araa, 60, 159, \dodoi{10.1146/annurev-astro-052920-125632}

\bibitem[{M.~C. {Wyatt} {et~al.}(2020){Wyatt}, {Kral}, \& {Sinclair}}]{Wyatt2020}
{Wyatt}, M.~C., {Kral}, Q., \& {Sinclair}, C.~A. 2020, \bibinfo{title}{{Susceptibility of planetary atmospheres to mass-loss and growth by planetesimal impacts: the impact shoreline},} \mnras, 491, 782, \dodoi{10.1093/mnras/stz3052}

\bibitem[{J. Yang \& R. Hu(2024)Yang \& Hu}]{yang_chemical_2024}
Yang, J., \& Hu, R. 2024, \bibinfo{title}{Chemical mapping of temperate sub-{Neptune} atmospheres: {Constraining} the deep-interior {H2O}/{H2} using the atmospheric {CO2}/{CH4},} \url{http://arxiv.org/abs/2406.01955}

\bibitem[{J. Yang {et~al.}(2026)Yang, Kempton, \& Savel}]{yang_sub-neptunes_2026}
Yang, J., Kempton, E. M.-R., \& Savel, A.~B. 2026, \bibinfo{title}{Sub-{Neptunes} as {Soot} {Factories}: {Deep} {Atmosphere} {Hydrocarbon} {Formation} and {Quenching} as the {Origin} of {Sub}-{Neptune} {Aerosol} {Trends},} \dodoi{10.48550/arXiv.2604.11919}

\bibitem[{R.~V. Yelle(2004)Yelle}]{yelle_aeronomy_2004}
Yelle, R.~V. 2004, \bibinfo{title}{Aeronomy of extra-solar giant planets at small orbital distances,} Icarus, 170, 167, \dodoi{10.1016/j.icarus.2004.02.008}

\bibitem[{E.~D. Young {et~al.}(2023)Young, Shahar, \& Schlichting}]{young_earth_2023}
Young, E.~D., Shahar, A., \& Schlichting, H.~E. 2023, \bibinfo{title}{Earth shaped by primordial {H2} atmospheres,} Nature, 616, 306, \dodoi{10.1038/s41586-023-05823-0}

\bibitem[{E.~D. Young {et~al.}(2025)Young, Werlen, Marcum, Stixrude, \& Dullemond}]{young_differentiation_2025}
Young, E.~D., Werlen, A., Marcum, S.~P., Stixrude, L., \& Dullemond, C.~P. 2025, \bibinfo{title}{Differentiation, the {Exception}, {Not} the {Rule}: {Evidence} for {Full} {Miscibility} in {Sub}-{Neptune} {Interiors},} The Planetary Science Journal, 6, 251, \dodoi{10.3847/PSJ/ae1012}

\bibitem[{S.~N. {Yurchenko} {et~al.}(2011){Yurchenko}, {Barber}, \& {Tennyson}}]{Yurchenko2011}
{Yurchenko}, S.~N., {Barber}, R.~J., \& {Tennyson}, J. 2011, \bibinfo{title}{{A variationally computed line list for hot NH$_{3}$},} \mnras, 413, 1828, \dodoi{10.1111/j.1365-2966.2011.18261.x}

\bibitem[{S.~N. {Yurchenko} {et~al.}(2020){Yurchenko}, {Mellor}, {Freedman}, \& {Tennyson}}]{Yurchenko2020}
{Yurchenko}, S.~N., {Mellor}, T.~M., {Freedman}, R.~S., \& {Tennyson}, J. 2020, \bibinfo{title}{{ExoMol line lists - XXXIX. Ro-vibrational molecular line list for CO$_{2}$},} \mnras, 496, 5282, \dodoi{10.1093/mnras/staa1874}

\bibitem[{S.~N. {Yurchenko} {et~al.}(2024){Yurchenko}, {Owens}, {Kefala}, \& {Tennyson}}]{Yurchenko2024}
{Yurchenko}, S.~N., {Owens}, A., {Kefala}, K., \& {Tennyson}, J. 2024, \bibinfo{title}{{ExoMol line lists - LVII. High accuracy ro-vibrational line list for methane (CH$_{4}$)},} \mnras, 528, 3719, \dodoi{10.1093/mnras/stae148}

\bibitem[{S.~N. {Yurchenko} \& J. {Tennyson}(2014){Yurchenko} \& {Tennyson}}]{Yurchenko2014}
{Yurchenko}, S.~N., \& {Tennyson}, J. 2014, \bibinfo{title}{{ExoMol line lists - IV. The rotation-vibration spectrum of methane up to 1500 K},} \mnras, 440, 1649, \dodoi{10.1093/mnras/stu326}

\bibitem[{S.~N. {Yurchenko} {et~al.}(2026){Yurchenko}, {Barnfield}, {Bowesman}, {Brady}, {Guest}, {Kefala}, {Ni}, {Perri}, {Smola}, {Solokov}, {Tao}, \& {Tennyson}}]{Yurchenko2026}
{Yurchenko}, S.~N., {Barnfield}, M.~G., {Bowesman}, C.~A., {et~al.} 2026, \bibinfo{title}{{ExoMol line lists ─ LXIII. ExoMol line lists for 12 isotopologues of CO$_{2}$},} \mnras, 545, staf2135, \dodoi{10.1093/mnras/staf2135}

\bibitem[{E. Zlimen {et~al.}(2026)Zlimen, Schlichting, \& Rogers}]{zlimen_reconciling_2026}
Zlimen, E., Schlichting, H.~E., \& Rogers, J.~G. 2026, Reconciling {Giant} {Impacts} with the {Radius} {Valley}: {The} {Role} of {Silicate}-{Hydrogen} {Miscibility} for {Sub}-{Neptunes}, arXiv, \dodoi{10.48550/arXiv.2610.38852}

\end{thebibliography}

\end{document}